\documentclass[universe,review,submit,moreauthors,pdftex]{Definitions/mdpi_mod}

\firstpage{1} 
\pubvolume{1}
\issuenum{1}
\articlenumber{0}
\pubyear{2021}
\copyrightyear{2020}
\datereceived{} 
\dateaccepted{} 
\datepublished{} 
\hreflink{https://doi.org/} % If needed use \linebreak
\usepackage{slashed}
\usepackage[T1]{fontenc}

\newcommand{\f}[2]{\frac{#1}{#2}}

\newcommand{\la}{\langle}

\newcommand{\ra}{\rangle}

\newcommand{\Oc}{{\cal O}}
\newcommand{\tr}{{\rm tr}}
\renewcommand{\Re}{{\rm Re}\,}

\newcommand{\de}{\partial}

\renewcommand{\linenumbers}{}

\makeatletter
\def\blfootnotetext{\gdef\@thefnmark{*}\@footnotetext}
\makeatother

\Title{Localization of Dirac Fermions in Finite-Temperature Gauge Theory}

\TitleCitation{Localization of Dirac Fermions in Finite-Temperature Gauge Theory}

\Author{Matteo Giordano $^{1,}$*\orcidA{} and Tam\'as G.~Kov\'acs $^{1,2}$\orcidB{}}

\AuthorNames{Matteo Giordano and and Tam\'as G.~Kov\'acs}

\AuthorCitation{Giordano, M.; Kov\'acs, T.G.}
\address{%
  $^{1}$ \quad ELTE E\"otv\"os Lor\'and University, Institute for
  Theoretical Physics, P\'azm\'any P\'eter s\'et\'any 1/A, H-1117,
  Budapest,
  Hungary\\
  $^{2}$ \quad Institute for Nuclear Research, Bem t\'er 18/c, H-4026,
  Debrecen, Hungary}

\corres{Correspondence: giordano@bodri.elte.hu}

\abstract{It is by now well established that Dirac fermions coupled to
non-Abelian gauge theories can undergo an Anderson-type localization
transition. This transition affects eigenmodes in the lowest part of
the Dirac spectrum, the ones most relevant to the low-energy physics
of these models. Here we review several aspects of this phenomenon,
mostly using the tools of lattice gauge theory. In particular, we
discuss how the transition is related to the finite-temperature
transitions leading to the deconfinement of fermions, as well as to
the restoration of chiral symmetry that is spontaneously broken at low
temperature. Other topics we touch upon are the universality of the
transition, and its connection to topological excitations (instantons)
of the gauge field and the associated fermionic zero modes. While the
main focus is on Quantum Chromodynamics, we also discuss how the
localization transition appears in other related models with different
fermionic contents (including the quenched approximation), gauge
groups, and in different space-time dimensions. Finally, we offer some
speculations about the physical relevance of the localization
transition in these models.} 


\begin{document}

\section{Introduction}

Quantum Chromodynamics (QCD) is currently our best microscopic
description of strong interactions. As is well known, QCD is a gauge
theory with gauge group SU(3), coupling six ``flavors'' of {\it
  quarks}, which are spin-$\f{1}{2}$ Dirac fermions transforming in
the fundamental representation of the group, to the eight spin-1 gauge
bosons (known as {\it gluons}) associated with the local SU(3) symmetry.
Despite their apparently simple form, the interactions of quarks and
gluons, as dictated by the gauge principle and encoded in the Dirac
operator, give rise to a wide variety of phenomena.  Most notably, the
low-energy properties of strongly interacting matter are largely
determined by the phenomena of confinement and chiral symmetry
breaking (see, e.g.,
Refs.~\cite{Greensite:2003bk,Greensite:2011zz,Chandrasekharan_Wiese2004,
  Faber:2017alm}). At zero temperature, quarks and gluons are in fact
confined within hadrons by a linearly rising potential, up to
distances where a quark-antiquark pair can be created out of the
vacuum. Furthermore, there is an approximate chiral symmetry
associated with the lightest quarks, which in the limit of exactly
massless quarks is broken spontaneously. This determines most of the
properties of light hadrons, once the effects of the explicit breaking
by the light quark masses is taken into account.

Confinement and chiral symmetry breaking persist also at nonzero but
low temperature and densities. It is well established that, at
vanishing chemical potential, QCD undergoes a finite-temperature
transition to a deconfined, chirally restored phase ({\it quark-gluon
  plasma}), around $T_c\approx 155~{\rm MeV}$~\cite{Borsanyi:2010bp,
  Bazavov:2016uvm}. This transition is a rapid but analytic
crossover~\cite{Aoki:2006we}, with both confining and chiral
properties of the theory changing dramatically in a relatively narrow
interval of temperatures.

In particular, the confining properties are determined by the fate of
an approximate $\mathbb{Z}_3$ center symmetry, i.e., a symmetry under
gauge transformations which are periodic in time up to an element of
the group center. At low temperature, center symmetry is only
explicitly and mildly broken by the presence of quarks; at higher
temperatures, instead, center symmetry is strongly broken
spontaneously by the ordering of the Polyakov loop, i.e., the holonomy
of the gauge field along a straight path winding around the temporal
direction. In the ``quenched'' limit of infinitely heavy quarks, QCD
reduces to SU(3) pure gauge theory, where center symmetry is an exact
symmetry of the action, realized at low temperature \textit{\`a la}
Wigner-Weyl in the Hilbert space of the system, while spontaneously
broken at high temperatures. Around the same temperature at which the
Polyakov loop starts becoming ordered in QCD, corresponding roughly to
the spontaneous breaking of center symmetry, also the chiral
properties of the system change radically. The approximate order
parameter of chiral symmetry, i.e., the chiral condensate
$\la \bar{\psi}\psi\ra$, decreases rapidly around $T_c$, corresponding
to the disappearance of the effects of spontaneous breaking in the
chiral limit. The net effect is an effective restoration of chiral
symmetry, up to the explicit breaking due to the quark masses.

While the existence and the nature of the finite-temperature
transition are by now well established, the mechanisms of confinement
and chiral symmetry breaking, and similarly of deconfinement and
chiral symmetry restoration, are not fully understood yet; nor is the
apparently close relation between these two phenomena that are in
principle completely unrelated. An important role in the breaking and
restoration of chiral symmetry is played by the topological properties
of the gauge field configurations.

It might be possible to understand the formation of a chiral
condensate in the low temperature phase of QCD in terms of instantons
and the associated zero modes~\cite{Diakonov:1984vw,Diakonov:1985eg,
  Diakonov:1995ea,Smilga:1992yp,
  Janik:1998ki,Osborn:1998nm,Osborn:1998nf,GarciaGarcia:2003mn}. It is
well known that there is an exact zero mode of the Dirac operator
associated with an isolated instanton or anti-instanton, and with
their finite-temperature versions known as
calorons~\cite{Harrington:1978ve,Harrington:1978ua,Kraan:1998kp,
  Kraan:1998pm,Kraan:1998sn,Lee:1997vp,Lee:1998vu,Lee:1998bb}. Typical
gauge field configurations can be interpreted as a more or less dense
medium of instantons and anti-instantons. For a sufficiently high
density of topological objects, the associated zero modes will
strongly mix and form a finite band of near-zero Dirac modes, which in
turn gives rise to a finite condensate via the Banks-Casher
relation~\cite{Banks:1979yr}. At higher temperatures the density of
topological objects decreases, and so do the density of near-zero
Dirac modes and the chiral condensate, until the symmetry is
effectively restored. This is the ``disordered medium scenario'' for
chiral symmetry breaking. The mechanism of confinement is understood
less clearly, and various proposals have been put forth: we invite the
interested reader to consult
Refs.~\cite{Greensite:2003bk,Greensite:2011zz}.

Relatively recently, a third phenomenon has been found to take place
in correspondence with deconfinement and chiral symmetry restoration
in QCD, namely the localization of the low-lying eigenmodes of the
Dirac operator~\cite{Gockeler:2001hr,Gattringer:2001ia,
  GarciaGarcia:2005vj,GarciaGarcia:2006gr,Gavai:2008xe,Kovacs:2009zj,
  Bruckmann:2011cc,Kovacs:2010wx,Kovacs:2012zq,Giordano:2013taa,
  Nishigaki:2013uya,Giordano:2014qna,Ujfalusi:2015nha,Dick:2015twa,
  Cossu:2016scb,Holicki:2018sms}.  Localization is a widely studied
subject in condensed matter physics, since Anderson's work on the
absence of diffusion in random lattice
system~\cite{Anderson:1958vr}. In his seminal paper, Anderson showed
how the presence of disorder causes the spatial localization of energy
eigenmodes. For electrons in a disordered medium, such as a conductor
with impurities, localized modes appear at the band edge, beyond a
``mobility edge'' separating extended and localized modes. As the
amount of impurities/disorder increases, the mobility edge moves
towards the band center, eventually leading to all modes becoming
localized, and to the conducting sample turning into an insulator. It
is outside the scope of this paper (and frankly quite a Herculean
task) to provide an exhaustive account of the developments in the
theory of Anderson localization, and we refer the interested reader to
the reviews~\cite{thouless1974electrons,lee1985disordered,
  kramer1993localization,Evers:2008zz,anderson50}.

It has been shown that a similar localization phenomenon takes place
for the low-lying modes of the Dirac operator in QCD above the
pseudocritical temperature~\cite{GarciaGarcia:2005vj,
  GarciaGarcia:2006gr,Kovacs:2009zj,Kovacs:2012zq,Giordano:2013taa,
  Cossu:2016scb,Holicki:2018sms}: up to a critical point in the
spectrum, i.e., the analogue of the ``mobility edge'', low modes are
spatially localized on the scale of the inverse
temperature~\cite{Kovacs:2012zq,Cossu:2016scb}. The mobility edge
depends on the temperature $T$, and its extrapolation towards the
confined phase vanishes at a temperature compatible with
$T_c$~\cite{Kovacs:2012zq,Holicki:2018sms}. At the mobility edge, a
second-order phase transition takes place in the spectrum, which has
been shown to be a genuine Anderson
transition~\cite{Giordano:2013taa,Nishigaki:2013uya,
  Giordano:2014qna,Ujfalusi:2015nha}.

A disorder-driven transition, such as the Anderson transition in the
Dirac spectrum, obviously needs a source of disorder.  This was first
identified in the local fluctuations of the topological charge
density, treated as a dilute ensemble of pseudoparticles
(calorons)~\cite{Gockeler:2001hr,Gattringer:2001ia,GarciaGarcia:2005vj,
  GarciaGarcia:2006gr}. As already mentioned above, these topological
objects individually support localized zero modes of the Dirac
operator; since they overlap, the corresponding modes mix and shift
away from zero, but for a dilute ensemble this effect is small and
modes remain localized and near zero.  While evidence was produced
supporting the connection between localization and topological
objects, it turned out that not all localized modes could be explained
this way~\cite{Bruckmann:2011cc}, and at least another source of
disorder was needed.  This was identified in the fluctuations of the
Polyakov loop~\cite{Bruckmann:2011cc}, a hypothesis supported by
numerical
results~\cite{Bruckmann:2011cc,Cossu:2016scb,Holicki:2018sms} and by
the critical properties found at the mobility
edge~\cite{Giordano:2013taa,Nishigaki:2013uya,Giordano:2014qna,
  Ujfalusi:2015nha}. This led to the so-called ``sea/islands picture''
of localization, proposed in Ref.~\cite{Bruckmann:2011cc} and further
elaborated in
Refs.~\cite{Giordano:2015vla,Giordano:2016cjs,Giordano:2016vhx}: Dirac
eigenmodes tend to localize on ``islands'' of Polyakov loop
fluctuations away from its ordered value, which form an extended
``sea'' in the deconfined phase.  The sea/islands picture requires
only the existence of a phase with ordered Polyakov loop in order for
localization to appear, and so leads one to expect the localization of
the low Dirac modes in a generic gauge theory with a deconfinement
transition. This has been verified in a variety of
models~\cite{Kovacs:2010wx,
  Giordano:2016nuu,Giordano:2016vhx,Kovacs:2017uiz,Bruckmann:2017ywh,
  Giordano:2019pvc,Vig:2020pgq,Bonati:2020lal,Baranka:2021san},
including ones without topology, thus providing further support to the
sea/islands picture.  This also clearly suggests a strong connection
between localization of the low Dirac modes and deconfinement.

The relation between localization and chiral symmetry restoration has
received less attention, mostly because of the intrinsic difficulty of
studying gauge theories with massless fermions, where chiral symmetry
is exact.  It is, however, clearly established that localization of
the low modes is accompanied by evident changes in the spectral
density at the low end of the spectrum.  In
Refs.~\cite{Edwards:1999zm,Kovacs:2017uiz,Vig:2020pgq,Vig:2021oyt} it
was observed that in the quenched theory a peak of near-zero
modes of topological origin forms, followed by a spectral range with
low mode density. A similar peak of localized modes was observed in
Ref.~\cite{Dick:2015twa} in the presence of dynamical fermions. (The
presence of this peak was discussed before in
Ref.~\cite{Alexandru:2015fxa}, and has been studied recently in
Refs.~\cite{Ding:2020xlj,Kaczmarek:2021ser}, although the localization
properties of the eigenmodes are not studied there.)  It is shown in
Ref.~\cite{Vig:2021oyt} that the near-zero peak can be explained in
terms of a dilute instanton/anti-instanton gas and the associated zero
modes. Recently, it has been proposed that the presence of a finite
density of near-zero localized modes in the chiral limit can lead to
the disappearance of the finite-temperature massless excitations
predicted by the finite-temperature version of Goldstone's
theorem~\cite{giordano_GT_lett}.

While the presence of localization at high temperature, and its
connection with the finite-temperature transition in QCD and in other
gauge theories are by now fairly well established, the physical
meaning of localization of Dirac modes, and a detailed understanding
of the aforementioned connection with deconfinement and chiral
symmetry restoration, have proved to be quite elusive. The hope is
that a better understanding of localization can shed light on the
mechanisms of confinement and chiral symmetry breaking, and on the
finite-temperature deconfining and chirally restoring transition.

In this review we will discuss developments in the study of the
localization of Dirac modes in finite-temperature gauge
theories. While the main focus is on QCD, we will discuss also other
models, showing in particular that the connection between localization
and deconfinement is a general phenomenon. As this review is aimed
both at the particle physics and condensed matter communities (and
expected to disappoint them both), we provide brief introductions to
the subjects of finite-temperature QCD and Anderson localization in an
attempt to bridge the gaps. Older results were already reviewed in
Refs.~\cite{Giordano:2014qna,Giordano:2018iei}.

Localization in gauge theories also occurs in a few other contexts,
albeit those correspond to physical situations very different from the
one discussed in the present review. For this reason here we do not
discuss them in detail, and only mention them for completeness.  A
non-exhaustive list of references includes the following papers. For
results concerning the relation between the localization properties of
the low Dirac modes and the topological structure of the vacuum in
gauge theories, we refer to
Refs.~\cite{Zakharov:2006vt,deForcrand:2006my,
  Ilgenfritz:2007xu,Hollwieser:2011uj,Ilgenfritz:2013oda} and
references therein. The role played by localization in the Aoki phase
of quenched QCD with Wilson fermions, especially concerning the fate
of Goldstone modes, is studied in
Refs.~\cite{Golterman:2003qe,Golterman:2005fe}.  Localization
properties of the eigenmodes of the covariant Laplacian in Yang-Mills
theories are studied in Refs.~\cite{Greensite:2005yu,
  Greensite:2006ns}.

The plan of this paper is as follows. In Section \ref{sec:FTQCD} we
review finite-temperature QCD and related issues. In Section
\ref{sec:loc_general} we review the topic of localization and Anderson
transitions in some generality. In Section \ref{sec:QCD} we discuss
localization in QCD at finite temperature. The disordered medium
scenario and the sea/islands picture, providing mechanisms for
localization, are discussed in Section \ref{sec:mech}. Localization in
gauge theories other than QCD is discussed in Section
\ref{sec:othergt}. Finally, in Section \ref{sec:concl} we draw our
conclusions and show some prospects for the future.

\section{QCD at finite temperature}
\label{sec:FTQCD}

In strongly interacting systems such as QCD localization takes place
as the systems cross from the hadronic to the high-temperature
quark-gluon plasma state. To put localization in QCD in the proper
context, in the present section we summarize some basic facts about
this finite temperature transition.  For an introduction to gauge
theories at finite temperature we refer the reader to the literature
(see, e.g., Refs.~\cite{Rothe:1992nt_new,Montvay:1994cy}).

In an extended sense, QCD is a gauge theory with $N_f$ flavors of
quarks transforming in the fundamental representation of the gauge
group SU($N_c$) and interacting via the corresponding gauge field, the
excitations of which are the gluons. In a stricter sense, in QCD $N_f$
is fixed to six, the number of known quark flavors, and $N_c=3$. At finite
temperature, the theory is formally defined by the Euclidean partition
function
   
\begin {equation}
    \label{eq:qcd_part}
  Z_{\rm QCD} = \int [dA]\,e^{-S_{\rm YM}[A]} \prod_f\det \left(\slashed{D}[A] + m_f\right) \,,
\end{equation}
where the product runs over the quark flavors with masses $m_f$, $A$
is the gauge field, $S_{\rm YM}$ is the Euclidean Yang-Mills action
and

\begin{equation}
\slashed{D}[A]=\sum_{\mu=1}^4 \gamma_\mu(\partial_\mu+igA_\mu)
\end{equation}
is the Euclidean Dirac operator, with $\gamma_\mu$ the Euclidean,
Hermitean gamma matrices and $g$ the coupling constant. The temporal
direction is compactified to a circle of size equal to the inverse
temperature, and periodic boundary conditions in the temporal
direction are imposed on the gauge fields. In this form of the
partition function the quark fields, appearing in the action quadratically,
have been explicitly integrated out, resulting in the quark
determinant. The Dirac operator is an anti-Hermitean operator with
purely imaginary spectrum, which is furthermore symmetric about zero
thanks to the property $\{\gamma_5,\slashed{D}\}=0$.

It is instructive to consider the theory as a function of its
parameters, that in reality are fixed by the observed properties of
hadrons. The only such parameters of QCD are the quark
masses.\footnote{\label{foot:running} In principle, there is also the
  gauge coupling, but it turns out not to be a freely adjustable
  parameter, instead it runs with the energy scale. See
  Ref.~\cite{Weinberg:1996kr}, ch.~18.} In particular,
the low-energy properties of light hadrons are completely determined
by the masses of the lightest quarks, the $u$ and $d$ quark, and to
some extent the heavier $s$ quark. The other three known quark flavors
are so heavy that they have little influence on the low energy
physics.

Quark masses are also important parameters from a theoretical point of
view, because they crucially influence some symmetries of the
system. Even though in nature these are only approximate symmetries,
considering them helps to better understand the finite temperature
transition of QCD.  In an imaginary world with two massless quark
flavors, i.e., when $m_u=m_d=0$, QCD would have an exact
${\rm SU}(2)_V \times {\rm SU}(2)_A \times {\rm U}(1)_A$ chiral
symmetry. Here ${\rm SU}(2)_V$ is a rotation in the two-dimensional
$(u,d)$ flavor-space that acts identically on all the Dirac
components. In contrast, the flavor non-singlet axial symmetry
${\rm SU}(2)_A$ not only mixes the two flavors, but also transforms
the left and right Dirac components with opposite phases. Finally, the
${\rm U}(1)_A$ flavor-singlet axial symmetry acts trivially in flavor
space and rotates the left and right Dirac components with opposite
phases. Even though these are all symmetries of the classical
Lagrangian, after quantization the ${\rm U}(1)_A$ part of the symmetry
is anomalously broken. Furthermore, at zero temperature the
${\rm SU}(2)_A$ axial symmetry is spontaneously broken. The emerging
three Goldstone bosons are the analogues of the pions, and the order
parameter of the symmetry breaking is the light quark condensate
$\langle \bar{\psi} \psi \rangle $.  Finally, the vector part of the
symmetry ${\rm SU}(2)_V$ remains intact even for finite, but equal
quark masses.

In reality, the nonzero and non-equal masses of the $u$ and $d$ quarks
explicitly break these symmetries; however, the spontaneous and anomalous
breaking inherited from the massless theory both turn out to be much stronger
than this explicit breaking. In fact, in an imaginary world with zero $u$ and
$d$ quark masses, the low-energy properties of the light hadrons would be much
the same as they are in the real world. The only important exceptions would be
the pions, which in that case would be exact Goldstone bosons with zero mass.

If one imagines changing the quark masses, the other interesting limit
is the one in which quarks are much heavier than in reality. In
particular, in the limit of infinitely heavy quarks the quark
determinant in the path integral completely decouples and the
back-reaction of the quarks on the gauge field disappears. This is the
so-called quenched theory. In this limit QCD has a different exact
symmetry, the symmetry group being the center of the gauge group, in
the case at hand $\mathbb{Z}_3$.  The symmetry transformation in
question is a gauge transformation that is singular along a spacelike
hypersurface, and its singularity is characterized by an element of
the center $\mathbb{Z}_3$. Recalling that the system is finite in the
temporal direction with periodic boundary conditions for the gauge
field, this symmetry transformation is a gauge transformation that is
not periodic in the temporal direction (hence singular), and it
multiplies by the same $\mathbb{Z}_3$ center element all the
holonomies (gauge parallel transporters) going around the system in
the temporal direction. The holonomies wrapping around the system in
the temporal direction along a straight path are also called Polyakov
loops.

Gauge invariant local gluonic quantities are defined in terms of
holonomies around small loops, and those never wrap around the
system. These types of loops cross the hypersurface where the gauge
transformation is singular the same number of times in both
directions. As a result, the $\mathbb{Z}_3$ factors along such a loop
always cancel, and gauge invariant local quantities are invariant with
respect to the $\mathbb{Z}_3$ center transformations. In contrast,
fermionic quantities are affected, since such a singular gauge
transformation essentially introduces an extra $\mathbb{Z}_3$ twist
for the temporal boundary condition of the fermions through the
covariant derivative in the Dirac operator. The boundary condition
affects the spectrum of the Dirac operator and also its determinant
that appears in the path integral. At low temperatures where the
correlation length is much smaller than the temporal extent of the
system, and Polyakov loops fluctuate locally with little correlation,
the temporal boundary condition has only a small impact on the Dirac
determinant. Consequently, the quarks only mildly break the
$\mathbb{Z}_3$ symmetry. However, at high temperature, where the
correlation length becomes comparable to or larger than the temporal
size of the system, and the Polyakov loops tend to align with each
other, this picture changes drastically.

To understand exactly how that happens, let us first recall that in
finite temperature quantum field theory the temporal boundary
condition for fermions is antiperiodic. For free massless fermions
this implies a gap in the spectrum of the Dirac operator equal to the
first Matsubara frequency. If by a singular gauge transformation (as
defined above) we introduce an additional $\mathbb{Z}_3$ twist in the
boundary condition then the gap will decrease, because the twist $\pi$
corresponding to the antiperiodic boundary condition will decrease to
$\pi\pm2\pi/3 =\pm\pi/3~({\rm mod}~2\pi)$. In the interacting theory
there is no gap in the spectrum, but through this mechanism the low
end of the spectrum is much denser when the spatially averaged
Polyakov loop is in the complex center sectors (i.e., close to one of
the complex center elements) than when it is in the real one (i.e.,
close to the identity), where the effective twist comes only from the
antiperiodic boundary condition. As a result, the fermion determinant,
that disfavors larger low-mode density, strongly favors the real
Polyakov loop sector, and for finite quark mass this is the only
sector that contributes to the path integral. This is how fermions
explicitly break the $\mathbb{Z}_3$ center symmetry and select the
real Polyakov loop sector out of the three sectors that would be
equivalent in their absence. This mechanism is at work also at low
temperature, but much less effective there since the average Polyakov
loop fluctuates around zero.\footnote{This type of argument first
  appeared in Ref.~\cite{Stephanov:1996he} to explain the difference
  in the chiral condensate observed in the various center sectors of
  quenched lattice QCD in the deconfined
  phase~\cite{Chandrasekharan:1995gt}.}

Now going back to the quenched theory, at zero and low temperature,
the exact $\mathbb{Z}_3$ symmetry of its Lagrangian remains intact,
while above a critical temperature this symmetry is spontaneously
broken and its order parameter, the trace of the Polyakov loop,
develops a nonzero expectation value. In fact, in the quenched theory,
the logarithm of the expectation value of the Polyakov loop is
proportional to the gauge field energy it costs to insert an
infinitely heavy static quark in the system. In this way, the
vanishing of the Polyakov loop in the low-temperature phase shows that
no free quarks can exist there, so quarks are confined into
hadrons. In contrast, the nonzero expectation value of the Polyakov
loop in the high temperature phase implies that quarks are not
confined there.

The nature of the finite-temperature transition in extended QCD is
governed by the chiral and the $\mathbb{Z}_3$ symmetries, which---as
we have already seen---depend on the quark masses. In the quenched
limit (infinite quark mass) lattice simulations have shown the
transition to be weakly first
order~\cite{Francis:2015lha,Lucini:2003zr}, and this behavior persists
for large enough, but finite quark masses.  If the quarks become
lighter, the transition weakens and for intermediate quark masses
there is a wide region where it is only a crossover. In particular,
the light-quark masses in nature fall in this
range~\cite{Aoki:2006we}. For even smaller quark masses, the
transition is again expected to become a true phase transition, but
its order depends on the number of light quark flavors. For two light
flavors (and physical strange quark mass) it is expected to be second
order, whereas for three light flavors a first order phase transition
is anticipated.  However, the presence of these phase transitions,
previously predicted based on an epsilon
expansion~\cite{Pisarski:1983ms} (see also~\cite{Pelissetto:2013hqa}
for the role played by the ${\rm U}(1)_A$ anomaly), have not yet been
confirmed by lattice simulations, because simulations close to the
chiral limit are technically challenging.

Most of the results discussed in this review are based on numerical
calculations on the lattice.  Lattice field theory is a
nonperturbative approach to the quantization of quantum field
theories, based on the discretization of the relevant path integrals
that define the theory in the path-integral approach. We provide here
only a very brief introduction to this subject, referring the
interested reader to the extensive literature (e.g., the
books~\cite{Creutz:1984mg,Rothe:1992nt_new,Montvay:1994cy,DeGrand:2006zz,
  Gattringer:2010zz}). In the lattice approach to gauge theories
devised by Wilson~\cite{Wilson:1974sk}, the SU(3) gauge fields of QCD
are replaced by unitary SU(3) matrices ({\it link variables})
associated with the links of a finite hypercubic lattice. In continuum
language, these correspond to the parallel transporters of the gauge
fields along the paths connecting neighbouring lattice points. After a
suitable discretization of the gauge action, the relevant path
integrals are obtained by integrating over the gauge fields, which in
practice means integrating the link variables over the group manifold
with the invariant (Haar) group measure. The desired, continuum field
theory is obtained (if this is possible) by properly tuning the
parameters in the action, so that the correlation length of the system
in lattice units diverges, and the system ``forgets'' about the
underlying lattice. For pure gauge theories, the only available
parameter is the lattice inverse gauge coupling (usually denoted by
$\beta$), which ceases to be a freely adjustable parameter and turns
instead into a measure of the lattice spacing.\footnote{As the lattice
  spacing corresponds to the inverse of the largest energy attainable
  on the lattice, this indicates that the gauge coupling runs with the
  energy scale, see footnote \ref{foot:running}, and
  Ref.~\cite{Creutz:1984mg}, ch.~13.}

The approach outlined above is easily generalized to other gauge
theories based on different gauge groups, by simply replacing the
SU(3) link variables and the corresponding Haar measure with elements
of the relevant gauge group and the corresponding Haar measure. The
inclusion of fermions instead is not straightforward, especially for
what concerns the implementation of chiral symmetry. Nonetheless,
there are several viable discretization of the Dirac operator, which
are expected to all lead to the same results in the continuum limit.
Since they appear below in Section \ref{sec:QCD}, we mention Wilson
fermions, staggered fermions (possibly rooted), domain wall fermions,
overlap fermions, and twisted mass fermions (see
Ref.~\cite{DeGrand:2006zz,Gattringer:2010zz} and references therein
for details). We finally mention that several improvement schemes
exist that bring the system closer to the continuum limit, i.e., that
reduce the effects due to the finiteness of the lattice spacing. Such
schemes exist both for the gauge action and for the fermionic
determinant (see Ref.~\cite{DeGrand:2006zz,Gattringer:2010zz} and
references therein for details).

\end{paracol}
\begin{figure}[t]	
\widefigure
  \centering
    \includegraphics[width=0.5\textwidth]{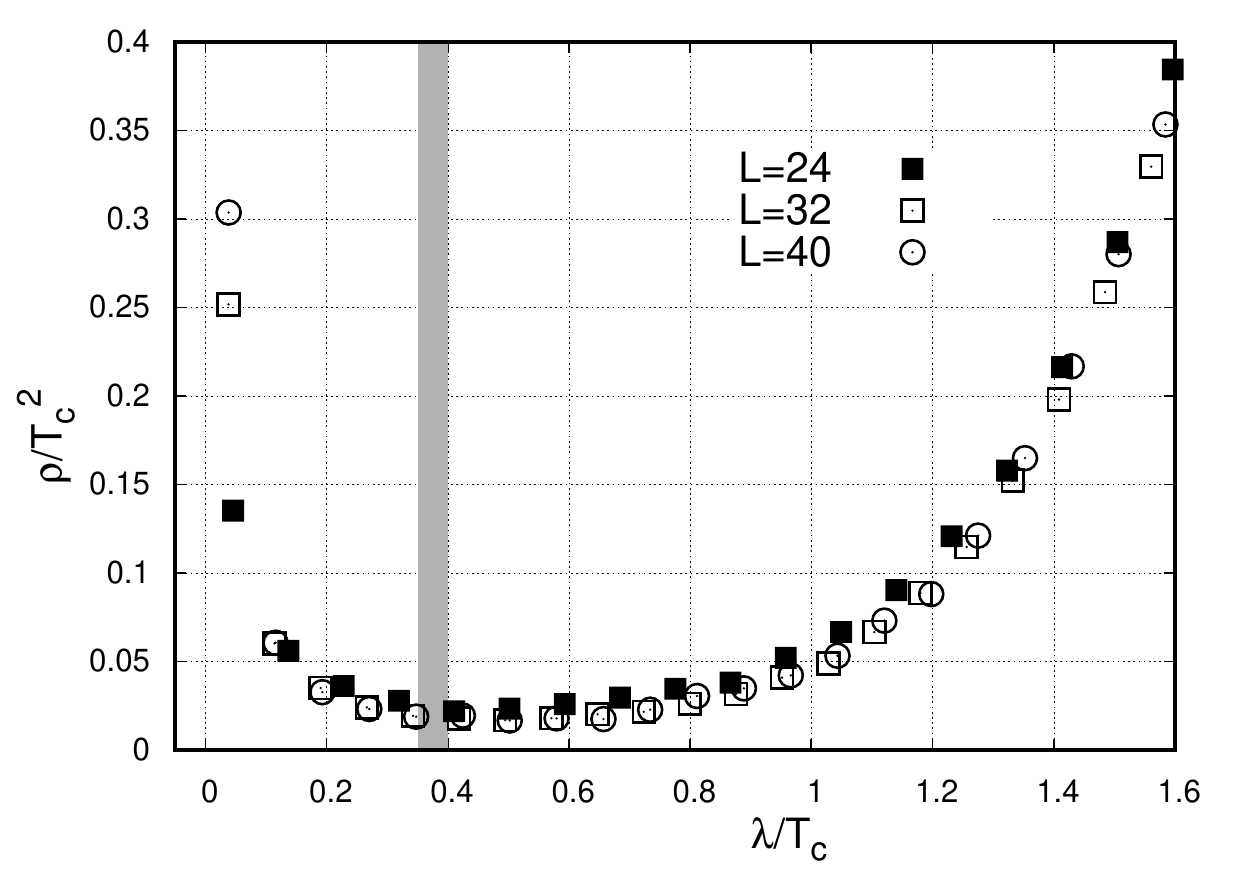}
    \caption{The spectral density, Eq.~\eqref{eq:spec_dens} (here
      normalized by the volume), of the overlap Dirac operator in
      quenched QCD (i.e., SU(3) pure gauge theory), just above the
      phase transition at $T=1.045T_c$. The grey band indicates the
      point separating the lowest, topological modes from the rest
      (its width equals the corresponding uncertainty). From
      Ref.~\cite{Vig:2021oyt}.\newline
      {\footnotesize Figure adapted from R.{\'A}.~Vig
      and
      T.G.~Kov{\'a}cs, % \textit{Ideal topological gas in the high
        % temperature phase of SU(3) gauge theory},
      arXiv:2101.01498
      (2021), and used under a
      \href{https://creativecommons.org/licenses/by/4.0}{CC-BY 4.0}
      license.}
      \label{fig:specdens}}
\end{figure}  
\begin{paracol}{2}
\linenumbers
\switchcolumn

\subsection{Finite-temperature transition, Dirac spectrum, and
  localization---an overview }
\label{sec:ftdlo}

We have seen that in the two extreme cases, the quenched limit and the
chiral limit, two different symmetries, the $\mathbb{Z}_3$ center
symmetry and the ${\rm SU}(2)_A$ axial symmetry govern the
transition. The respective order parameters, the Polyakov loop and the
quark condensate, signal spontaneous breaking of the symmetry in the
high temperature phase for the $\mathbb{Z}_3$ symmetry and in the low
temperature phase for the ${\rm SU}(2)_A$ axial symmetry. In nature,
both symmetries are only approximate, the transition is a crossover
and the order parameters have only inflection points in the crossover
region.  It also follows that in real QCD there is no sharply defined
transition temperature. In contrast, regardless of the quark mass, the
localization transition, i.e., the appearance of the first localized
modes at the low edge of the Dirac spectrum, occurs at a sharply
defined critical temperature. Moreover---as anticipated in the
Introduction, and as we will see below in Sections \ref{sec:QCD} and
\ref{sec:othergt}---the localization transition occurs in the
temperature range of the deconfining and chiral crossover in the case
of real QCD, and exactly at the deconfining temperature in the
quenched limit. This suggests that there might be a connection between
the thermodynamic (chiral and deconfining) transitions on the one
hand, and the localization transition on the other hand.

Besides the coincidence of their respective critical or pseudocritical
temperatures, these phenomena are also connected through the degrees
of freedom playing the most important role in their respective
dynamics. When the system crosses into the high temperature phase, the
spectral density of the Dirac operator around zero drops considerably,
exactly vanishing in the chiral limit. This is how the chiral
symmetry, spontaneously broken at low temperature, is restored above
the transition. Indeed, through the Banks-Casher
relation~\cite{Banks:1979yr}, the order parameter of chiral symmetry
breaking, the quark condensate, is proportional in the chiral limit to
the spectral density at zero, and generally strongly sensitive to the
low end of the spectrum. Lower spectral density also means that
eigenmodes close to each other in the spectrum are less likely to be
mixed by fluctuations of the gauge field, which might lead to
localization at the low end of the Dirac spectrum.

The spectral density, however, is not the only important parameter
that influences localization. In the quark mass regions numerically
explored so far, where the transition is either a crossover (near and
below the physical values of the light quark masses) or a true phase
transition governed by (approximate) center symmetry (heavy quark
limit), the spectral density does not immediately drop to zero at the
(pseudo)critical temperature. In particular, in the quenched limit
just above the transition a narrow but tall spike at zero appears in
the spectral density (see Fig.~\ref{fig:specdens}). This is due to
near-zero modes associated with a dilute gas of calorons and
anticalorons, local fluctuations of the topological
charge~\cite{Vig:2021oyt}. Even though the spectrum is dense in the
spike, eigenmodes there are localized~\cite{Kovacs:2019txb}. A similar
peak of near-zero modes is also found for physical, near-physical, and
below-physical light-quark
masses~\cite{Alexandru:2015fxa,Dick:2015twa,Ding:2020xlj,Kaczmarek:2021ser}. For
near-physical masses these modes are found to be
localized~\cite{Dick:2015twa}, and most likely this persists as the
mass is decreased.

\end{paracol}
\begin{figure}[t]	
\widefigure
  \centering
    \includegraphics[width=0.5\textwidth]{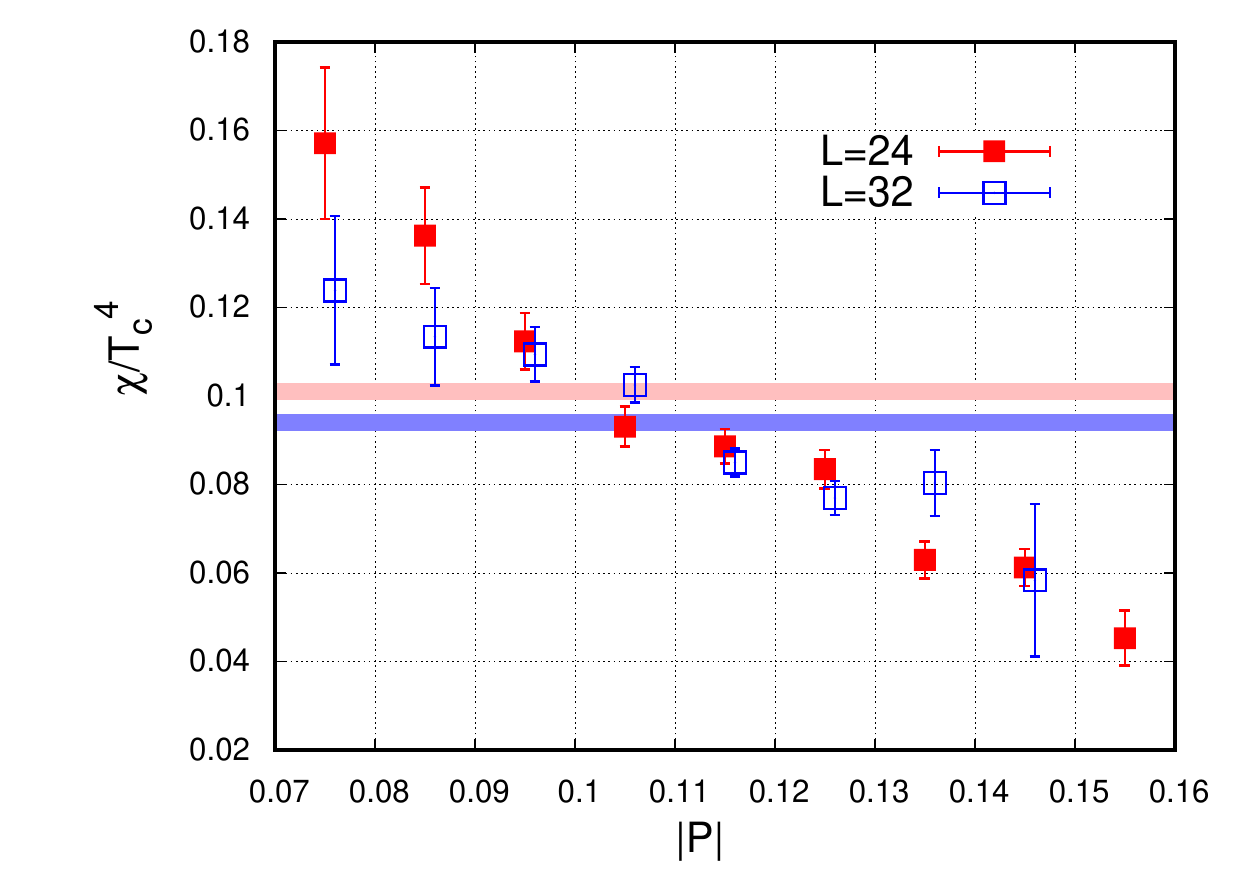}
    \caption{The dependence of the topological susceptibility on the
      value of the spatially averaged Polyakov loop in quenched
      lattice simulations. The susceptibility was computed by dividing
      the configurations in sets according to the spatially averaged
      Polyakov loop. The averages for the whole ensembles (in two
      volumes) are shown by the horizontal bands, the widths
      indicating the uncertainties. From
      Ref.~\cite{Vig:2021oyt}.\newline
      {\footnotesize Figure adapted
        from R.{\'A}.~Vig and
        T.G.~Kov{\'a}cs, % \textit{Ideal topological gas in the high
        % temperature phase of SU(3) gauge theory},
      arXiv:2101.01498 (2021), and used under a
      \href{https://creativecommons.org/licenses/by/4.0}{CC-BY 4.0}
      license.}
    \label{fig:chi_vs_plp}}
\end{figure}  
\begin{paracol}{2}
\linenumbers
\switchcolumn

Recently, the spike in the spectral density received another
interpretation. It was argued that it signals the appearance of a new,
previously undiscovered ``phase'' of QCD, intermediate between the
low-temperature confined and the high-temperature deconfined
phase~\cite{Alexandru:2019gdm}. In a more recent paper the same
authors studied a newly defined infrared dimension $d_{\rm IR}$ of the
eigenmodes in the low end of the spectrum of the chirally symmetric
overlap Dirac operator. They concluded that the exact zero modes have
$d_{\rm IR}=3$, and in the spectral peak $d_{\rm IR}$ changes rapidly
but smoothly from 2 to 1 as one moves up in the
spectrum~\cite{Alexandru:2021pap}. This behavior persists up to the
bulk of the spectrum, where the spectral density, together with the
infrared dimension $d_{\rm IR}$ of the modes starts to increase
again. This nontrivial change in the infrared dimension all happens
in the region where based on the spectral statistics and the scaling
of the participation ratio with the volume, the eigenmodes are thought
to be localized. It would be interesting to further investigate how
$d_{\rm IR}$ relates to the usual fractal dimension $D_2$ (see
Eq.~\eqref{eq:mfexp}), and what kind of spatial structure in the
eigenmodes gives rise to this nontrivial behavior. This could also
depend on the chiral and locality properties of the particular
discretization of the Dirac operator.

Topological fluctuations and the localization of the eigenmodes are
both intimately related to fluctuations of the Polyakov loop, the
order parameter of the quenched transition. The spatial localization
of low Dirac eigenmodes is found to strongly correlate with local
fluctuations of the Polyakov loop away from its symmetry-breaking
equilibrium
value~\cite{Bruckmann:2011cc,Cossu:2016scb,Holicki:2018sms}. This
gives rise to the sea/islands picture of localization that we will
discuss in Section \ref{sec:mech_seaislands} of the present paper in
more details. Localization on calorons and localization on Polyakov
loop fluctuations are, however, not mutually exclusive, as calorons
always contain large fluctuations of the Polyakov loop. In fact,
within a caloron, the Polyakov loop wraps around the gauge group in a
topologically nontrivial way. The connection between calorons and
Polyakov loop fluctuations is also shown by the strong correlation
between the Polyakov loop and the topological susceptibility that can
be observed close to the transition in the high temperature phase (see
Fig.~\ref{fig:chi_vs_plp}).

The finite temperature transition of QCD is a result of the interplay
of all those mechanisms that we just discussed, involving the Polyakov
loop, the topological fluctuations and the spectral density of the
Dirac operator around zero. Since localization is intimately related
to all these aspects, it might hold the key to a better intuitive
understanding of the dynamics of the transition.

\section{Localization and Anderson transitions}
\label{sec:loc_general}

In a classic paper~\cite{Anderson:1958vr}, Anderson showed that a
sufficiently large amount of disorder in a lattice system prevents
quantum-mechanical diffusion. Working in the one-particle
tight-binding approximation, and mimicking the effect of disorder by
supplementing the tight-binding Hamiltonian with a random potential on
the lattice sites, Anderson showed that all the eigenfunctions of the
system are localized for sufficiently strong disorder (i.e., for a
sufficiently broad distribution for the random potential).

A practical example of this situation is a ``dirty'' crystal where
some of the lattice atoms are replaced by impurities. Anderson's
results imply that all the electron eigenstates become localized for a
sufficiently large concentration of impurities. This prevents electron
diffusion and the associated transport phenomena; in particular, the
d.c.\ conductivity at zero temperature
vanishes~\cite{halperin1967,mott1967}. Localization then provides a
possible mechanism for a disorder-induced metal-insulator transition
(MIT).

Anderson's original arguments were later scrutinized and clarified by
several authors~\cite{ziman1969,anderson1970,thouless1970,mott1970,
  economou1970,economou1972,abouchacra1973}. Since then, the topic of
disorder-induced localization, or {\it Anderson localization}, has
been extensively studied in the condensed matter community, and it is
impossible for us to provide here a comprehensive survey, or even do
justice to the related literature. In this section we limit ourselves
to a short review of the main aspects of Anderson localization,
especially those relevant to gauge theories, discussed in the next
Section. We invite the interested reader to consult the
reviews~\cite{thouless1974electrons,lee1985disordered,
  kramer1993localization,Evers:2008zz,anderson50}.

\end{paracol}
\begin{figure}[t]	
\widefigure
  \centering
    \includegraphics[width=0.5\textwidth]{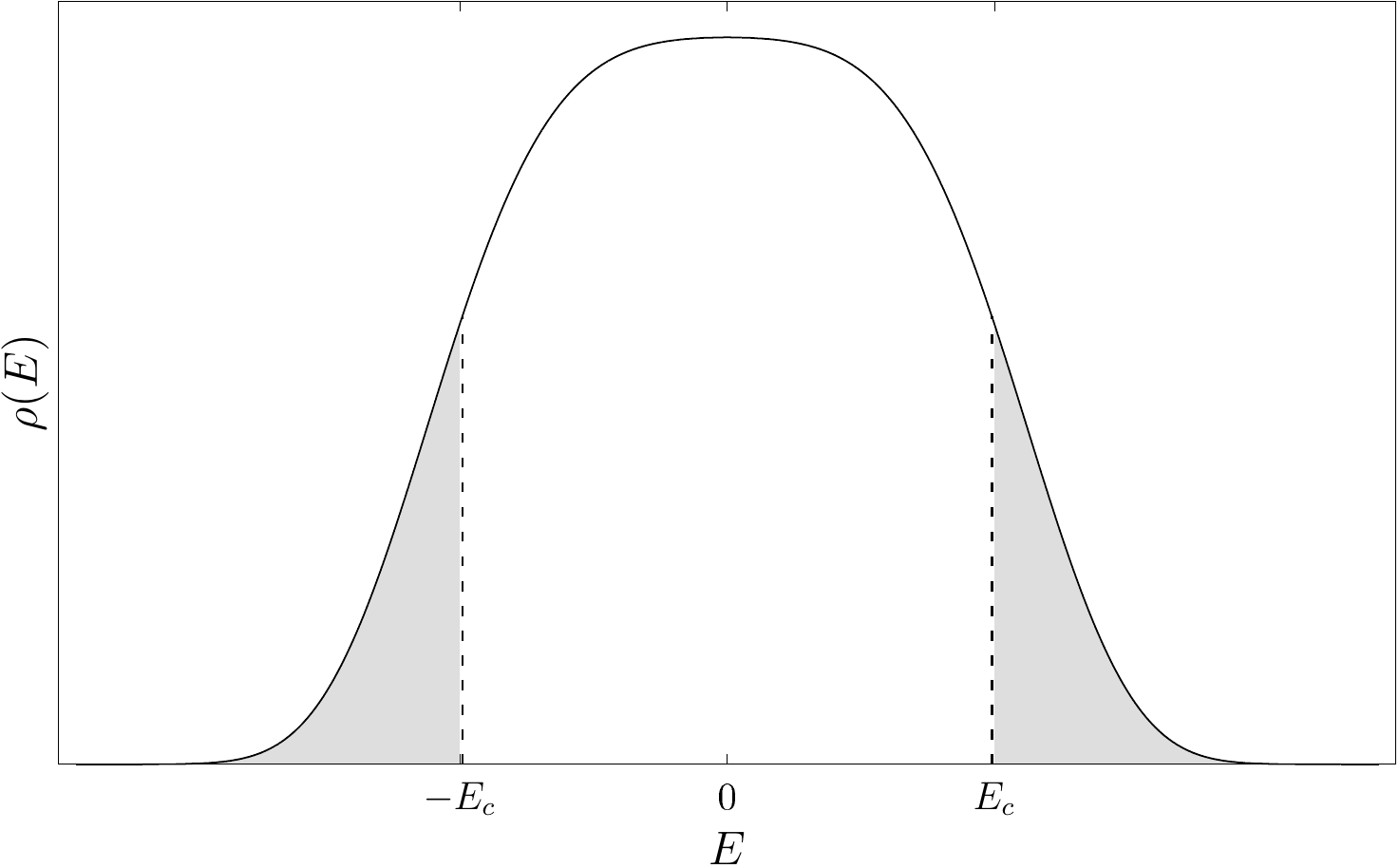}
  \caption{Sketch of the density of states $\rho$ (see
    Eq.~\eqref{eq:spec_dens}) as a function of energy $E$ in the
    Anderson model, Eq.~\eqref{eq:oAM}. Localized modes are present in
    the shaded region beyond the mobility edges
    $\pm E_c$.\label{fig:AMband}}
\end{figure}  
\begin{paracol}{2}
\linenumbers
\switchcolumn

\subsection{The Anderson model}
\label{sec:loc_AM}

In its simplest form, the (orthogonal) Anderson model Hamiltonian reads 

\begin{equation}
  \label{eq:oAM}
  H_{\vec{x}, \vec{y}}^{{\rm AM}} = \varepsilon_{\vec{x}} \delta_{\vec{x}, \vec{y}} + 
  \sum_{\mu=1}^3   (\delta_{\vec{x}+\hat{\mu},\vec{y}} +
  \delta_{\vec{x}-\hat{\mu},\vec{y}})\,,
\end{equation}
where $\vec{x}, \vec{y}$ label the sites of a simple cubic lattice
with lattice vectors $\hat{\mu}$, $\mu=1,2,3$, and
$\varepsilon_{\vec{x}}$ is a random on-site potential, with uniform
probability distribution in the interval $[-\f{W}{2},\f{W}{2}]$. The
lattice spacing and the hopping energy are set to 1 for
simplicity. The width $W$ of the distribution is a measure of the
amount of disorder in the system, with $W=0$ corresponding to a
perfectly pure crystal. In this case, for a lattice of side $L$ with
periodic boundary conditions the eigenstates of $H^{{\rm AM}}$ are
plane waves with wave vectors $\vec{p}=\f{2\pi \vec{k}}{L}$, with
$k_\mu=0,1,\ldots, L-1$.  However, as soon as even a small amount of
disorder is put into the system, i.e., $W\neq 0$, the eigenmodes
$\psi(\vec{x})$ at the band edge become exponentially localized, i.e.,
$|\psi(\vec{x})|^2 \sim e^{-|\vec{x}-\vec{x}_0|/\xi}$ for $E$ beyond
critical energies $\pm E_c(W)$ called ``mobility
edges''~\cite{mott1967} (see Fig.~\ref{fig:AMband}).  As the amount of
disorder $W$ in the system increases, the mobility edge moves towards
the band center. Eventually, for $W$ larger than a critical disorder,
$W_c$, all the modes become localized. If Eq.~\eqref{eq:oAM} describes
the conduction band of an electron in some ``dirty'' crystalline
system, for large enough $W$ the Fermi energy will lie in the
localized part of the band; d.c.\ transport then takes place through
hopping of electrons from one localized state to another, which has an
exponentially small probability of happening, and in the limit of
infinite size leads to the absence of charge transport. As the amount
of impurities increases past the critical value, the system then
undergoes a metal-to-insulator transition.

\subsection{Anderson transitions}
\label{sec:loc_AT}

When the energy of the modes crosses the mobility edge $E_c(W)$ at
fixed $W$, or equivalently when the disorder in the system crosses the
energy-dependent critical disorder $W_c(E)$ at fixed mode energy $E$,
the nature of the eigenmodes of $H^{{\rm AM}}$ changes from
delocalized to localized.  As argued in Ref.~\cite{locgangoffour}, in
three dimensions the associated transition is a second-order phase
transition ({\it Anderson transition}), with divergent correlation
length $\xi(E) \sim |E-E_c|^{-\nu}$ or $\xi(W) \sim |W-W_c|^{-\nu}$,
where the same exponent $\nu$ is expected.

This prediction is based on the so-called {\it scaling theory of
  localization} (see Ref.~\cite{lee1985disordered} for an
introduction): first proposed in Ref.~\cite{locgangoffour} based on
previous ideas exposed in
Refs.~\cite{edwards1972numerical,thouless1974electrons,
  licciardello1975conductivity,wegner1976electrons}, it was later put
on a firmer basis through a field-theoretical description of
disordered systems and Anderson
transitions~\cite{wegner1979mobility,efetov1980,efetov1983supersymmetry}
(see Ref.~\cite{Evers:2008zz} for a full list of references).  The
basic idea is that the change in the conductance $G(L)$ of the
system\footnote{The {\it conductance} $G(L)$ for a $d$-dimensional
  (hyper)cubic sample of linear size $L$ equals
  $G(L) = \sigma L^{d-2}$ where $\sigma$ is the {\it conductivity} of
  the system.} as its size $L$ is increased is controlled only by the
localized or delocalized nature of the energy eigenmodes, which in
turn is measured by the conductance itself, as a proxy for the
disorder in the system. This implies a scaling behavior of the
conductance, $\f{d\ln G(L)}{d\ln L} = \beta(G(L))$. Using the
asymptotics of the $\beta$ function obtained from localized or
delocalized modes is then enough to show that in three dimensions
there is an unstable fixed point (in the renormalization-group sense),
and so a mobility edge in the energy spectrum and a phase transition
at some critical amount of disorder (see
Fig.~\ref{fig:scaling}). 
In one dimension no Anderson transition is expected as all modes are
localized in the presence of disorder~\cite{mott1961,borland1963}, while the
situation in two dimensions is more complicated (see below).

\end{paracol}
\begin{figure}[t]	
\widefigure
  \centering
  \includegraphics[width=0.4\textwidth]{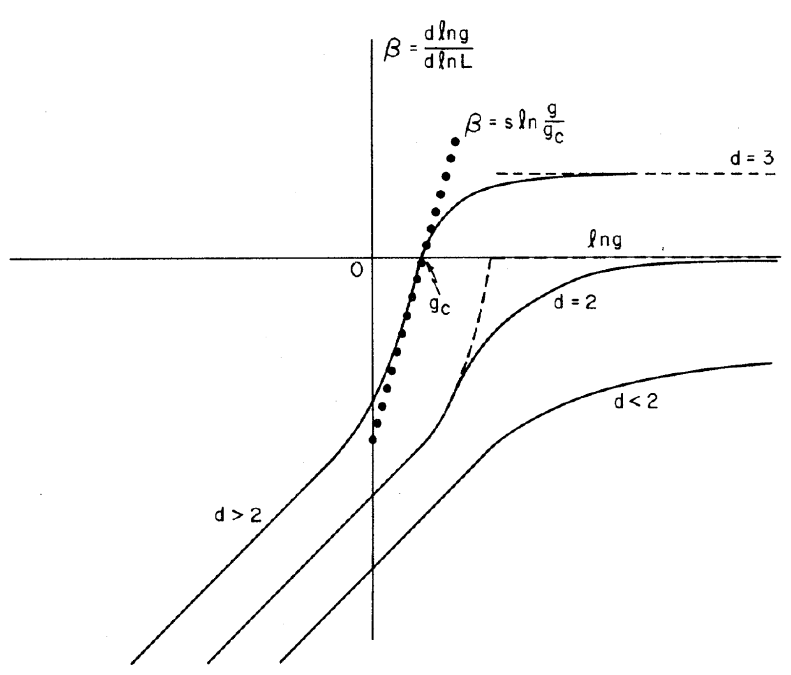}
  \caption{The scaling function $\beta(g)$ against the dimensionless
    conductance $g=\f{2\hbar G}{e^2}$ in various dimensions. From
    Ref.~\cite{locgangoffour}.
    \newline {\footnotesize Reprinted figure with permission from
      E.~Abrahams, P.W.~Anderson, D.C.~Licciardello, and
      T.V.~Ramakrishnan, Phys.\ Rev.\ Lett.\ 42, 673 (1979). Copyright
      (1979) by the American Physical Society.}
\label{fig:scaling}}
\end{figure}
\begin{paracol}{2}
\linenumbers
\switchcolumn

The Anderson model that we just described, Eq.~\eqref{eq:oAM}, is but
the simplest disordered Hamiltonian in three dimensions, and can be
generalized in various ways. One can consider different probability
distributions for the on-site disorder (e.g., the Lloyd
model~\cite{Lloyd_1969}), add off-diagonal disorder by making also the
hopping terms
random~\cite{theodorou1976,antoniou1977,economou1977localization},
increase the range of interaction by adding more hopping terms (e.g.,
Ref.~\cite{inui1994}), and so on.  However, according to the general
theory of the renormalization group (see, e.g.,
Ref.~\cite{cardy_1996}), critical properties at a second-order phase
transition are shared by systems in the same {\it universality class},
determined only by general properties such as the dimensionality and
the symmetries of the system.\footnote{The dimensionality and the
  symmetry class do not always determine uniquely the universality
  class of the Anderson transition: see Ref.~\cite{Evers:2008zz}.}

From a technical point of view, the Anderson model is a model of
(sparse) random matrices. The properties of these models are the
subject of Random Matrix Theory (RMT)~\cite{mehta2004random,
  Guhr:1997ve,Verbaarschot:2000dy}. For random systems, the relevant
symmetry classification has been provided by
Dyson~\cite{dyson1962threefold}, later extended by
Verbaarschot~\cite{Verbaarschot:1994qf}, and completed by Altland and
Zirnbauer~\cite{Altland:1996zz,Altland:1997zz,zirnbauer1996riemannian,Zirnbauer:2010gg}
(see also Refs.~\cite{Evers:2008zz,Chiu_2016}).  The main {\it
  symmetry classes}, relevant to the models discussed in this review,
are determined by the existence (or not) of an antiunitary symmetry
operator $T$ (``time reversal'') commuting with the Hamiltonian,
$[T,H]=0$, and further specified by whether $T^2=1$ or $T^2=-1$. If
$T$ exists and $T^2=1$, the system is in the {\it orthogonal class}
($O$): this is the case for the model in Eq.~\eqref{eq:oAM}. If $T$
does not exist, the system is in the {\it unitary class} (U). Perhaps
the simplest example of a system in this class is the so-called
unitary Anderson model (UAM),

\begin{equation}
  \label{eq:uAM}
   H_{\vec x, \vec y}^{{\rm UAM}} = \varepsilon_{\vec x} \delta_{\vec x, \vec y} + 
  \sum_{\mu=1}^3 
  (\delta_{\vec x+\hat\mu,\vec y} +
  \delta_{\vec x-\hat\mu,\vec y}){e^{i\phi_{\vec x, \vec y}}}\,,
\qquad \phi_{\vec y, \vec x} = -\phi_{\vec x, \vec y}\,,
\end{equation}
which includes also off-diagonal disorder in the form of random phases
$\phi_{\vec x, \vec y}$ in the hopping terms, mimicking the presence
of a random magnetic field. Finally, if $T$ exists and $T^2=-1$, the
system is in the {\it symplectic class} (S).  This classification is
complete as far as the statistical properties in the bulk of the
spectrum are concerned.

A refined classification is needed if one wants to discuss statistical
spectral properties near the origin.  In this case, one has to
consider whether also a ``particle-hole'' symmetry exists, realized in
terms of an antiunitary operator $C$ obeying $\{C,H\} =0$, and if so
whether $C^2=\pm 1$.  This gives rise to nine different combinations.
The eight combinations obtained when at least $T$ or $C$ exists
correspond to eight different symmetry classes. If both $T$ and $C$
exist, it automatically follows that a unitary operator $\Gamma=TC$
exists, anticommuting with the Hamiltonian, $\{\Gamma , H\}=0$, and
satisfying $\Gamma^2=1$.  However, a $\Gamma$ satisfying this property
can exist also if $T$ and $C$ are both absent.  In this case there are
two further symmetry classes, corresponding to whether such a $\Gamma$
exists or not, for a total of ten.  The classification is summarized
in Tab.~\ref{tab:symclass}.  In particular, if $\Gamma$ exists and
commutes with $T$ (if this also exists), the system belongs to one of
the chiral classes (chO, chU, and chS).  Examples of systems of this
type are provided by certain lattice models with random hopping terms,
and no on-site potential, on bipartite lattices.

\begin{table}[t]
  \centering
  \begin{tabular}{ccc|ll}
    \hline    \hline
    $T$ & $C$ & $\Gamma$ & class  \\
    \hline    \hline
        & & & \multicolumn{2}{l}{Wigner-Dyson classes} \rule{0pt}{9pt}\\
    \hline
    0 & 0 & 0 & A & (unitary) \rule{0pt}{9pt}\\
    $+$ & 0 & 0 & AI &(orthogonal) \\
    $-$ & 0 & 0 & AII &(symplectic) \\
    \hline
        & & &  \multicolumn{2}{l}{chiral classes}\rule{0pt}{9pt}\\
    \hline
    0 & 0 & 1 & AIII &(chiral unitary) \rule{0pt}{9pt}\\
    $+$ & $+$ & 1 & BDI &(chiral orthogonal) \\
    $-$ & $-$ & 1 & CII &(chiral symplectic) \\
    \hline
        & & & \multicolumn{2}{l}{Bogoliubov-de Gennes classes}\rule{0pt}{9pt}\\
    \hline
    0 & $-$ & 0 & C & \rule{0pt}{9pt}\\
    $+$ & $-$ & 1 & CI &\\
    0 & $+$ & 0 & D &\\
    $-$ & $+$ & 1 & DIII & \\
    \hline    \hline
  \end{tabular}
  \caption{Symmetry classes of random matrix ensembles. Entries
    corresponding to time-reversal ($T$) and particle-hole ($C$) symmetry
    indicate whether the symmetry is absent ($0$) or, if present, what
    is its square ($\pm$); entries corresponding to chiral symmetry
    ($\Gamma$) indicate whether it is absent or present ($0$ or $1$).}
  \label{tab:symclass}
\end{table}

\subsection{Detecting localization: eigenmode observables}
\label{sec:loc_detect_em}

A convenient way to study the localization properties of the
eigenmodes of a random lattice Hamiltonian and how they change along
the spectrum is by means of the {\it inverse participation ratios}
(IPRs),\footnote{The participation ratio (see below) was introduced in
  Ref.~\cite{bell1970}; its inverse as a measure of localization is
  discussed in Ref.~\cite{thouless1974electrons}.}

\begin{equation}
  \label{eq:IPR}
  {\rm IPR}_q \equiv \sum_{x} |\psi(x)|^{2q}\,,
\end{equation}
where it is assumed that eigenmodes obey the usual normalization
condition, i.e., $\sum_{x} |\psi(x)|^{2}=1$, and $x$ now labels the
sites of the relevant lattice, assumed finite and of volume $V=L^d$,
with $L$ the linear size and $d$ the dimensionality. Unless specified
otherwise, in the following both the term and the notation IPR,
without subscript, will be used to refer specifically to the case
$q=2$.  For modes extended throughout the whole system, one has
qualitatively $|\psi_{\rm ext}(x)|^2\sim 1/V$, and so
${\rm IPR}_q \sim V^{1-q}$: after averaging over the possible
realizations of disorder, which will be denoted with $\la\ldots\ra$,
and taking the large-volume limit, one has then
$\la {\rm IPR}_q\ra \to 0$ as $V\to \infty$ (for $q>1$).  For modes
localized in a region of volume $V_0$ one has
$|\psi_{\rm loc}(x)|^2\sim 1/V_0$ inside the localization region and
negligible outside, and so ${\rm IPR}_q \sim V_0^{1-q}$: one has then
$\la {\rm IPR}_q\ra \to {\rm const.}$ as $V\to \infty$. At the
mobility edge, instead, the scaling of ${\rm IPR}_q$ with the volume
depends on $q$ in a highly nontrivial way. One has in general

\begin{equation}
  \label{eq:mfexp}
{\rm IPR}_q  \sim L^{-(q-1)D_q}\,,
\end{equation}
where $D_q=d$ for extended modes and $D_q=0$ for localized modes,
while at criticality $D_q$ is not a constant. This reflects the
multifractal nature of eigenmodes at
$E_c$~\cite{wegner1980inverse,Castellani_1986}, and leads to define a
set of multifractal exponents characterizing the critical behavior at
the Anderson transition (see Ref.~\cite{Evers:2008zz}).

Closely related to the IPR is the {\it participation ratio},

\begin{equation}
  \label{eq:PR}
  {\rm PR} \equiv \f{1}{V}{\rm IPR}^{-1}  \qquad \left(= \f{1}{V} {\rm IPR}_2^{-1}\right)\,,
\end{equation}
which measures the fraction of the system effectively occupied by the
mode. For localized modes one has in the infinite volume limit
$\la {\rm PR}\ra \to 0$, while for delocalized modes extended
throughout the system one finds $\la {\rm PR}\ra \to$
a~nonzero~constant.  Another equivalent way to measure the
localization properties is to use the mode ``size'', i.e.,
$V\cdot {\rm PR} = {\rm IPR}^{-1}$, which as $V\to \infty$ (after
averaging over the disorder) remains constant for localized modes and
diverges for delocalized modes. For systems with nontrivial spin
and/or internal degrees of freedom, the eigenvectors
$\psi_{\alpha,c}(x)$ possess extra spin ($\alpha$) and/or internal
indices ($c$). In these cases it is convenient to employ a definition
of the IPR which is invariant under spacetime and internal (unitary)
rotations, i.e.,

\begin{equation}
  \label{eq:gaugeinvIPR}
  {\rm IPR} = \sum_{x}\left(\sum_{\alpha,c}
    |\psi_{\alpha,c}(x)|^2\right)^2
  = \sum_{x}\left(\psi(x)^\dag\psi(x)\right)^2\,,
\end{equation}
where the normalization condition $\sum_{x}\psi(x)^\dag\psi(x)=1$ is
understood.  For example, for eigenmodes of the continuum, Wilson, or
overlap Dirac operators, $\alpha=1,\ldots, 4$ is the Dirac index, and
$c=1,\ldots, N_c$ is the gauge group (``color'') index; for eigenmodes
of the staggered operator $\alpha$ is absent but $c$ is present.

\subsection{Detecting localization: eigenvalue observables}
\label{sec:loc_detect_ev}

Another useful tool to detect localization are the statistical
properties of the eigenvalues $\lambda_i$ of a random lattice
Hamiltonian, which are closely related to the localization properties
of its eigenvectors~\cite{altshuler1986repulsion}. Localized modes are
in fact expected to be sensitive only to local fluctuations in the
disorder, and so the corresponding eigenvalues are expected to
fluctuate independently. More precisely, after removal of
non-universal, model-dependent features by means of the so-called {\it
  unfolding} procedure~\cite{mehta2004random} (see below), the
unfolded eigenvalues corresponding to localized modes should obey
Poisson statistics. Delocalized modes, on the other hand, are expected
to be mixed easily by fluctuations in the disorder, and so the
corresponding unfolded spectrum should behave like that of a dense
random matrix, and display the statistics of the Gaussian ensemble of
Random Matrix
Theory~\cite{mehta2004random,Guhr:1997ve,Verbaarschot:2000dy} in the
appropriate symmetry class.

\end{paracol}
\begin{figure}[t]	
\widefigure
  \centering
  \includegraphics[width=0.5\textwidth]{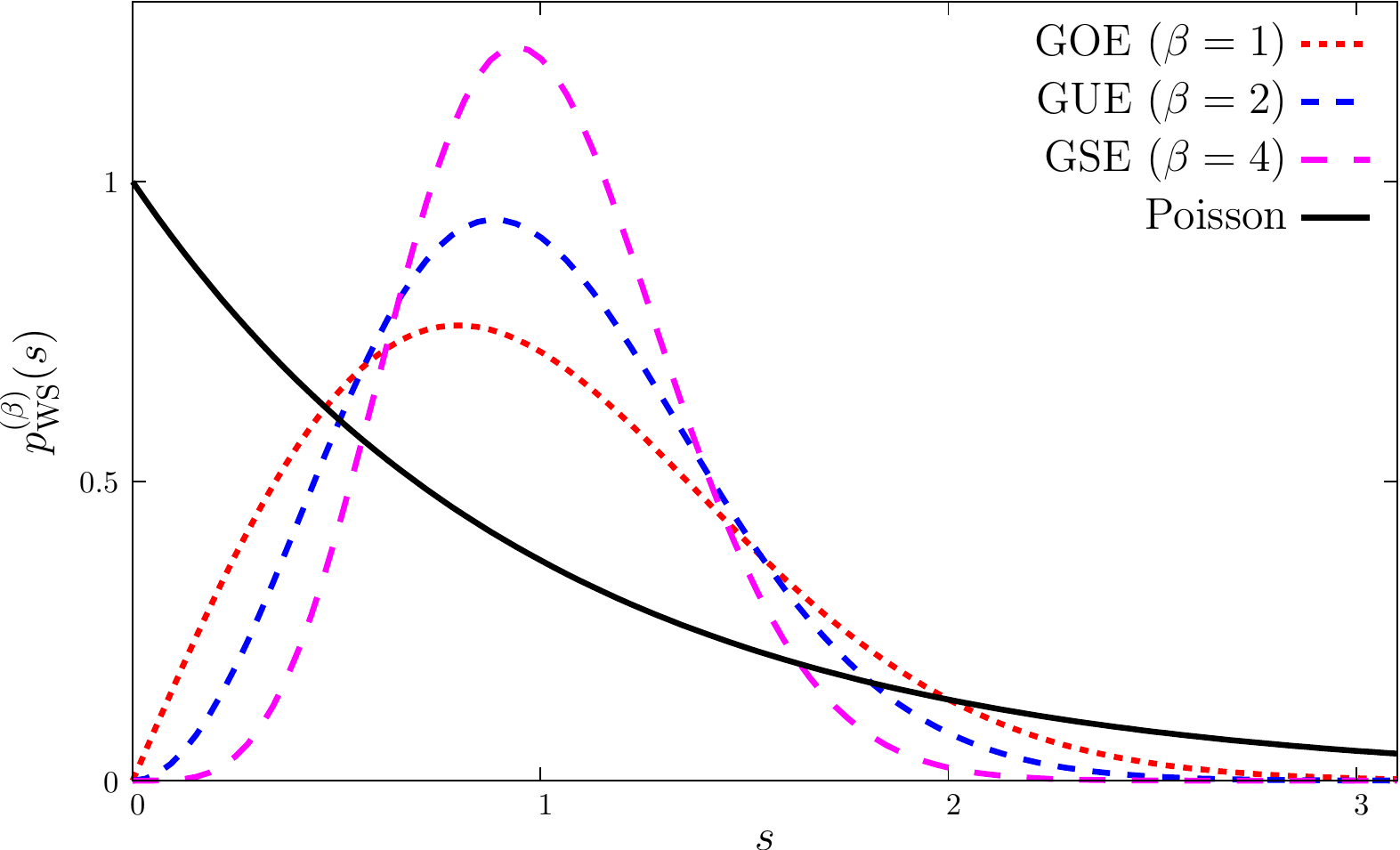}
  \caption{Wigner surmise for the various symmetry classes. The
    exponential distribution for Poisson statistics is also shown.\label{fig:pulsd}}
\end{figure}  
\begin{paracol}{2}
\linenumbers
\switchcolumn

Unfolding is a monotonic mapping of the eigenvalues $\lambda_i$ that
makes the spectral density equal to 1 throughout the spectrum. The
spectral density is defined as

\begin{equation}
  \label{eq:spec_dens}
 \rho(\lambda)\equiv \la \sum_i\delta(\lambda-\lambda_i)\ra \,.
\end{equation}
The unfolded eigenvalues $x_i$ are given by

\begin{equation}
  \label{eq:unfolding}
  \lambda_i \to x_i=\int^{\lambda_i}d\lambda \,\rho(\lambda)\,,
\end{equation}
and it is easy to see that they have unit density,
$\bar{\rho}(x) = \f{d\lambda}{dx}\rho(\lambda)=1$. For random matrix
models with dense matrices, the bulk statistical properties of the
unfolded spectrum are expected to be universal (i.e., to not depend on
the details of the model) and uniform throughout the spectrum. This has
been proved rigorously for a large class of matrix ensembles (see
Refs.~\cite{erdos2012,taovu2011} and references therein). One can then
determine these properties in the exactly solvable Gaussian ensembles
in the various symmetry classes (orthogonal, unitary,
symplectic)~\cite{mehta2004random}. In particular, the probability
distribution of the unfolded spacings $s_i = x_{i+1}-x_i$, or {\it
  unfolded level spacing distribution} (ULSD), $p_{\rm ULSD}(s)$, can
be obtained exactly, although not in closed form. A good approximation
for the ULSD is provided by the so-called {\it Wigner surmise} in the
appropriate symmetry class~\cite{Guhr:1997ve} (see Fig.~\ref{fig:pulsd}),

\begin{equation}
  \label{eq:WS}
  p_{\rm WS}^{(\beta)}(s) = a_\beta s^\beta e^{-b_\beta s^2}\,,
\end{equation}
where $\beta$ is the {\it Dyson index} of the Gaussian ensemble, and one has for the
various symmetry classes\footnote{Notice that by construction one has for
  a generic random matrix ensemble
  $\int_0^\infty ds\,p_{\rm ULSD}(s) = \int_0^\infty ds\,p_{\rm
    ULSD}(s) s =1$. This follows from the fact that the average
  spacing equals the inverse of the spectral density, which is 1 for
  the unfolded spectrum.}

\begin{equation}
  \label{eq:WS2}
  \begin{aligned}
    \text{orthogonal}:& &&&    \beta &=1\,,  &&& a_1&= \f{\pi}{2}\,, &&&   b_1&= \f{\pi}{4}\,, \\
    \text{unitary}:& &&&    \beta &=2\,,  &&& a_2&=\f{32}{\pi^2}\,, &&&   b_2&=\f{4}{\pi}\,, \\
    \text{symplectic}:& &&& \beta &=4\,, &&&
    a_4&=\f{262144}{729\pi^3}\,, &&&   b_4&=\f{64}{9\pi}\,. \\
  \end{aligned}
\end{equation}
For chiral classes, the same ULSD is found as for the corresponding
non-chiral classes. 

For independent eigenvalues, obeying Poisson statistics, the ULSD is the
exponential distribution,
   
\begin{equation}
  \label{eq:Pois}
  p_{\rm Poisson}(s) = e^{-s}\,.
\end{equation}
Both for localized and delocalized modes, exact analytical results are
then available for the statistical properties of the unfolded
spectrum, and the transition from one type of modes to the other can
be easily monitored across the spectrum. This allows in particular to
identify the mobility edge, where a different, critical statistics is
expected instead of Poisson or RMT
statistics~\cite{al1988repulsion,Shklovskii:1993zz}. Various families
of random matrix models have been developed to describe the critical
statistics~\cite{Muttalib:1993zz,Moshe:1994gc,Mirlin:1996zz,canali1996,kravtsov1997,nishigaki1998level,Nishigaki:1999zz,GarciaGarcia:2000zc}.

We mention in passing an alternative approach to the study of
universal statistical properties of the spectrum, based on the use of
the ratio of consecutive level spacings~\cite{Oganesyan_2007}. Since
this ratio is independent of the local spectral density, this approach
has the advantage of not requiring any unfolding, and has been shown
to provide more precise results than those obtained from the unfolded
spectrum in a variety of many-body systems (see Ref.~\cite{Atas_2013}
and references therein).

\subsection{Finite-size scaling at the Anderson transition}
\label{sec:loc_fss}

The mobility edge and the correlation-length critical exponent $\nu$
can be obtained by means of a finite-size scaling
study~\cite{Shklovskii:1993zz} of the average $\Oc_L(\lambda)$ of
suitable observables built out of the unfolded spectrum, and computed
locally in the spectrum, for lattices of linear size $L$. Local
statistics are defined formally as

\begin{equation}
  \label{eq:local_stats}
 \Oc_L(\lambda) \equiv \f{1}{\rho(\lambda)}\la \sum_i
 \delta(\lambda-\lambda_i) \Oc(\lambda_i)\ra \equiv \la \Oc\ra_\lambda \,,
\end{equation}
where $\Oc(\lambda_i)$ is some function of the eigenvalues (e.g., the
level spacing $\Delta\lambda_i=\lambda_{i+1}-\lambda_i$, or the unfolded level spacing
$s_i = x_{i+1}(\lambda_{i+1})-x_{i}(\lambda_{i})$ and its powers), and
the average is over the ensemble. The last equality sets an alternative
notation for local averages.

In practice, for a finite sample of disorder realizations obtained
numerically, local statistics are computed by dividing the spectrum in
small bins and averaging over modes within those, as well as over the sample.
Unfolding is done by first fitting some smooth
$\rho_{\rm average}(\lambda)$ to the numerical data and then applying
Eq.~\eqref{eq:unfolding}. Alternatively, one can sort the eigenvalues
of the sample by magnitude and replace them by their rank divided by
the number of disorder realizations.  If one is interested only in
unfolded level spacings, one can divide $\Delta\lambda_i$ by the
average level spacing $\la \Delta\lambda\ra_\lambda $ in the relevant
spectral region, $s_i = \f{\Delta\lambda_i}{\la
  \Delta\lambda\ra_\lambda}$ (notice that in the infinite-volume,
infinite-statistics limit one has
$\la \Delta\lambda\ra_\lambda = 1/\rho(\lambda)$).

For a finite-size scaling study, convenient observables are obtained
from the unfolded level spacing distribution, $p_{\rm ULSD}(s)$,
defined above. Commonly used are the second moment,
$ \la s^2\ra =\int_0^\infty ds\,p_{\rm ULSD}(s) s^2$, and the
integrated probability distribution
$I_{s_0} = \int_0^{s_0}ds\,p_{\rm ULSD}(s)$. As the system size grows,
$\Oc_L(\lambda)$ tends to its value for Poisson statistics in spectral
regions where modes are localized, and to its value for (the
appropriate) RMT statistics in spectral regions where modes are
delocalized. Near the mobility edge $\lambda_c$, renormalization-group
arguments and the one-parameter scaling
hypothesis~\cite{locgangoffour} imply that $\Oc_L(\lambda)$ depends on
$\lambda$ and $L$ only through the combination $\xi(\lambda)/L$, where
$\xi$ is the correlation length,\footnote{On the insulator side of the
  transition, $\xi$ can be identified with the localization length,
  while on the metallic side it can be related to the
  conductivity~\cite{kramer1993localization}.} that diverges at
$\lambda_c$ like $\xi(\lambda)\sim |\lambda-\lambda_c|^{-\nu}$. Since
$\Oc_L(\lambda)$ is analytic in $\lambda$ for finite $L$, it must then
take the form $\Oc_L(\lambda)=
f((\lambda-\lambda_c)L^{1/\nu})$. Corrections to one-parameter scaling
due to irrelevant operators can also be included, and the
corresponding critical exponents be
measured~\cite{slevin1999corrections} (see
Ref.~\cite{kramer2010finite} for an introduction).  The goodness of
one-parameter scaling can be visualized by means of the so-called
``shape analysis''~\cite{varga1995shape}, obtained by plotting one
spectral observable against another. If the scaling hypothesis is
correct, only $\xi/L$ should determine the statistical properties of
the spectrum, and so points corresponding to different $\lambda$ and
system sizes should all lie on a single curve, corresponding to a path
in the space of probability distributions connecting RMT and Poisson
going through the critical statistics.\footnote{Shape analysis in QCD
  is discussed in Refs.~\cite{Nishigaki:2013uya,Giordano:2014qna}.}
Thanks to the persistence of a remnant of multifractality near the
mobility edge~\cite{cuevas_2007}, one can apply similar finite-size
scaling techniques also to the study of eigenmodes near criticality,
in order to obtain the multifractal exponents~\cite{rodriguez_2009},
as well as the correlation-length exponent
$\nu$~\cite{rodriguez_2010}.

\subsection{Anderson transitions in specific models: analytic
  predictions and numerical results}
\label{sec:loc_AT2}

Critical properties at the Anderson transition have been extensively
studied by means of numerical simulations in the case of the
conventional symmetry classes (O, U, and S), see
Refs.~\cite{romer2003numerical,Evers:2008zz,kramer2010finite,ujfalusi2015finite}
and references therein.  According to the scaling theory of
localization~\cite{locgangoffour}, a second-order Anderson transition
is expected in all the conventional classes in three dimensions.  The
existence of these Anderson transitions has been confirmed
numerically, and measurements of the correlation length critical
exponent $\nu$ have shown that the three classes belong to different
universality
classes~\cite{slevin1999corrections,slevin1997anderson,asada2005anderson,
  rodriguez2011multifractal,ujfalusi2015finite,lindinger2017multif} 
(see Tab.~\ref{tab:loc_nures}).
The expected nontrivial multifractal structure has also been
found~\cite{rodriguez_2009,rodriguez_2010,rodriguez2011multifractal,
  ujfalusi2015finite,lindinger2017multif}.  Universality has been
explicitly demonstrated for the orthogonal class using different
disorder distributions~\cite{slevin1999corrections}, and for the
unitary class using different
Hamiltonians~\cite{ujfalusi2015finite,lindinger2017multif}.

%\end{paracol}
\begin{table}[t]
%  \widetable
  \centering
  \begin{tabular}{c|c|c|c}
    \hline\hline    symmetry class & method & $\nu$ & reference \\ \hline\hline
    \multirow{3}{*}{orthogonal} & localization length of quasi-1d bar
                                            &  $1.57_{-0.02}^{+0.02}$ &    %$1.57(2)$ &
                                                                                        \cite{slevin1999corrections}\rule{0pt}{12pt} \\
                                   & multifractal finite-size scaling &  $1.590_{-0.011}^{+0.012}$ % (1.579,1.602)
                                                    &  \cite{rodriguez2011multifractal}\rule{0pt}{12pt} \\
                                   & multifractal finite-size scaling &
                                                                        $1.595_{-0.013}^{+0.014}$ %(1.582..1.609)
                                                    & \cite{ujfalusi2015finite}\raisebox{-6pt}{\rule{0pt}{18pt}}\\ \hline
    \multirow{3}{*}{unitary} & localization length of quasi-1d bar &
                                                                     $1.43_{-0.04}^{+0.04}$
                                                    & %1.43(4) &
                                                                 \cite{slevin1997anderson}\rule{0pt}{12pt}\\
                                   & multifractal finite-size scaling & $1.437_{-0.011}^{+0.011}$
                                                        %(1.426..1.448)
                                    & \cite{ujfalusi2015finite}\rule{0pt}{12pt}\\
                   & multifractal finite-size scaling & $1.446_{-0.006}^{+0.006}$%  (1.440, 1.452)
                                    & \cite{lindinger2017multif}\raisebox{-6pt}{\rule{0pt}{18pt}}\\ \hline
    \multirow{2}{*}{symplectic} & localization length of quasi-1d bar
                            & $1.375_{-0.016}^{+0.016}$ & %1.375(16) &
                                           \cite{asada2005anderson}\rule{0pt}{12pt} \\
                   & multifractal finite-size scaling &
                                                        $1.383_{-0.024}^{+0.029}$
                                                        % (1.359..1.412)
                                    &
                                      \cite{ujfalusi2015finite}\raisebox{-6pt}{\rule{0pt}{18pt}}\\\hline\hline 
  \end{tabular}
  \caption{Correlation-length critical exponent for Anderson
    transitions in the conventional symmetry classes.}
  \label{tab:loc_nures}
\end{table}
% \begin{paracol}{2}
% \linenumbers
% \switchcolumn

In two dimensions, the predictions of the scaling theory of
localization depend strongly on the details of the model. Absence of
an Anderson transition is predicted in the orthogonal Anderson model,
where all modes are expected to be localized for nonzero disorder,
while an Anderson transition is predicted in the symplectic case (Ando
model)~\cite{hikami1980,WEGNER1989663}.  The inclusion of topological
effects in the field-theoretical description of disordered systems led
one to expect an Anderson transition also in the theory of the integer
Quantum Hall Effect~\cite{PRUISKEN1984277}, which belongs to the
unitary class (see Ref.~\cite{huckestein1995scaling} for a
review). While numerical evidence qualitatively supported this
idea~\cite{slevin2009critical}, significant quantitative discrepancies
between different microscopic models were observed (see
Ref.~\cite{Dresselhaus:2021fbx} for a summary), in contrast with the
expected universality of the transition. A better understanding of the
field theory describing the critical point was obtained only recently,
in terms of a conformal field theory deformed only by marginal
perturbations, that emerge from the spontaneous breaking of the
replica (super)symmetry of the relevant nonlinear sigma
model~\cite{Zirnbauer:2018ooz}. This proposal is quantitatively
supported by numerical results, and can explain the apparent numerical
discrepancies~\cite{Dresselhaus:2021fbx}.  A transition between
localized and delocalized modes was observed in the two-dimensional
unitary Anderson model~\cite{xie1998kosterlitz}. This transition is a
disorder-induced transition of topological
(Berezinski{\u\i}-Kosterlitz-Thouless~\cite{Berezinsky:1970fr,
  Berezinsky:1972fr, Kosterlitz:1973xp}) type, with exponentially
divergent correlation length,
$\log\xi \sim |\lambda-\lambda_c|^{-1/2}$, in contrast to the usual
second-order transition. For the unitary Anderson model there are
conflicting theoretical predictions: while perturbative contributions
lead to all states being localized (see, e.g.,
Ref.~\cite{WEGNER1989663,Evers:2008zz}), the inclusion of
nonperturbative terms can possibly lead to the presence of an Anderson
transition (see references cited in Ref.~\cite{xie1998kosterlitz}). A
similar transition of topological type was also observed in a model
for disordered graphene with strong long-range
impurities~\cite{zhang2009localization}, belonging to the orthogonal
class in two dimensions.

For our purposes, it is important to discuss the effect of
off-diagonal disorder on localization.  In the orthogonal class,
theoretical arguments~\cite{antoniou1977,economou1977localization}
suggest that off-diagonal disorder alone cannot localize modes at the
band center; only increasing the on-site disorder leads eventually to
localization of all the modes. This is confirmed by numerical results
in three dimensions for the orthogonal Anderson model with random
hopping~\cite{weaire1977numerical,cain1999off,biswas2000off,Evangelou_2003}. The
mobility edges $\pm E_c$ separating extended and localized modes move
towards the band center as the on-site disorder $W$ is increased, and
all modes are localized for $W>W_c$.  In the absence of diagonal
disorder ($W=0$) for bipartite lattices, this model belongs to the
chiral orthogonal class.  The critical exponent $\nu$ characterizing
the Anderson transition at $E_c\neq 0$ when $W=0$ is found to be in
agreement with that of the (non-chiral) orthogonal class, as well as
with the one characterizing the transition at $E=0$ as the critical
on-site disorder $W_c$ is reached~\cite{cain1999off,biswas2000off}.

The origin $E=0$ is singled out when the system has chiral symmetry
(which is always the case when the lattice is bipartite and only
off-diagonal disorder is present). In two dimensions, theoretical
arguments predict critical behavior of modes at $E=0$ (i.e., modes are
extended but not fully
delocalized)~\cite{wegner1979largen,oppermann1979disordered,gade1991n,
  gade1993anderson,inui1994,fabrizio2000anderson,motrunich2002,mudry2003},
while all other modes are localized. Numerical results indicate that
modes are indeed critical at
$E=0$~\cite{soukoulis1982study,eilmes1998two,xiong2001power,motrunich2002,
  markovs2007critical,schweitzer2012scaling,markovs2010logarithmic},
but an Anderson transition to localized modes can also
appear~\cite{motrunich2002,bocquet2003}. It has been argued that such
an Anderson transition can be present due to non-perturbative,
topological effects~\cite{Konig_2012}.  In three-dimensional models
with chiral symmetry, Anderson transitions at the origin ($E_c=0$) are
expected to show critical properties differing from those of the
correponding conventional class (and from those at $E_c\neq 0$).
Ref.~\cite{Garcia_Garcia_2006} studied the Anderson transition at
$E\sim 0$ in a chiral unitary model with purely off-diagonal disorder,
finding multifractal exponents differing from those of the
corresponding non-chiral class.  In
Refs.~\cite{Luo_2020,wang2021universality} the Anderson transition at
the origin was studied in two-band models with on-site disorder in the
chiral orthogonal and chiral unitary classes (as well as in other
non-conventional symmetry classes), finding correlation length
critical exponents differing from those of the corresponding
non-chiral classes (and not entirely universal).

The critical properties of Anderson transitions at $E_c\neq 0$ in
systems with chiral symmetry are instead not expected to differ from
those in the corresponding conventional classes.
Ref.~\cite{takaishi2018localization} provides evidence of localization
near the band center in a chiral unitary model mimicking fermions in a
background of correlated spins with antiferromagnetic coupling in
three dimensions, with the same critical properties as the non-chiral
class; in two dimensions states near the band center seem instead to
remain extended.  Anticipating the results discussed in the following
Sections,
Refs.~\cite{GarciaGarcia:2005vj,Giordano:2013taa,Nishigaki:2013uya,
  Ujfalusi:2015nha,Kovacs:2017uiz,Giordano:2019pvc} provide examples
of models in the chiral unitary class, both in three and two
dimensions, displaying Anderson transitions at finite energy, and
showing the same critical properties at the mobility edge as the
corresponding non-chiral classes.

\section{Localization and deconfinement in QCD at finite temperature}
\label{sec:QCD}

The study of localization in QCD was initially motivated by the idea
that the spontaneous breaking of chiral symmetry could have a similar
origin as conductivity in a disordered
medium~\cite{Diakonov:1984vw,Diakonov:1985eg,Diakonov:1995ea,Smilga:1992yp,
  Janik:1998ki,Osborn:1998nm,Osborn:1998nf,GarciaGarcia:2003mn}.  The
basic idea of the disordered medium scenario is that the near-zero
modes responsible for the breaking of chiral symmetry originate from
the mixing of the zero modes associated with overlapping instantons (or,
more precisely, calorons at finite temperature).  At finite
temperature these zero modes are exponentially localized on the scale
of the inverse temperature. If instantons/calorons overlap
sufficiently, mixing of the corresponding zero modes will transform
the zero eigenvalues into a near-zero band of levels, and lead to
delocalized eigenmodes~\cite{Diakonov:1995ea}.\footnote{The
  possibility of localization taking place in QCD was mentioned in
  Ref.~\cite{Halasz:1995vd}.}  This is analogous, for example, to the
Anderson-Mott insulator-metal transition\footnote{{\it Mott
    transitions} are MITs driven by electron-electron interactions, in
  contrast to the disorder-driven Anderson transition. In
  Anderson-Mott transitions both interactions and disorder play an
  important role.} driven by the impurity concentration in doped
semiconductors (see, e.g., Ref.~\cite{zvyagin2006}). Starting from the
chirally symmetric phase and decreasing $T$, the overlap of calorons
increases and eventually leads to a finite density of near-zero,
delocalized modes. This leads to expect a localization-delocalization
transition in the near-zero region.

As we will see below, this scenario is most likely only a part of the
story, and overlooks the important role played by deconfinement in
localizing the low Dirac modes. Instead of sticking to the disordered
medium scenario, we prefer to adopt a more general point of view,
looking at the Dirac operator as a random matrix, ignoring initially
any relation with topological objects and deconfinement.  After a few
introductory remarks on this approach, we review the available results
regarding localization and Anderson transitions in various lattice
approximations for QCD, following a chronological order. Some of the
results discussed here deal with the pure gauge theory, sometimes for
$N_c=2$, and are included in this Section mostly for historical
reasons. A summary of the results for QCD proper and organized by
topic is provided at the end of the Section.

\subsection{The Dirac operator as a random matrix}
\label{sec:QCD_RM}

The Dirac operator in the background of fluctuating gauge fields can
be intepreted as a sparse random matrix, and the properties of its
eigenvalues and eigenvectors can be studied with the machinery
discussed in the previous Section. For the continuum anti-Hermitian
Dirac operator, $-i\slashed{D}$ can be formally treated as the
Hamiltonian of a disordered system, with disorder provided by the
gauge fields. If an Anderson transition is present in its spectrum,
its critical properties are expected to be determined by the symmetry
class of the Dirac operator and by the dimensionality of the
space-time over which it is defined.

Concerning the symmetry class, the four-dimensional Dirac operator for
fundamental fermions in SU($N_c$) theories belongs to the chiral
unitary class for $N_c>2$, and to the chiral orthogonal class for
$N_c=2$.\footnote{The detailed symmetry classification of the Dirac
  operator in various dimensions is provided in
  Refs.~\cite{Verbaarschot:1994qf} (four dimensions),
  Refs.~\cite{Verbaarschot:1994ip,Magnea:1999iv, Magnea:1999ey} (three
  dimensions), and Ref.~\cite{Kieburg:2014eca} (two dimensions).}  The
spectral correlations of Dirac eigenvalues in QCD ($N_c=3$) are then
expected to display GUE-type bulk statistics,\footnote{Bulk statistics
  are not affected by the chiral symmetry, and should not be confused
  with the microscopic statistics near $\lambda=0$, which, in
  contrast, are affected if a nonzero density of modes is present. In
  the chirally broken phase, where
  $ \Sigma = \f{T}{V} \pi \rho(0)\neq 0$, with $V$ the volume, the
  statistical properties of the microscopic spectrum
  $z_i \equiv \lambda_i \Sigma \f{V}{T} $ near $\lambda=0$ are
  described by the microscopic correlations of the chGUE. See
  Ref.~\cite{Verbaarschot:2000dy} for a detailed review.} as long as
the corresponding eigenvectors are delocalized.\footnote{RMT is
  expected to govern correlations up to some characteristic separation
  scale between eigenvalues (``Thouless energy''), both for
  microscopic and bulk
  statistics~\cite{Verbaarschot:1995yi,Osborn:1998nm,Osborn:1998nf}. For
  the role played by fluctuations in the ensemble in determining this
  scale in the case of bulk statistics, see Ref.~\cite{Guhr:1998pa}.}
If localized modes are present, they are expected to obey Poisson
statistics, regardless of the symmetry class.

The discretization of the Dirac operator on a lattice is known to be
tricky due to the doubling problem (see
Refs.~\cite{Creutz:1984mg,Rothe:1992nt_new,Montvay:1994cy,DeGrand:2006zz,
  Gattringer:2010zz}), and in some cases its chiral properties are
changed (see Ref.~\cite{Halasz:1995vd} and
Ref.~\cite{Verbaarschot:2000dy}, sec.~5.2.1).  For staggered
fermions~\cite{Kogut:1974ag,Banks:1975gq,Susskind:1976jm} a remnant of
the continuum chiral symmetry preserves the chiral nature of the
symmetry class, and the staggered Dirac operator belongs to the chiral
unitary class, as the continuum operator, for $N_c\ge 3$. For $N_c=2$
the symmetry class is instead changed to the chiral symplectic
one.\footnote{The symmetry class of the staggered operator is actually
  independent of the spacetime dimension.}  Overlap
fermions~\cite{Narayanan:1993sk,Narayanan:1993ss,
  Neuberger:1997fp,Neuberger:1998wv} possess an exact lattice chiral
symmetry, and belong to the same symmetry class as their continuum
counterpart. More precisely, since the overlap operator is not
anti-Hermitean, this is true for its anti-Hermitean part. It is then
understood, unless specified otherwise, that the imaginary part of the
overlap eigenvalues is considered in the following. For the low modes
this is an adequate approximation, that becomes exact in the continuum
limit. Moreover, since the unfolded spectrum is unaffected by any
monotonic mapping, the statistical properties of the low modes are
unchanged if one uses other types of projection on the imaginary axis
(i.e., eigenvalue magnitude, stereographic projection).\footnote{We
  note in passing that for adjoint fermions in four dimensions the
  relevant class is the chiral symplectic class for all $N_c$ in the
  continuum, and on the lattice with overlap fermions; for staggered
  fermions it is instead the chiral orthogonal class for all
  $N_c$~\cite{Verbaarschot:2000dy,Bruckmann:2008xr}, independently of
  the dimension.}

Concerning the dimensionality of the problem, in finite-temperature
field theory the temporal size of the system is fixed (in physical
units) in the thermodynamic limit, and only the size of the $d$
spatial directions is sent to infinity in the thermodynamic limit. For
$d+1$ spacetime dimensions, the dimensionality of the disordered
system described by the Dirac operator in a gauge-field background is
then equal to $d$, while the temporal direction can be technically
seen as an internal degree of freedom.

 \end{paracol}
\begin{figure}[t]
\widefigure
  \centering
  \includegraphics[width=0.32\textwidth]{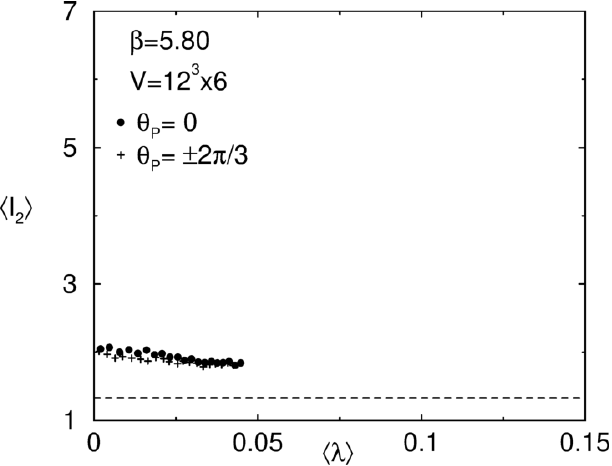}
  \hfil  \includegraphics[width=0.32\textwidth]{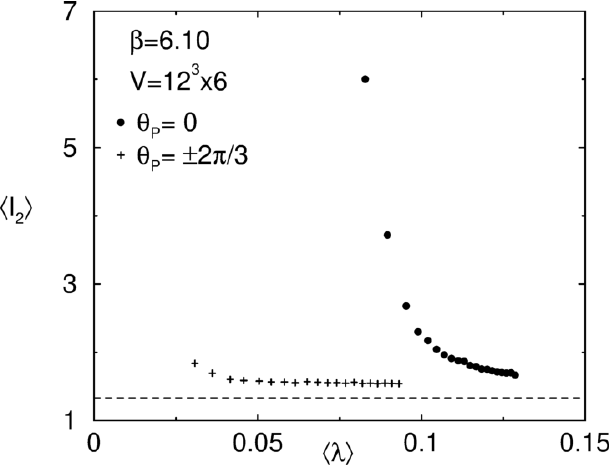}
  \hfil    \includegraphics[width=0.32\textwidth]{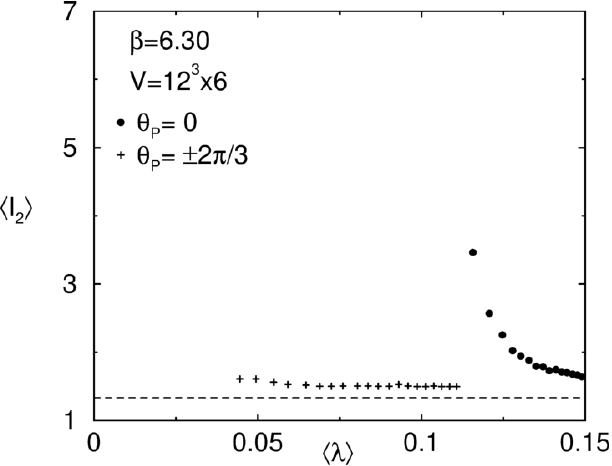}
  \caption{IPR of the low staggered modes on quenched configurations
    below (\textbf{left}), slightly above (\textbf{center}), and well
    above (\textbf{right}) the deconfinement transition for real
    ($\theta_P=0$) and complex ($\theta_P=\pm \f{2}{3}$) Polyakov loop
    sectors.  From Ref.~\cite{Gockeler:2001hr}.  \newline
    {\footnotesize Reprinted figure with permission from
      M.~G{\"o}ckeler, P.E.L.~Rakow, A.~Sch{\"a}fer, W.~S{\"o}ldner,
      and T.~Wettig, Phys.\ Rev.\ Lett.\ 87, 042001 (2001). Copyright
      (2001) by the American Physical Society.}  \label{fig:ipr_gock}}
  \end{figure}
\begin{paracol}{2}
\linenumbers
\switchcolumn

\subsection{Numerical results on the lattice}
\label{sec:QCD_num}

The disordered medium scenario was investigated in
Ref.~\cite{Gockeler:2001hr} by means of numerical simulations of
quenched QCD on the lattice on both sides of the finite-temperature
transition. They used a single spatial volume, employed the staggered
discretization of the Dirac operator, and studied the rotation- and
gauge-invariant version of the IPR of an eigenmode $\psi$,
Eq.~\eqref{eq:gaugeinvIPR}. They observed that in the physical
$\mathbb{Z}_3$ sector (real Polyakov loop sector) the IPR of the
lowest modes was considerably larger above the transition than below
the transition (see Fig.~\ref{fig:ipr_gock}). Moreover, above $T_c$ it
was larger in the real sector than in the complex Polyakov loop
sectors, where it does not change much across the
transition. Sensitivity to the Polyakov loop sector is equivalent to
sensitivity to the temporal boundary conditions, and shows that the
low modes cannot be localized in the temporal direction on a scale
much shorter than the temporal size.\footnote{From this observation,
  Ref.~\cite{Gockeler:2001hr} concluded that modes are actually
  extended in the temporal direction. This is actually not necessary:
  localization in the temporal direction on a scale comparable with
  the temporal size is sufficient for modes to be sensitive to the
  boundary conditions.} This suggests the presence of a localization
transition in the physical sector, with spatially localized low modes
at high temperature.  Evidence for some of the localized modes being
related to calorons was also provided.  More evidence for localization
of the low modes in the real sector appeared in
Ref.~\cite{Gattringer:2001ia}, where more volumes and a chirally
improved discretization of the Dirac operator were used. The volume
scaling of the IPR of the low modes in the real sector was found to be
in qualitative agreement with that expected for localized modes.

 \end{paracol}
\begin{figure}[t]
  \widefigure
  \centering
  \includegraphics[width=0.45\textwidth]{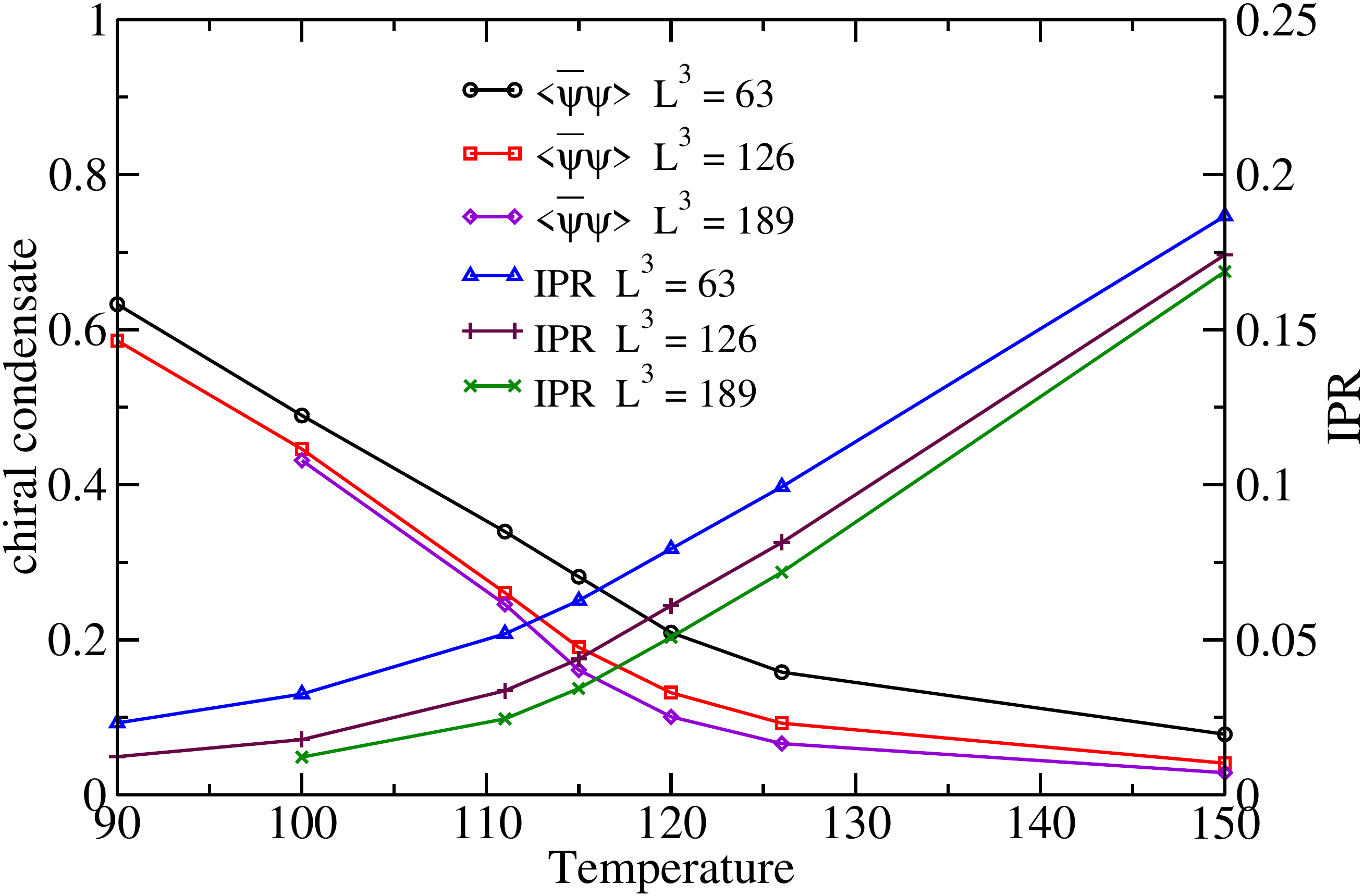}\hfil
  \includegraphics[width=0.435\textwidth]{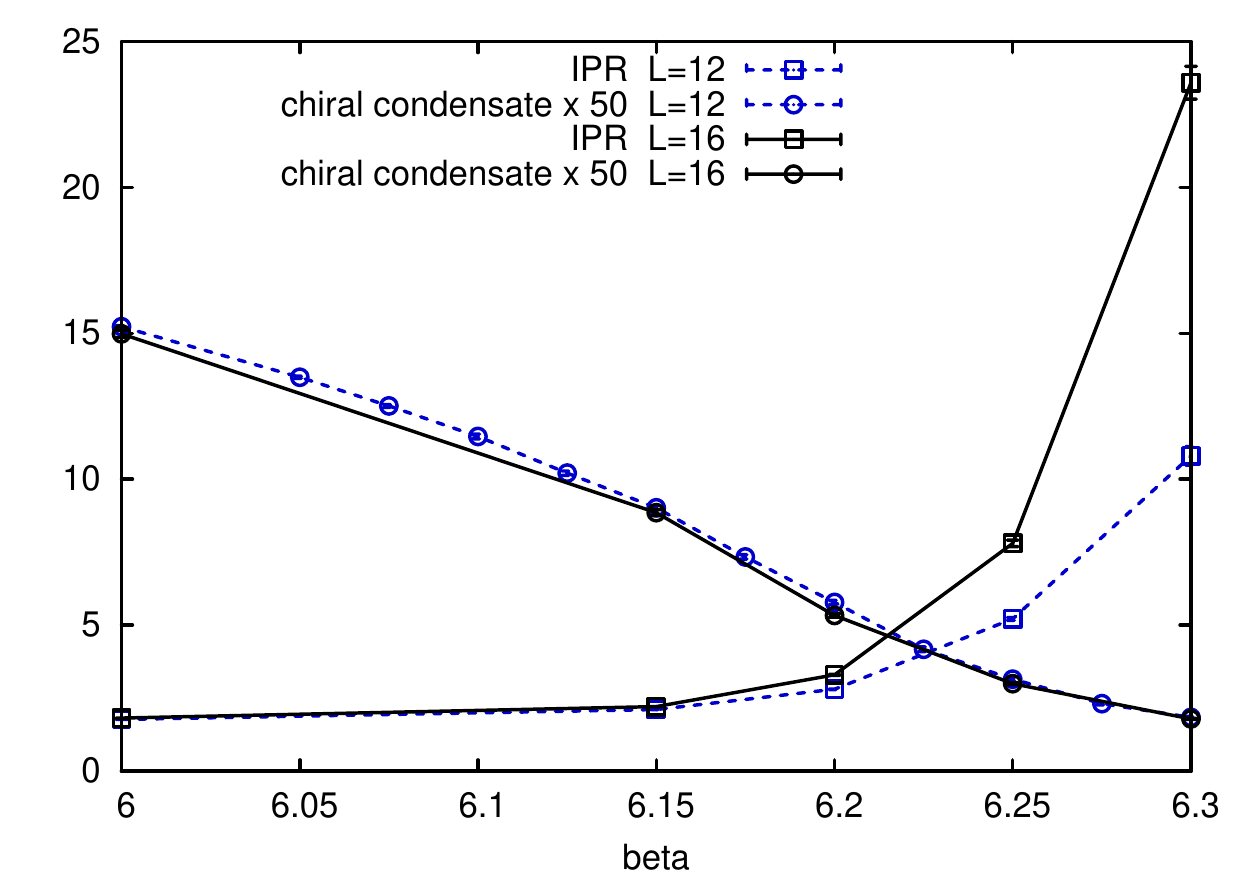}
  \caption{\textbf{Left}: IPR of the lowest Dirac mode and chiral
    condensate in an ILM model for QCD with two massless fermions for
    various system sizes (in ${\rm fm}^3$). From
    Ref.~\cite{GarciaGarcia:2005vj}.  \textbf{Right}: IPR (here times
    the volume $V=L^3$) of the low Dirac modes and chiral condensate in 2+1 flavor
    QCD with staggered fermions. From Ref.~\cite{GarciaGarcia:2006gr}.
    \newline {\footnotesize Reprinted from Nucl.\ Phys.\ A, 770,
      A.M.~Garc\'ia-Garc\'ia and J.C.~Osborn, ``Chiral phase
      transition and Anderson localization in the Instanton Liquid
      Model for QCD'', Pages 141--161, Copyright (2006), with
      permission from Elsevier.  Reprinted figure with permission from
      A.M.~Garc\'ia-Garc\'ia and J.C.~Osborn, Phys.\ Rev.\ D 75,
      034503 (2007). Copyright (2007) by the American Physical
      Society.}
    \label{fig:ipr_cc_ggo}}
\end{figure}
\begin{paracol}{2}
\linenumbers
\switchcolumn

The disordered medium scenario was investigated further by
Garc\'ia-Garc\'ia and Osborn in
Refs.~\cite{GarciaGarcia:2005vj,GarciaGarcia:2006gr}. In
Ref.~\cite{GarciaGarcia:2005vj} they considered an Instanton Liquid
Model (ILM) for the QCD vacuum (see Ref.~\cite{Schafer:1996wv}), and
studied the behavior of the instantonic zero modes. Changing the
temperature, and so the spatial extension of the zero modes, they
observed the appearance of a mobility edge near the origin, both in
the quenched approximation and in the presence of fermions. In the
quenched case, the multifractal properties of the near-zero modes at
the transition were found to be consistent with those of the 3d
unitary Anderson transition (see
Refs.~\cite{Evers:2008zz,ujfalusi2015finite,lindinger2017multif}). With
two massless flavors, the mobility edge appears at the same
temperature where the chiral condensate shows a drop (see
Fig.~\ref{fig:ipr_cc_ggo}, left). Although the thermodynamic limit was
not studied, this was taken as an indication that localization of the
low modes coincides with the chiral transition.

A test of the disordered medium scenario in a more realistic context
was presented in Ref.~\cite{GarciaGarcia:2006gr}. There the
localization properties of the near-zero modes were studied on the
lattice in quenched QCD, i.e., pure gauge SU(3) theory, and
``unquenched'' QCD, i.e., with 2+1 flavors of dynamical quarks of
relatively large masses, leading to heavier-than-physical
pions~\cite{Bernard:2005mf,Aubin:2004wf}.  The one-loop Symanzik
improved gauge action was used, and the
Asqtad-improved~\cite{Orginos:1998ue,Toussaint:1998sa,Lepage:1997id,Lepage:1998vj}
staggered discretization was employed for the lattice Dirac operator.
In both cases, they found indications of critical (i.e.,
volume-independent) spectral statistics (from the second moment of the
ULSD) at a temperature $T_c^{\rm loc}$, where also ${\rm IPR}\cdot V$
starts increasing with the volume (for the unquenched case see
Fig.~\ref{fig:ipr_cc_ggo}, right). In the quenched case, indications
of a vanishing spectral density and of an increase of the Polyakov
loop are found at a similar temperature $T_c^{{\rm dec}/\chi}$,
identified with the deconfinement temperature (in the physical
$\mathbb{Z}_3$ sector). In the unquenched case, $T_c^{\rm loc}$ is
close to the crossover temperature $T_c^{\chi}$ obtained from the
chiral susceptibility. Although the use of small lattices does not
allow a full quantitative assessment, these indications suggest that
an Anderson transition takes place near the origin of the spectrum as
the system crosses over from the low-temperature to the
high-temperature phase, with the low-lying Dirac modes turning from
delocalized to localized, and the formation of a mobility edge that
separates them from delocalized modes in the bulk of the spectrum.

\end{paracol}
\begin{figure}[t]
  \widefigure
  \centering
  \includegraphics[width=0.4\textwidth]{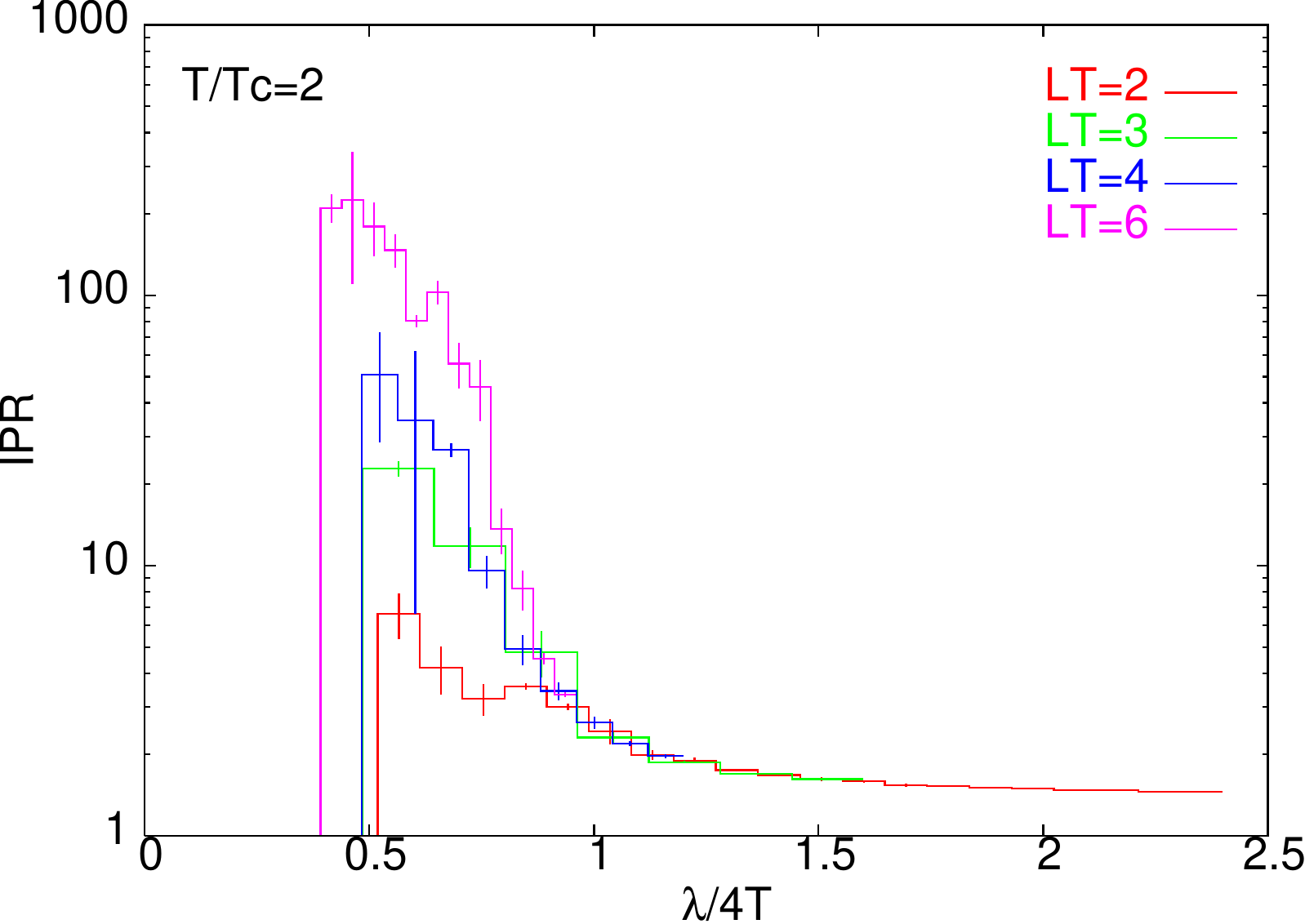}\hfil
  \includegraphics[width=0.4\textwidth]{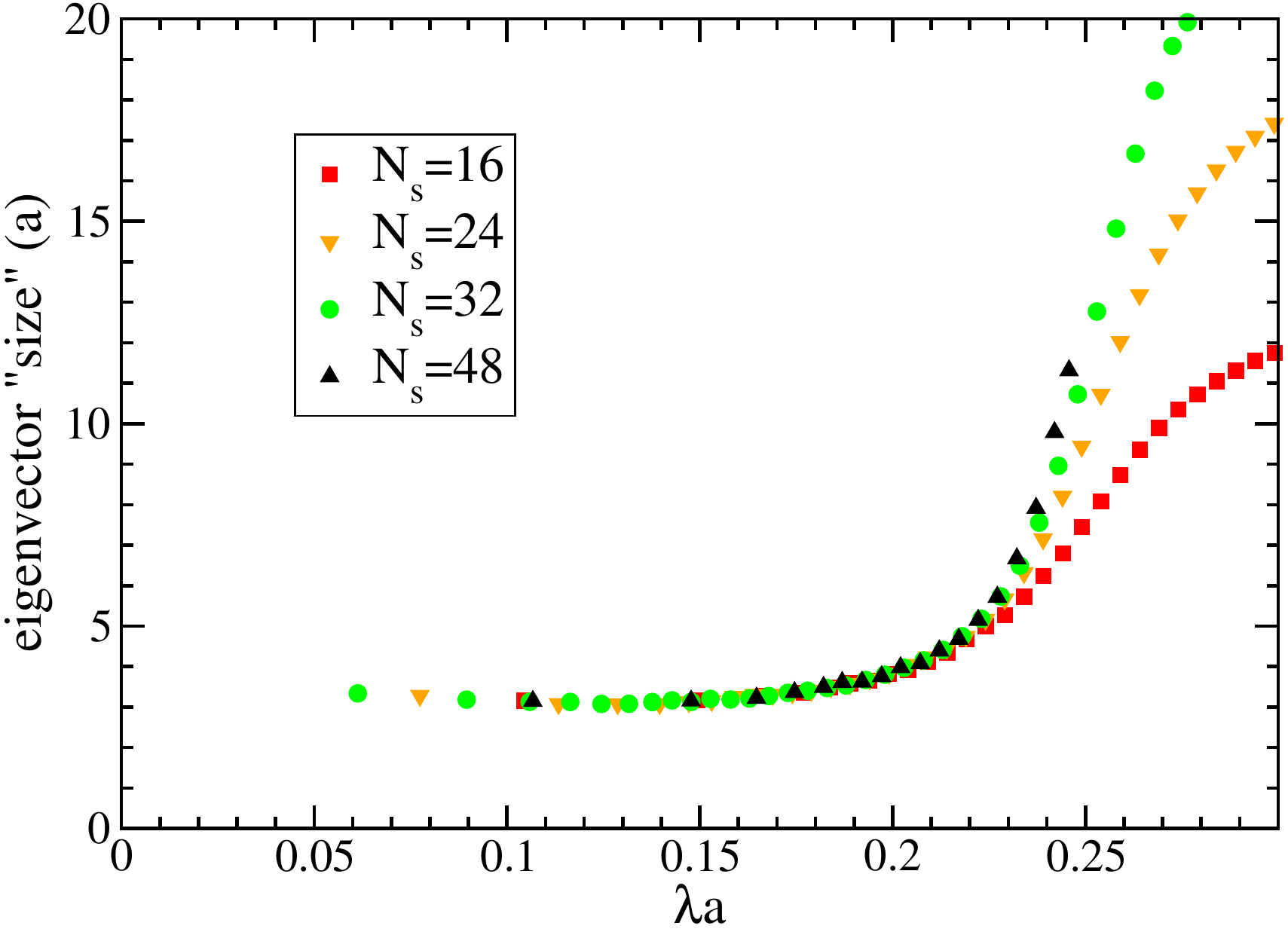}
  \caption{\textbf{Left}: IPR (here times the volume $V=L^3$) of low
    staggered modes for various volumes at $T=2T_c^{2f}$ in QCD with
    two flavors of dynamical staggered fermions. From
    Ref.~\cite{Gavai:2008xe}. \textbf{Right}: Mode size for low staggered
    modes in the background of quenched SU(2) configurations at
    $T=2.6T_c^{{\rm SU}(2)}$. From Ref.~\cite{Kovacs:2010wx}.
    \newline {\footnotesize Reprinted figure with permission from
      R.V.~Gavai, S.~Gupta, and R.~Lacaze, Phys.\ Rev.\ D 77, 114506
      (2008). Copyright (2008) by the American Physical Society.
      Reprinted figure with permission from T.G.~Kov\'acs and
      F.~Pittler, Phys.\ Rev.\ Lett.\ 105, 192001 (2010). Copyright
      (2010) by the American Physical Society.}
    \label{fig:ggl_kp}}
\end{figure}
\begin{paracol}{2}
\linenumbers
\switchcolumn

Studies of localization in QCD-like settings includes also the case of
two flavors of dynamical staggered fermions~\cite{Gavai:2008xe}, and
that of quenched two-color QCD (i.e., pure gauge SU(2) theory)
analyzing overlap~\cite{Kovacs:2009zj,Bruckmann:2011cc} and
staggered~\cite{Kovacs:2010wx} spectra. In the two-flavor three-color
case Ref.~\cite{Gavai:2008xe} found that ${\rm IPR}\cdot V$ of the low
modes was volume-independent below the transition temperature
$T_c^{2f}$, but above that it scaled with the volume in a manner
compatible with localization (see Fig.~\ref{fig:ggl_kp}, left).
Ref.~\cite{Kovacs:2009zj} shows evidence of absence of correlations in
the low-lying overlap spectrum, which is typical of localized modes,
at $T=2.6T_c^{{\rm SU}(2)}$, where $T_c^{{\rm SU}(2)}$ is the
deconfinement temperature of the pure gauge SU(2)
theory. Ref.~\cite{Kovacs:2010wx} shows clear evidence of localization
of the low-lying staggered modes, and of the presence of a mobility
edge separating them from delocalized bulk modes, again at
$T=2.6T_c^{{\rm SU}(2)}$. This is obtained by studying how (i) the
scaling with the lattice spatial volume of the spatial ``size''
$({\rm IPR}^{-1}/N_t)^{\f{1}{3}}$ of the eigenmodes, and (ii) the ULSD
of the corresponding eigenvalues change along the spectrum. The
spatial extension of the low modes is volume-independent, while higher
up in the spectrum it is seen to increase with the lattice size (see
Fig.~\ref{fig:ggl_kp}, right). Looking at the ULSD in different
spectral regions, it is observed that it matches the exponential
distribution of Poisson statistics for the lowest modes, changing
towards the symplectic Wigner surmise\footnote{\label{foot:class}For
  the fundamental representation of the gauge group SU(2) the
  staggered operator is in the symplectic class due to the property
  $\sigma_2 U \sigma_2 = U^*$ of SU(2) matrices $U$ (see
  Ref.~\cite{Halasz:1995vd} and Ref.~\cite{Verbaarschot:2000dy},
  sec.~5.2.1).}  as one moves towards the bulk. Finally, in
Ref.~\cite{Bruckmann:2011cc} the transition in the overlap spectrum
from localized to delocalized modes is studied via the ULSD at
$T=2.6T_c^{{\rm SU}(2)}$. A clear change from the exponential to the
orthogonal Wigner surmise is observed.\footnote{\label{foot:class1}The
  overlap operator is in the same symmetry class as the corresponding
  continuum operator, so the orthogonal class for fundamental fermions
  and gauge group SU(2) (see Ref.~\cite{Halasz:1995vd} and
  Ref.~\cite{Verbaarschot:2000dy}, sec.~5.2.1).} Moreover, assuming
that there are no strong interactions among instantons and
anti-instantons, it is argued that the instanton density is too low to
match the density of localized modes at this temperature. Indications
of correlations between localized modes and local fluctuations of the
Polyakov loop away from its ordered value (i.e., 1, in the physical
sector) are also reported (see Section \ref{sec:mech_seaislands} for a
detailed discussion).

\end{paracol}
\begin{figure}[t]
  \widefigure
  \centering
  \includegraphics[width=0.45\textwidth]{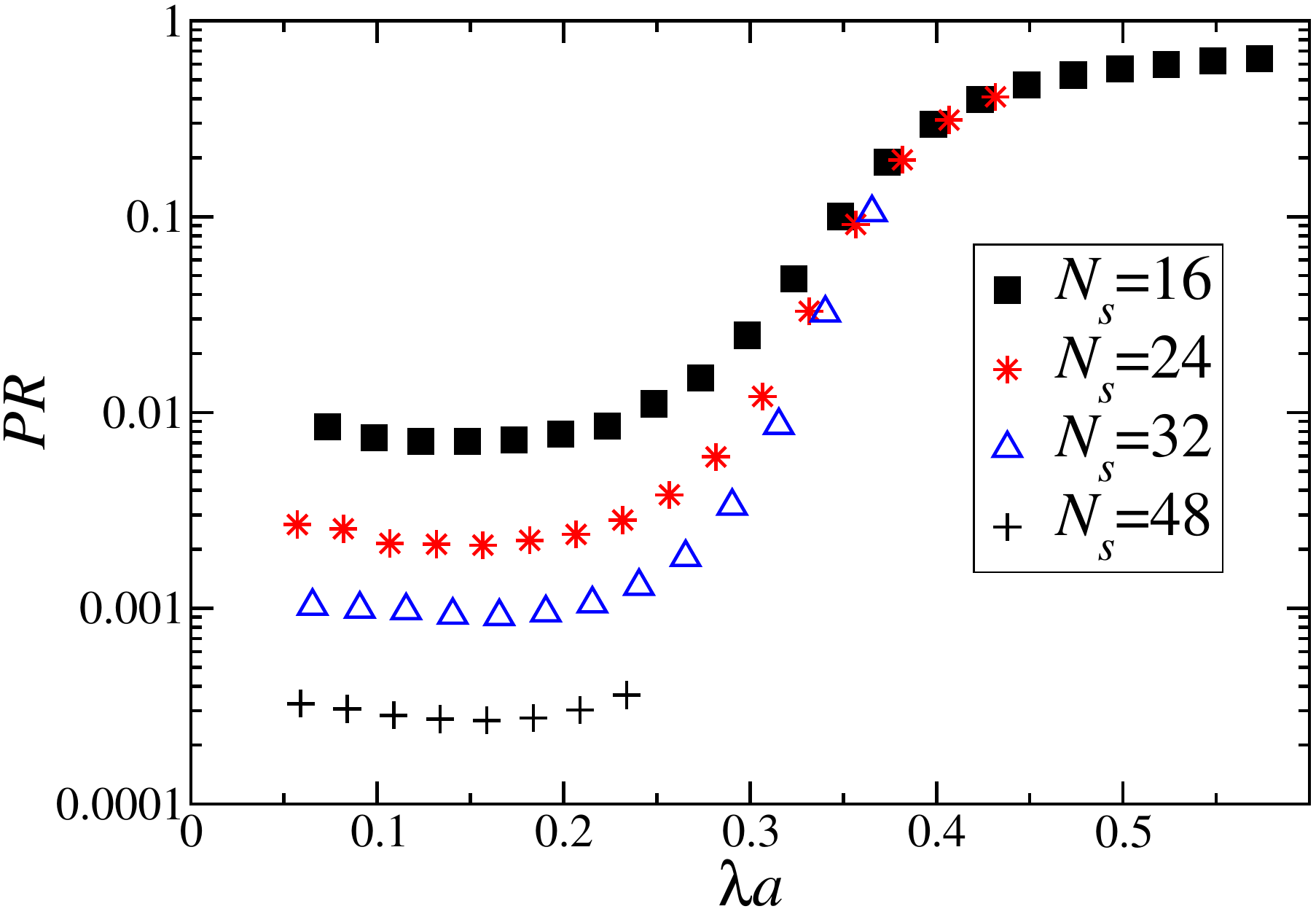}\hfil
  \includegraphics[width=0.445\textwidth]{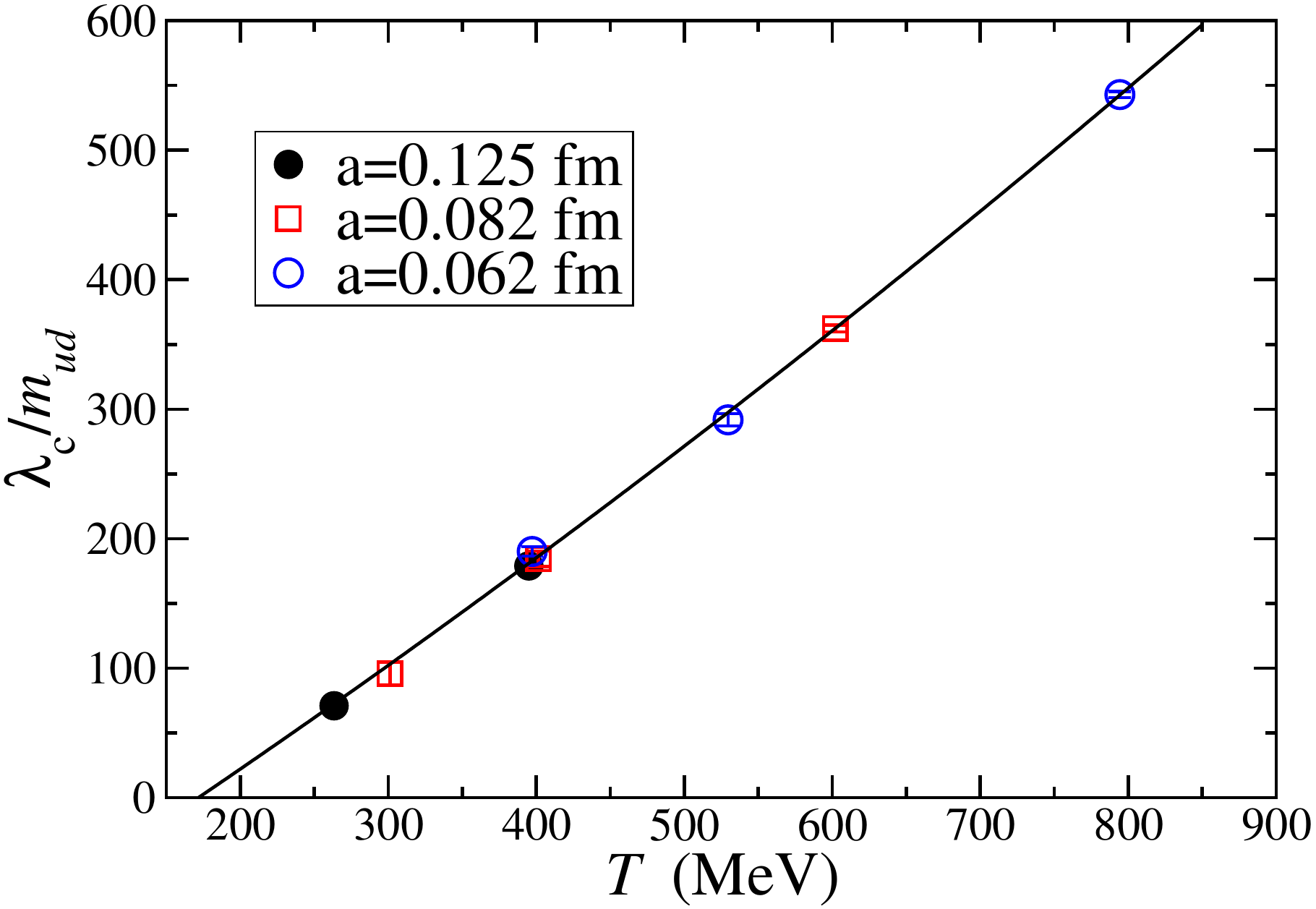}
  \caption{Average PR of the staggered Dirac modes as a function of
    the eigenvalue $\lambda$ for various volumes at
    $T=394\,{\rm MeV}$, $a=0.125\,{\rm fm}$ (\textbf{left}) and
    renormalized mobility edge $\lambda_c/m_{\rm ud}$ as a function of
    $T$ (\textbf{right}) in 2+1 flavor QCD with staggered fermions and
    physical quark masses. From Ref.~\cite{Kovacs:2012zq}.  \newline
    {\footnotesize Reprinted figures with permission from
      T.G.~Kov\'acs and F.~Pittler, Phys.\ Rev.\ D 86, 114515
      (2012). Copyright (2012) by the American Physical Society.}
    \label{fig:prlc_kp}}
\end{figure}
\begin{paracol}{2}
\linenumbers
\switchcolumn

A comprehensive study of localization in the high-temperature phase of
real-world QCD appeared in Ref.~\cite{Kovacs:2012zq}, using a
tree-level Symanzik improved gauge action and a two-level stout
smeared~\cite{Morningstar:2003gk} staggered fermion action for 2+1
quark flavors with physical mass~\cite{Aoki:2005vt}. Several volumes,
aspect ratios and lattice spacings were used, covering the temperature
range $1.7 T_c < T < 5T_c$ (here
$T_c=155~{\rm MeV}$~\cite{Borsanyi:2010bp}) with lattices of linear
size $2\, {\rm fm} \le L \le 6\, {\rm fm}$. Localized modes were
observed at the low end of the spectrum (see Fig.~\ref{fig:prlc_kp},
left). A temperature-dependent mobility edge $\lambda_c(T)$ separating
low-lying, localized modes from delocalized bulk modes was found in
the whole temperature range, studying how the spectral statistics
change along the spectrum from Poisson to unitary RMT type. More
precisely, $\lambda_c$ was estimated as the inflection point of the
variance of the ULSD,
$\la s^2\ra_\lambda-\la s\ra_\lambda^2 = \la s^2\ra_\lambda -1 $,
computed locally in the spectrum.  As $T$ increases, $\lambda_c(T)$
increases as well. Extrapolation to the continuum is studied at
$T=400~{\rm MeV}$. The mobility edge is expected to renormalize like a
quark mass, and the ratio $\lambda_c/m_{\rm ud}$ is indeed shown to be
independent of the lattice spacing within numerical uncertainties. The
localization length $l\equiv a\la{\rm IPR}^{-\f{1}{4}}\ra$ of the low
modes is also shown to extrapolate to a finite continuum limit, and
$lT$ is found to be between 0.7 and 0.9 for all the lattice
ensembles. A second-order polynomial fit to the RG-invariant quantity
$\lambda_c(T)/m_{\rm ud}$ shows that it extrapolates to zero at
$T_c^{\rm loc}=170~{\rm MeV}$ (see Fig.~\ref{fig:prlc_kp}, right),
which is within the temperature range where the system undergoes a
crossover from the low-temperature phase to the high-temperature
phase~\cite{Borsanyi:2010bp,Bazavov:2016uvm}. This is consistent with
localization of the low modes appearing as the system changes from
confined and chirally broken to deconfined and chirally restored. The
density of localized modes (number of modes per unit spatial volume)
is seen to increase with $T$ (see Fig.~\ref{fig:locmoddens} below).

The critical behavior of the eigenmodes at the mobility edge was
studied in
Refs.~\cite{Giordano:2013taa,Nishigaki:2013uya,Ujfalusi:2015nha}.  All
these references use the same setup as Ref.~\cite{Kovacs:2012zq} with
$N_t=4$ and $a=0.125~{\rm fm}$, corresponding to $T=2.6 T_c$.  In
Ref.~\cite{Giordano:2013taa} it was established, by means of a finite
size scaling analysis of the integrated ULSD $I_{s_0}$, that the
transition from localized to delocalized modes at the mobility edge is
indeed an Anderson transition (see Fig.~\ref{fig:scaling_MFE_gkp},
left). The critical exponent was found to be $\nu=1.43(6)$, in
agreement with the one obtained for the 3d unitary Anderson
model~\cite{slevin1997anderson} (see Tab.~\ref{tab:loc_nures}). In
Ref.~\cite{Nishigaki:2013uya} the critical eigenvalue statistics at
the mobility edge was studied in terms of the one-parameter family of
deformed random matrix ensembles of
Refs.~\cite{nishigaki1998level,Nishigaki:1999zz}. The critical
statistics was shown to be indeed volume-independent, and well
described by a deformed random matrix ensemble, with deformation
parameter consistent with the one found for the 3d unitary Anderson
model~\cite{slevin1997anderson,ujfalusi2015finite,lindinger2017multif}.
Finally, in Ref.~\cite{Ujfalusi:2015nha} the critical exponent $\nu$
and the multifractal exponents where studied using the finite size
scaling techniques for the eigenmode density developed in
Refs.~\cite{rodriguez_2009,rodriguez_2010}. All exponents were found
to be in agreement with those of the 3d unitary Anderson
model~\cite{ujfalusi2015finite,lindinger2017multif} (see
Fig.~\ref{fig:scaling_MFE_gkp}, right).

\end{paracol}
\begin{figure}[t]
  \widefigure
  \centering
  \includegraphics[width=0.45\textwidth]{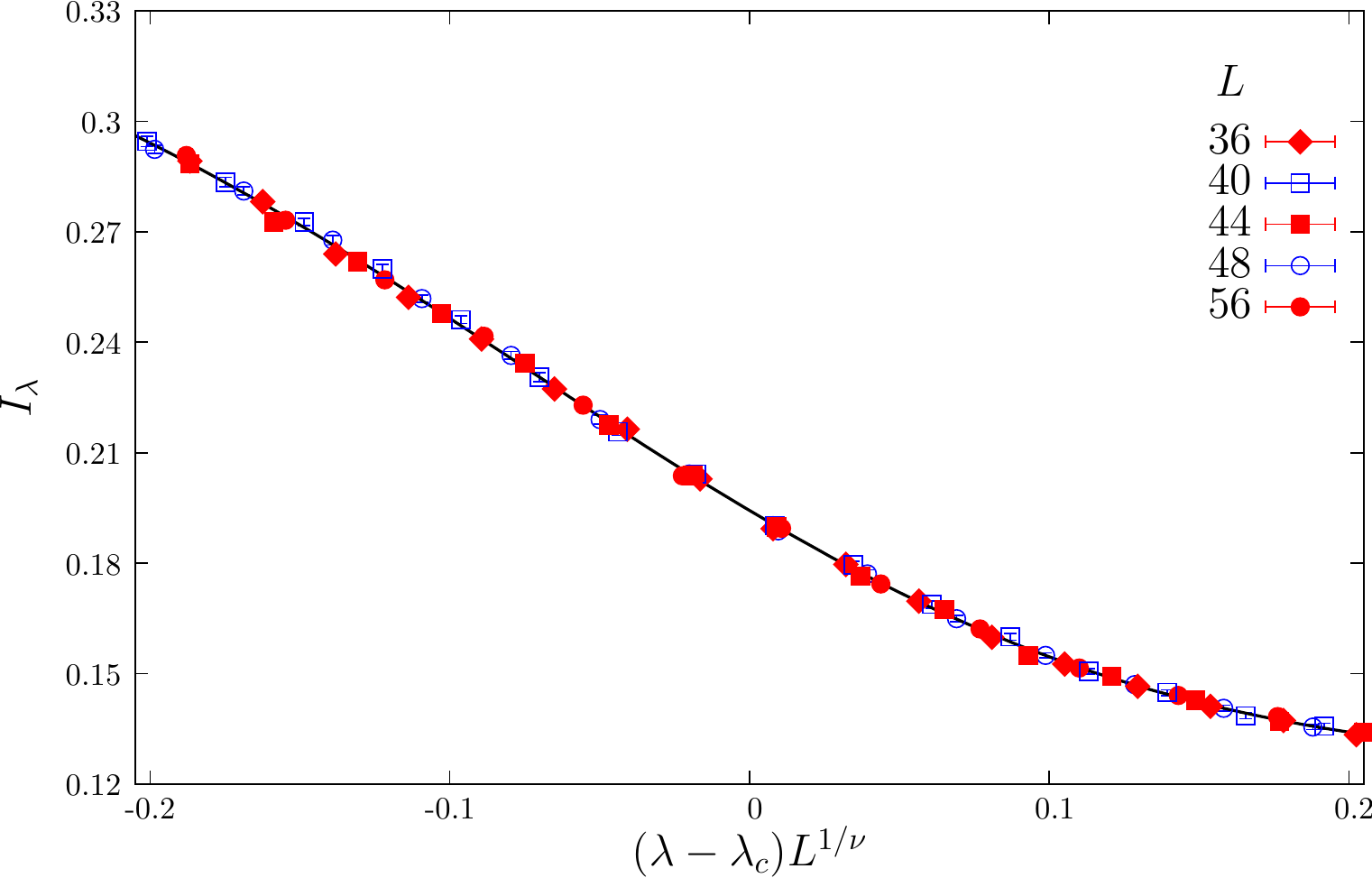}\hfil
  \includegraphics[width=0.4\textwidth]{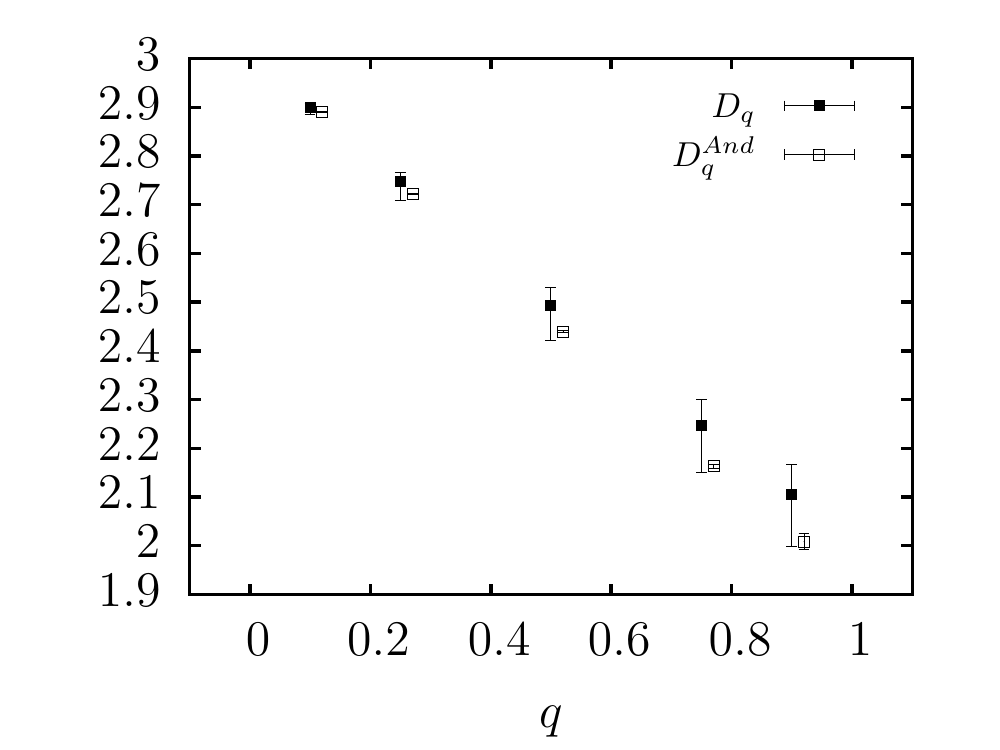}
  \caption{\textbf{Left}: scaling of $I_{s_0}$ near the mobility edge in
    2+1 QCD with staggered quarks, at $T=394\,{\rm MeV}$ and
    $a=0.125\,{\rm fm}$. The linear size $L$ of the lattice is
    expressed in lattice units.  Here $\lambda_c\simeq 0.3364$,
    $\nu\simeq 1.43$, and the critical value is
    $I_{s0}|_{\lambda_c} \simeq 0.194$.  From
    Ref.~\cite{Giordano:2013taa}. \textbf{Right}: multifractal exponents
    $D_q$, governing the scaling of ${\rm IPR}_q \sim L^{-D_q(q-1)}$
    at large linear size $L$ at the mobility edge, in 2+1 QCD with
    staggered quarks ($T=394\,{\rm MeV}$, $a=0.125\,{\rm fm}$) and in
    the 3d unitary Anderson model. From Ref.~\cite{Ujfalusi:2015nha}.
    \newline {\footnotesize Reprinted figure with permission from
      M.~Giordano, T.G.~Kov\'acs and F.~Pittler, Phys.\ Rev.\ Lett.\
      112, 102002 (2014). Copyright (2014) by the American Physical
      Society. Reprinted figure with permission from L.~Ujfalusi,
      M.~Giordano, F.~Pittler, T.G.~Kov\'acs and I.~Varga, Phys.\
      Rev.\ D 92, 094513 (2015).  Copyright (2015) by the American
      Physical Society.}
    \label{fig:scaling_MFE_gkp}}
\end{figure}
\begin{paracol}{2}
\linenumbers
\switchcolumn

While mostly focussed on the properties of the spectrum,
Ref.~\cite{Dick:2015twa} briefly discussed the localization properties
of the low Dirac modes in QCD near the crossover temperature. The
spectrum of the overlap operator was studied in the background of
gauge configurations generated with tree-level improved Symanzik gauge
action and 2+1 flavors of highly improved staggered quarks
(HISQ)~\cite{Follana:2006rc} with near-physical quark masses
($m_l/m_s=1/20$, $m_\pi= 160\,{\rm MeV}$). Evidence was found of a
small peak of localized near-zero modes at $T=1.2T_c$ and $T=1.5T_c$
(here $T_c=154~{\rm MeV}$). Localization was inferred from the
smallness of the PR; the volume scaling was not discussed. It was
suggested that near-zero modes in the peak correspond to an
approximate superposition of the exact zero modes associated with
instanton--anti-instanton pairs.  Comparison of the PR of zero and
near-zero modes shows however that only a fraction of near-zero modes
is compatible with this interpretation. The large fluctuations of the
PR of the near-zero modes suggests instead a large variability of the
number of topological lumps participating in the superposition.

Localization in two-flavor QCD was studied in
Ref.~\cite{Cossu:2016scb} using tree-level improved Symanzik gauge
action and dynamical M{\"o}bius domain-wall
fermions~\cite{Brower:2004xi,Brower:2005qw,Brower:2012vk} with stout
smearing, and looking at the spectrum of the Hermitian operator
$\gamma_5 D$, with $D$ the four-dimensional effective Dirac operator
of the five-dimensional domain-wall fermion. This operator is in the
chiral unitary class. The temperature range was
$0.9 T_c \le T \le 1.9 T_c$ with $T_c\simeq 175~{\rm MeV}$ the
deconfinement temperature estimated from the average Polyakov loop. A
range of bare quark masses, two spatial volumes and two temporal
extensions (in lattice units) were used. Above $T_c$, the scaling of
the PR of the eigenmodes shows that the lowest modes are localized,
while moving up in the spectrum modes become delocalized (see
Fig.~\ref{fig:PR_ch}, left). The size $v\equiv V\cdot {\rm PR}$ of the
low modes increases along the spectrum and decreases with $T$ (see
Fig.~\ref{fig:PR_ch}, right). The localization length
$l\equiv v^{\f{1}{4}}$ of the lowest (nonzero) mode was shown to be of
the order of the inverse temperature, $lT\sim 1.3$. The ULSD computed
locally in the spectrum was seen to change from Poisson-type to
RMT-type in the unitary class as one moves from the lowest modes
towards the bulk. By contrast, below $T_c$ RMT statistics was observed
everywhere in the spectrum.  Changing boundary conditions to periodic
in time, low modes were found to be delocalized with RMT statistics
also above $T_c$. A clear correlation between the spatial density
$\psi^\dag\psi(x)$ of the low modes and the local fluctuations of the
Polyakov loop $P(\vec{x})$ away from its ordered value was observed,
favoring sites with $\Re \tr P(\vec{x})$ close to $-1$, and becoming
stronger as $T$ increases (see Fig.~\ref{fig:dens_PL}
below). Correlation with action ($s(x)$) and topological charge
($q(x)$) densities was also observed, with localized modes favoring
sites with large $s$ and $q$, in particular ``(anti)self-dual'' sites
where $|q|/s \sim 1$. The overlap of the left- and right-chirality
components of the modes was seen to be the smallest for the lowest
modes, and to increase as one moves towards the bulk; it was also seen
to increase with temperature, and showed little dependence on the bare
quark mass.

\end{paracol}
\begin{figure}[t]
  \widefigure
  \centering
  \includegraphics[width=0.45\textwidth]{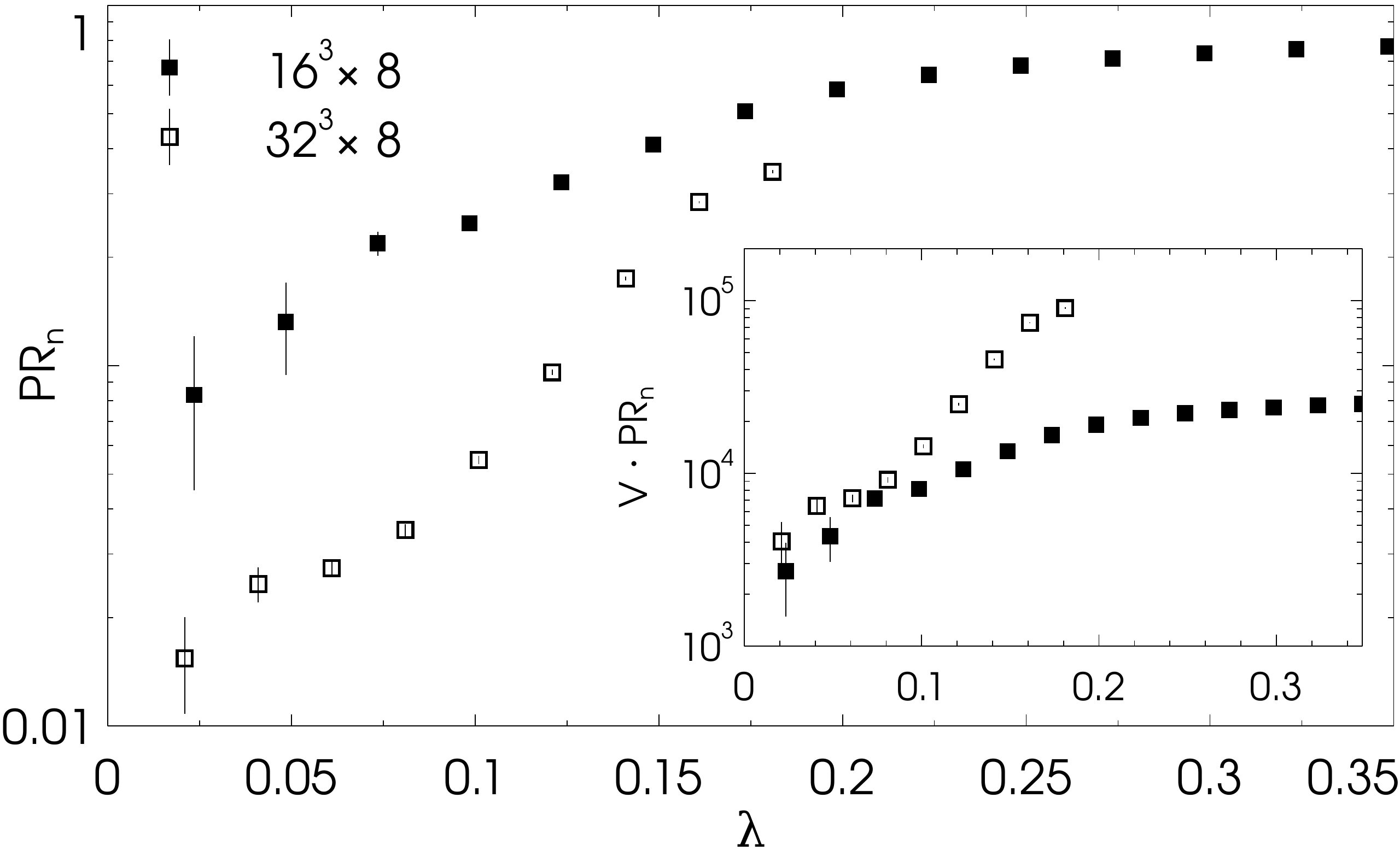}\hfil
  \includegraphics[width=0.45\textwidth]{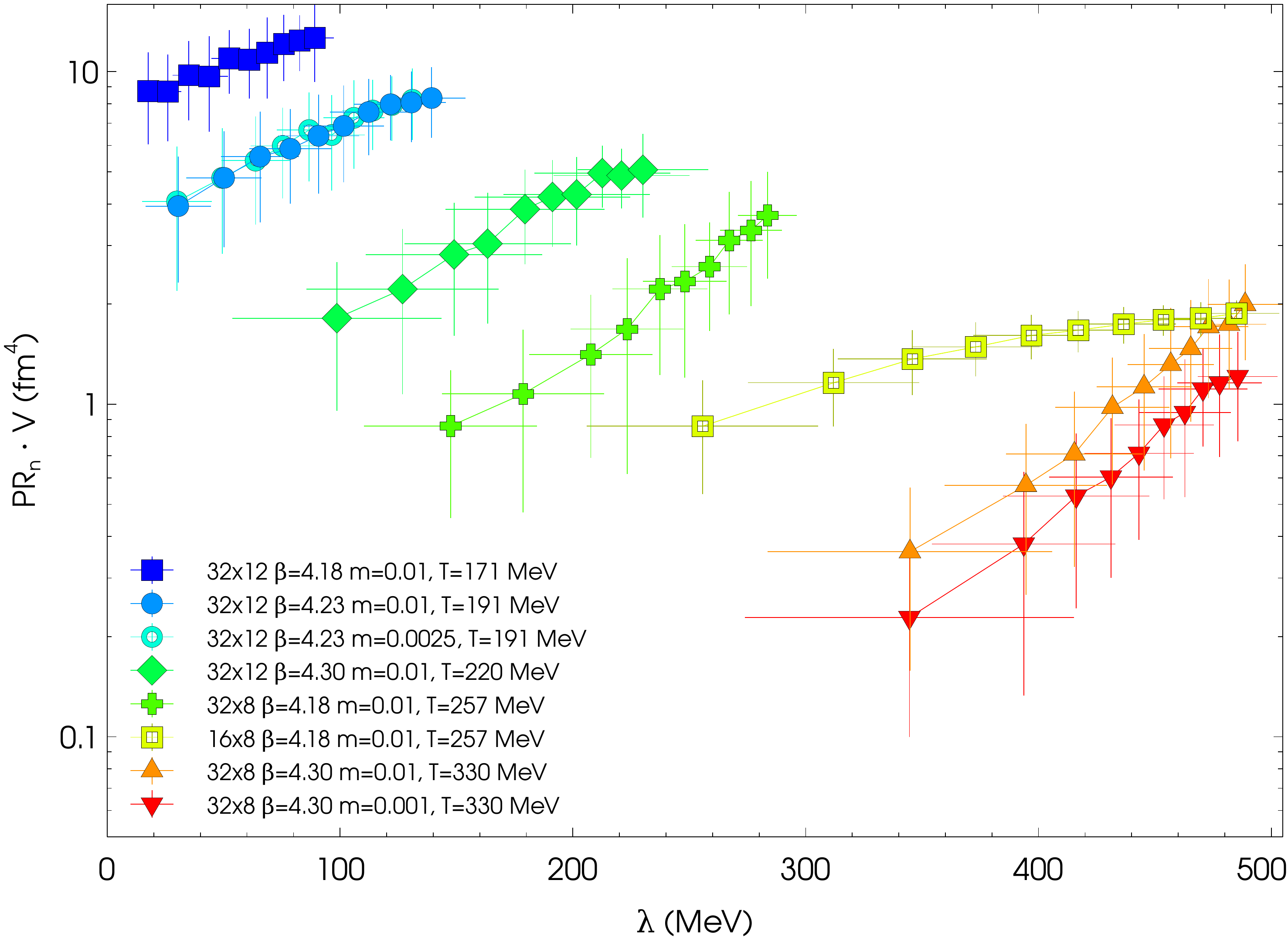}
  \caption{Scaling of the PR (and of $V\cdot {\rm PR}$ in the inset)
    for $\beta=4.18$, $N_t=8$ ($T=257\,{\rm MeV}\simeq 1.5 T_c$) and
    bare mass $m=0.01$ (\textbf{left}), and dependence on $T$ and $m$
    of the PR of the 10 lowest modes (\textbf{right}) in two-flavor
    QCD with M{\"o}bius domain-wall fermions. From
    Ref.~\cite{Cossu:2016scb}.  \newline {\footnotesize Figures
      adapted from G.~Cossu and
      S.~Hashimoto, % \textit{Anderson Localization in high temperature
      % QCD: Background configuration properties and Dirac eigenmodes},
    J.\ High Energy Phys.\ 06, 56 (2016), and used under a
    \href{https://creativecommons.org/licenses/by/4.0}{CC-BY 4.0}
    license.}
  \label{fig:PR_ch}}
\end{figure}
\begin{paracol}{2}
\linenumbers
\switchcolumn

In Ref.~\cite{Holicki:2018sms} (see also Ref.~\cite{Holicki:2019ufl}),
localization was studied in 2+1+1 flavor QCD with physical strange and
charm masses but heavy pions ($m_\pi\simeq 370\,{\rm MeV}$), looking
at the (stereographically projected) spectrum of the overlap operator
in the background of configurations generated with Iwasaki gauge
action and dynamical twisted-mass Wilson
fermions~\cite{Burger:2013hia,Burger:2015xda,Burger:2017xkz}. Above
$T_c\simeq 188~{\rm MeV}$, a very small PR is found for the lowest
modes, which increases to around 0.8 in the bulk. The position of the
mobility edge was estimated as the inflection point of the PR in the
spectrum. As a function of $T$, $\lambda_c(T)$ appears to be linear,
and its extrapolation vanishes at a temperature compatible with
$T_c$. A strong anticorrelation of the localized modes with
$\Re \tr P(\vec{x})$ was observed.

\subsection{Summary}
\label{sec:QCD_sum}

Let us summarize the results discussed in this Section, restricting to
QCD ``proper'', i.e., gauge group SU(3) and dynamical quarks. QCD-like
models and other gauge theories are discussed in detail below in
Section \ref{sec:othergt}.
\begin{itemize}
\item {\bf Low Dirac modes are localized in lattice QCD in the
    high-temperature phase}. More precisely, a large amount of
  evidence indicates that the low Dirac modes are localized in lattice
  QCD, for temperatures above the finite-temperature transition, for
  more or less physical quark content and masses, and different
  fermion discretizations~\cite{GarciaGarcia:2006gr,
    Kovacs:2012zq,Dick:2015twa,Cossu:2016scb,Holicki:2018sms}.  The
  available evidence suggests that localization is not a lattice
  artifact and survives the continuum limit: both the localization
  length and the renormalized mobility edge seem in fact to possess a
  continuum limit. Evidence is, however, limited to a single study, and
  at a single temperature~\cite{Kovacs:2012zq}.
\item {\bf Localization appears approximately at the transition}. As
  the transition is only a crossover, this statement can only be of
  qualitative nature. In all the cases discussed
  above~\cite{GarciaGarcia:2006gr,Kovacs:2012zq,
    Cossu:2016scb,Holicki:2018sms}, localization appears somewhere in
  the range of temperatures where the crossover takes place.
\item {\bf The localization length is of the order of the inverse
  temperature}~\cite{Kovacs:2012zq,Cossu:2016scb}.
\item {\bf An Anderson transition takes place in the Dirac spectrum in
  the high-tem\-per\-a\-ture phase}.
  More precisely, a mobility edge separating localized and delocalized
  modes in the spectrum is observed on the
  lattice~\cite{Kovacs:2012zq,Giordano:2013taa,Nishigaki:2013uya,
    Ujfalusi:2015nha,Cossu:2016scb,Holicki:2018sms}. For staggered
  fermions it has been shown that a genuine Anderson transition takes
  place at the mobility edge~\cite{Giordano:2013taa,
    Nishigaki:2013uya, Ujfalusi:2015nha}.
\item {\bf Localized modes correlate with local fluctuations in the
    confining and topological properties of the configurations}. More
  precisely, the spatial position of localized modes shows
  correlations with the local fluctuations of the Polyakov loop away
  from order~\cite{Cossu:2016scb,Holicki:2018sms}, as well as with
  positive fluctuations of the action density and of the magnitude of
  the topological charge density, especially at (anti)self-dual
  points~\cite{Cossu:2016scb}.
\end{itemize}
We now list a few remarks.
\begin{itemize}
\item As disordered systems, almost all the models discussed in this
  Section are in the 3d chiral unitary class.\footnote{The only
    exception is the SU(2) theory studied in
    Refs.~\cite{Kovacs:2009zj,Bruckmann:2011cc,Kovacs:2010wx}, which
    is in the 3d chiral orthogonal or chiral symplectic class
    depending on the fermion discretization, see footnotes
    \ref{foot:class} and \ref{foot:class1}.}  The appearance of
  localized modes at the band center contrasts with the delocalized
  nature of the band center in the 3d chiral orthogonal Anderson
  model~\cite{cain1999off,biswas2000off}. On the other hand, it agrees
  with what was found in the 3d chiral unitary Anderson
  model~\cite{Garcia_Garcia_2006}, and in the Anderson model with
  correlated disorder of Ref.~\cite{takaishi2018localization} in the
  same class.
\item The results of
  Refs.~\cite{Giordano:2013taa,Nishigaki:2013uya,Ujfalusi:2015nha}
  indicate that a genuine second-order Anderson transition is present
  in the staggered Dirac spectrum in high-temperature QCD, in the
  universality class of the 3d unitary Anderson model.  Since QCD is
  in the 3d chiral unitary class, this suggests that the Anderson
  transition at nonzero eigenvalue for the 3d chiral and non-chiral
  unitary classes belong to the same universality class.  This is not
  surprising, as chiral symmetry is not expected to play an important
  role in the bulk of the spectrum, but only near the origin, around
  which the spectrum is symmetric precisely due to chiral
  symmetry. Further support to the lack of any differences in the
  transition of the chiral and non-chiral model is given by the
  findings of Ref.~\cite{GarciaGarcia:2005vj} concerning the
  multifractal exponents in the ILM model for QCD, and by the critical
  statistics found in the Anderson model with correlated disorder of
  Ref.~\cite{takaishi2018localization}.  A different critical behavior
  is found instead when the Anderson transition is at the origin in 3d
  chiral
  models~\cite{Garcia_Garcia_2006,Luo_2020,wang2021universality}.
\item In the ILM model of Ref.~\cite{GarciaGarcia:2005vj}, both in the
  quenched and unquenched cases, a second mobility edge was observed
  higher up in the spectrum, moving towards the high end as the
  temperature is decreased. While this part of the spectrum is not
  representative of real QCD, as the model neglects nonzero modes at
  the outset, it is nonetheless possible that a similar localization
  mechanism at the high end of the spectrum applies in QCD as
  well.\footnote{Localized modes at the high end of the staggered
    Dirac spectrum have also been found in 2+1-dimensional
    $\mathbb{Z}_2$ gauge theory~\cite{Baranka:2021san}, see Section
    \ref{sec:othergt}.}
\item It is now clear that the Dirac spectral density does not vanish
  in the deconfined phase of pure gauge SU(3) theory, if one uses
  sufficiently fine lattices, or lattice discretizations of the Dirac
  operator with good chiral properties; instead, a peak is formed near
  the origin (see
  Refs.~\cite{Edwards:1999zm,Alexandru:2015fxa,Kovacs:2017uiz}). A
  sort of ``chiral transition'' still takes place at deconfinement,
  where the peak structure appears.
\item The disordered medium scenario requires that the densities
  of instantons and of localized modes match in the high-temperature
  phase. As observed in Ref.~\cite{Bruckmann:2011cc} and, in the pure
  gauge case, in Ref.~\cite{Kovacs:2019txb} (see Section
  \ref{sec:othergt}), the instanton density, obtained assuming an ideal
  (non-interacting) instanton gas approximation, is lower than the density of localized
  modes (number of modes per unit spatial volume). Moreover, the
  latter is seen to increase with $T$~\cite{Kovacs:2012zq}, while the
  instanton density decreases. This indicates that topology can only
  partially explain the localization of the low Dirac modes.
\item An alternative interpretation of localization in terms of
  topological objects was proposed in Ref.~\cite{Cossu:2016scb}. The
  authors suggest that localized modes favor regions where $L$-type
  (Kaluza-Klein) monopole-antimonopole pairs are located. These are
  one of the types of monopole constituents inside
  calorons~\cite{Kraan:1998sn}. This interpretation is supported by
  the correlation with Polyakov-loop fluctuations, action and
  topological density, and chirality. A direct identification of
  monopoles or a quantitative estimate of their density is, however,
  unavailable.

\end{itemize}

\section{Mechanisms for localization}
\label{sec:mech}

In this Section, we discuss in some detail the two mechanisms, or more
precisely the two sources of disorder, proposed so far to explain
localization of the low Dirac modes in QCD: the disordered medium
scenario, based on topology fluctuations; and the sea/islands picture,
based on fluctuations of the Polyakov loop. We have already briefly
discussed the disordered medium scenario in the previous Section; here
we discuss it again, both to keep this Section self-contained, and to
give more details.

\subsection{The disordered medium scenario}
\label{sec:mech_dms}

As is well known, the continuum Dirac operator in the background of a
gauge configuration of topological charge $Q$ has $n_\pm$ exact zero
modes of definite chirality $\pm 1$, with $Q=n_+-n_-$ (index theorem).
In particular, for instantons (resp.~anti-instantons) of topological
charge 1 (resp.~$-1$) one finds an exact zero mode of positive
(resp.~negative) chirality.  The same holds for the finite-temperature
generalization of instantons known as
calorons~\cite{Harrington:1978ve,Harrington:1978ua,Kraan:1998kp,
  Kraan:1998pm,Kraan:1998sn,Lee:1997vp,Lee:1998vu,Lee:1998bb}.\footnote{For
  a review of instantons and calorons we refer the reader to
  Refs.~\cite{Diakonov:1995ea,Schafer:1996wv,Diakonov:2009jq}.} The
zero modes supported by instantons (i.e., at $T=0$) are algebraically
localized, decaying like $1/R^3$ with the distance $R$ from the
instanton center. The zero modes supported by calorons are instead
fully delocalized in the temporal direction, and exponentially
localized in the spatial directions, decaying like $e^{-rT}$, with $r$
the spatial distance from the caloron center and $T$ the temperature
of the system.  For a dilute ensemble of these objects, their
associated zero modes are not exact Dirac eigenmodes any longer, due
to the fact that instantons/calorons overlap.  The low-lying Dirac
eigenmodes are instead linear combinations of these ``unperturbed''
zero modes,\footnote{In a first approximation, the nonzero unperturbed
  modes associated with topological objects can be neglected.}
obtained by diagonalizing the ``perturbed'' Dirac operator, which in
the zero-mode basis reads (see, e.g., Ref.~\cite{Schafer:1996wv})

\begin{equation}
  \label{eq:dismedscen1}
  i\slashed{D} =
  \begin{pmatrix}
    \mathbf{0} & T_{IA} \\
    T_{AI} &     \mathbf{0}
  \end{pmatrix}\,,
\end{equation}
with $T_{IA}$ and $T_{AI}=T_{IA}^\dag$ the matrices of the overlap
integrals of $i\slashed{D}$ between the unperturbed zero modes
associated with an instanton--anti-instanton pair.\footnote{Overlap
  integrals vanish for a pair of instantons or anti-instantons due the
  definite (and equal) chirality of the zero modes.}  For a dilute
ensemble the total topological charge $Q$ is expected to be simply
equal to the sum of the individual charges. Out of all the unperturbed
zero modes, $Q$ are preserved by topology despite %in spite of
mixing,\footnote{While the index theorem requires only $Q=n_+-n_-$, it
  is expected that only zero modes of one chirality appear in typical
  gauge configurations.} while the remaining ones are not protected by
topology and spread around $\lambda=0$ forming a band.  The extent of
this spreading and the resulting density of near-zero Dirac modes for
typical gauge configurations are dynamical issues, determined by the
typical density and size of topological objects.  In the quenched
case, a finite spectral density of near-zero modes is expected to
survive as long as a non-negligible density of topological objects
supports them. In the presence of dynamical fermions, the fermionic
determinant tends to suppress configurations with a higher density of
near-zero modes, and so suppresses topological excitations, with
respect to the quenched case, but a finite spectral density is still
possible.  In any case, the details of the dynamics, including
especially the temperature and the fermion masses, determine whether a
nonzero density of near-zero modes is formed, i.e., loosely speaking,
whether chiral symmetry is spontaneously broken.

\end{paracol}
\begin{figure}[t]
  \centering
  \includegraphics[width=0.45\textwidth]{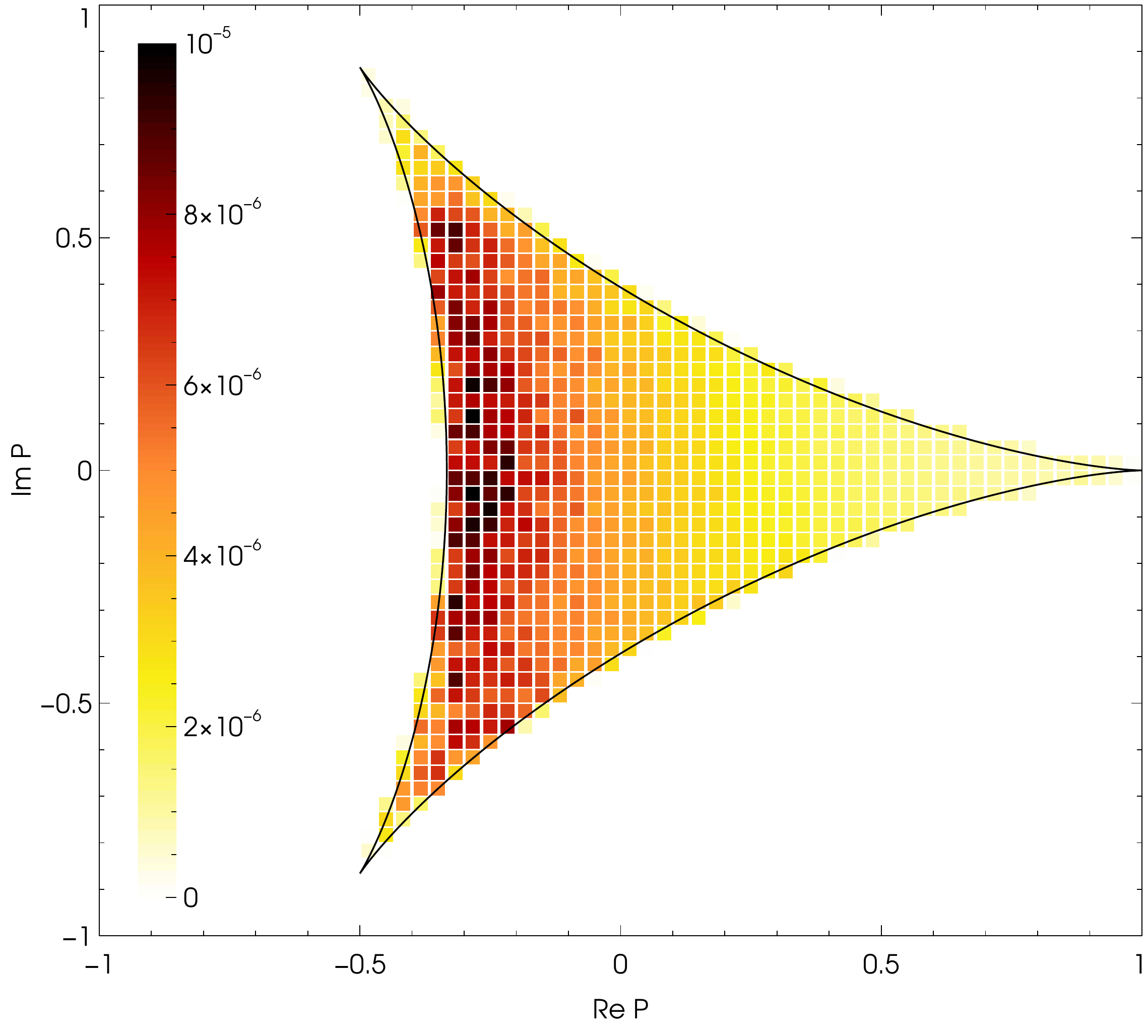}
  \caption{Density plot of the average local norm $\psi^\dag\psi$ of
    low Dirac modes in the Polyakov loop plane (here
    $ \text{\fontfamily{cmss}\selectfont P} = \f{1}{3}\tr\,P$) in the
    high-temperature phase. From Ref.~\cite{Cossu:2016scb}.  \newline
    {\footnotesize Figure adapted from G.~Cossu and
      S.~Hashimoto, % \textit{Anderson Localization in high temperature
      % QCD: Background configuration properties and Dirac eigenmodes},
    J.\ High Energy Phys.\ 06, 56 (2016), and used under a
    \href{https://creativecommons.org/licenses/by/4.0}{CC-BY 4.0}
    license.}
  \label{fig:dens_PL}}
\end{figure}
\begin{paracol}{2}
\linenumbers
\switchcolumn

For sufficiently low temperature, and not too many quark flavors, the
density of topological objects and the effect of mixing become strong
enough to overcome the repulsive effect of the fermionic determinant,
and chiral symmetry breaks spontaneously through the formation of a
nonzero density of near-zero modes. This is the disordered medium
scenario for chiral symmetry breaking~\cite{Diakonov:1984vw,
  Diakonov:1985eg,Diakonov:1995ea,Smilga:1992yp,Janik:1998ki,
  Osborn:1998nm,Osborn:1998nf,GarciaGarcia:2003mn}. The mixing of the
unperturbed modes is also expected to spread them out in space, over
topological objects that overlap non-negligibly with their original
location.  If the typical spatial distance $n^{-\f{1}{3}}$ between
topological objects is large compared to the typical spatial range
$1/T$ of the corresponding unperturbed zero modes, one expects the
resulting perturbed near-zero modes to remain localized on a few
objects only.  Here $n= \f{{\cal N}_{\rm top}}{V}$ is the spatial
density of ${\cal N}_{\rm top}$ calorons and anti-calorons in a finite
spatial volume $V$. At high temperatures both density and range are
small, and near-zero modes are expected to be localized. As the
temperature decreases, both density and range increase, with more and
more topological objects overlapping, and near-zero modes are expected
to eventually delocalize over the whole system~\cite{Diakonov:1995ea}.
It is reasonable to expect that delocalization will take place around
the same temperature as chiral symmetry breaking (in the loose sense
explained above). It should be clear, however, that finite spectral
density and delocalization of modes near the origin are not
automatically linked.\footnote{For example, modes are localized at the
  band center in the Anderson model above the critical disorder, but
  with finite spectral density; and in the near-zero spike found right
  above $T_c$ in QCD and pure gauge SU(3) theory.}

According to the scenario above, near-zero localized modes should be
associated with local lumps of topological charge. It is worth noting
that (anti)calorons in SU($N_c$) gauge theory are made up of $N_c$
(anti)monopole constituents~\cite{Kraan:1998sn}, and that when these
are well separated the associated zero mode is localized on a single
constituent; which one depends on the holonomy (Polyakov loop) of the
gauge field at asymptotic distance from the
core~\cite{GarciaPerez:1999ux,Chernodub:1999wg}.  For typical
high-temperature ordered configurations with Polyakov loop in the
trivial sector, the relevant constituent is the type-$L$ monopole,
which also has the largest action and topological charge densities, as
well as the smallest size (see Ref.~\cite{Diakonov:2009jq}). This
further characterizes the favorable locations for modes according to
the disordered medium scenario.

\end{paracol}
\begin{figure}[t]
  \widefigure
  \centering
    \includegraphics[width=0.45\textwidth]{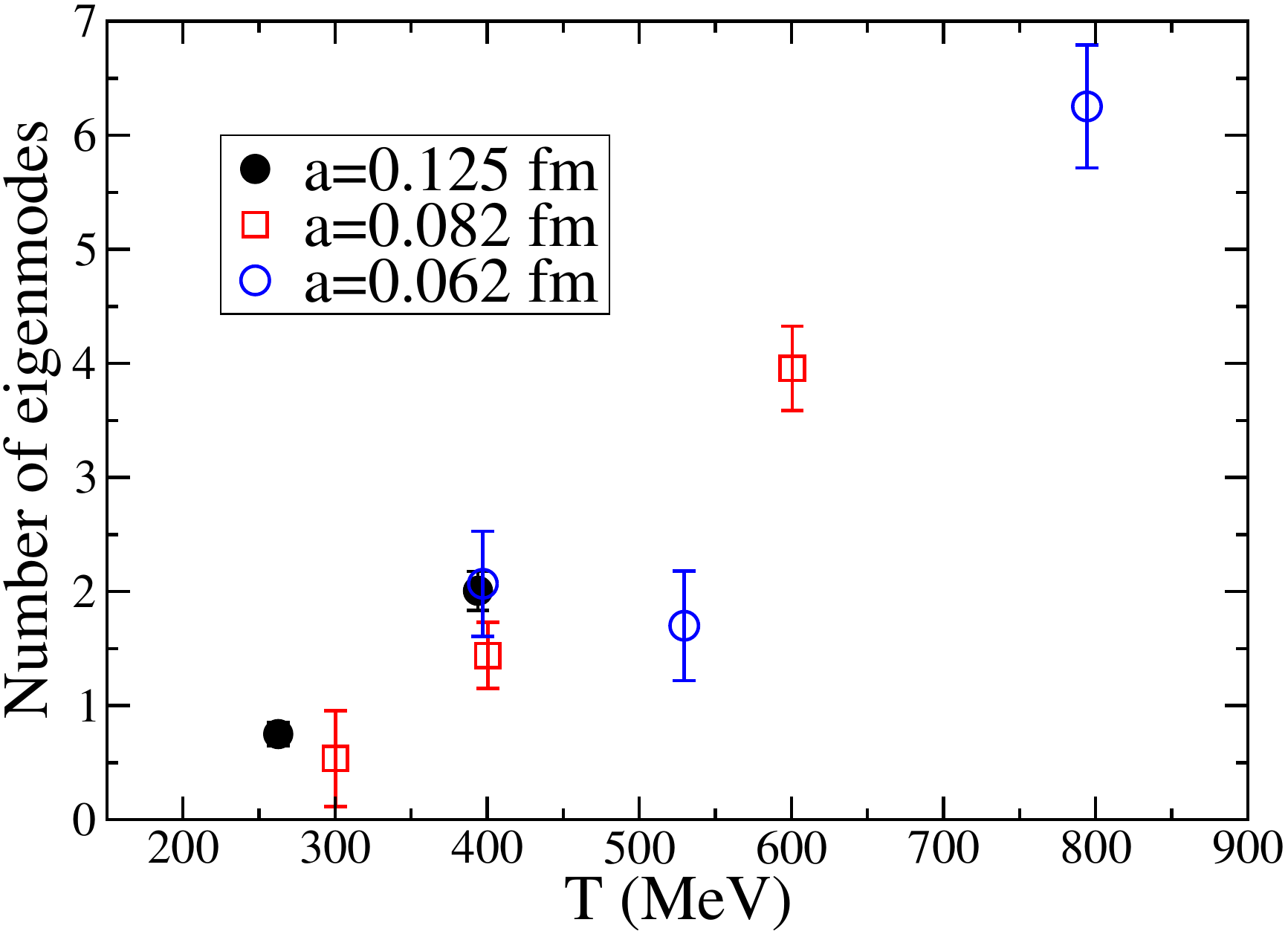}
  \caption{Density of localized modes (modes per cubic fermi) in
    high-temperature QCD.  From Ref.~\cite{Kovacs:2012zq}.  \newline
    {\footnotesize Reprinted figure with permission from T.G.~Kov\'acs
      and F.~Pittler, Phys.\ Rev.\ D 86, 114515 (2012). Copyright
      (2012) by the American Physical Society.}
    \label{fig:locmoddens}}
\end{figure}
\begin{paracol}{2}
\linenumbers
\switchcolumn

Refs.~\cite{Gockeler:2001hr,Gattringer:2001ia,Dick:2015twa,Cossu:2016scb,
  Kovacs:2019txb} provide evidence that {\it some} of the modes are
indeed localized on topological objects.  In particular,
Ref.~\cite{Cossu:2016scb} shows that the locations favored by some of
the localized low modes have all the features of $L$-type monopoles
and antimonopoles: large action and topological charge densities, near
(anti)self-duality, and near degeneracy of two eigenvalues of the
untraced Polyakov loop (see Fig.~\ref{fig:dens_PL}).\footnote{The
  claim of Ref.~\cite{Cossu:2016scb} is actually stronger: localized
  low modes do localize on $L-\bar{L}$ monopole-antimonopole pairs. We
  believe that this claim is not fully supported by the available
  evidence. On the one hand, while both selfdual and anti-selfdual
  points are clearly favored by localized modes, there is no clear
  evidence that these modes localize where selfdual and anti-selfdual
  points are spatially close. On the other hand, $L$-type
  anti(monopoles) are located at sites where a pair of the eigenvalues
  $(e^{i\phi_1},e^{i\phi_2},e^{-i(\phi_1+\phi_2)})$ of the untraced
  Polyakov loop is nearly degenerate and close to $-1$ (fluctuations
  of the degenerate pair around $-1$ correspond to fluctuations of the
  Polyakov loop at spatial infinity around 1), and while these sites
  are among the favorable localization points, sites with
  $\Re\tr P = -1$ but without eigenvalue degeneracy are at least
  equally (if not more) favorable, see Fig.~\ref{fig:dens_PL}.}  In
Refs.~\cite{Kovacs:2019txb,Vig:2021oyt} it is shown that for pure
gauge SU(3) theory the distribution of the number of near-zero modes
in the peak of the spectral density near zero (see
Fig.~\ref{fig:specdens}) is consistent with the distribution of a
dilute gas of topological objects. This suggests that the peak of
near-zero modes indeed originates from the zero modes associated with
topological objects. This provides further evidence supporting the
disordered medium scenario as a viable mechanism for localization, and
its close relation with spontaneous chiral symmetry breaking.

If localization were entirely due to mixing topological would-be zero
modes, then the density of localized modes would be equal to that of
the topological objects (calorons).\footnote{This observation applies
  also to the case in which the relevant objects are the $L$-type
  monopoles and antimonopoles, independently of them being part of
  calorons, as suggested in Ref.~\cite{Cossu:2016scb}.} Since above
the transition the density of calorons decreases sharply with
increasing temperature, we would expect the same behavior of the
density of localized modes. However, as shown in
Ref.~\cite{Kovacs:2012zq}, the density of localized modes actually
increases with temperature (see Fig.~\ref{fig:locmoddens}). In the
pure gauge case, no more than half of the localized modes seem to be
of topological origin for temperatures as low as
$1.03T_c$~\cite{Kovacs:2019txb} and as the temperature increases, this
fraction rapidly decreases. In Fig.~\ref{fig:topvsloc} we show the
temperature dependence of the fraction of localized modes that can be
associated with near-zero modes of topological
origin~\cite{Kovacs:2017uiz,Kovacs:2019txb}.  We show results obtained
with the overlap Dirac operator, and with the staggered Dirac operator
for three different values of the lattice spacing.  All the results
are consistent and show that with increasing temperature a rapidly
decreasing fraction of the localized modes are of topological
origin. We can conclude that while topology-related localized modes
may suffice to explain the near-zero peak, they cannot explain all the
remaining localized modes found in a typical high-temperature gauge
configuration, which therefore require a different supplementary
mechanism.

\end{paracol}
\begin{figure}[t]
  \widefigure
  \centering
    \includegraphics[width=0.50\textwidth]{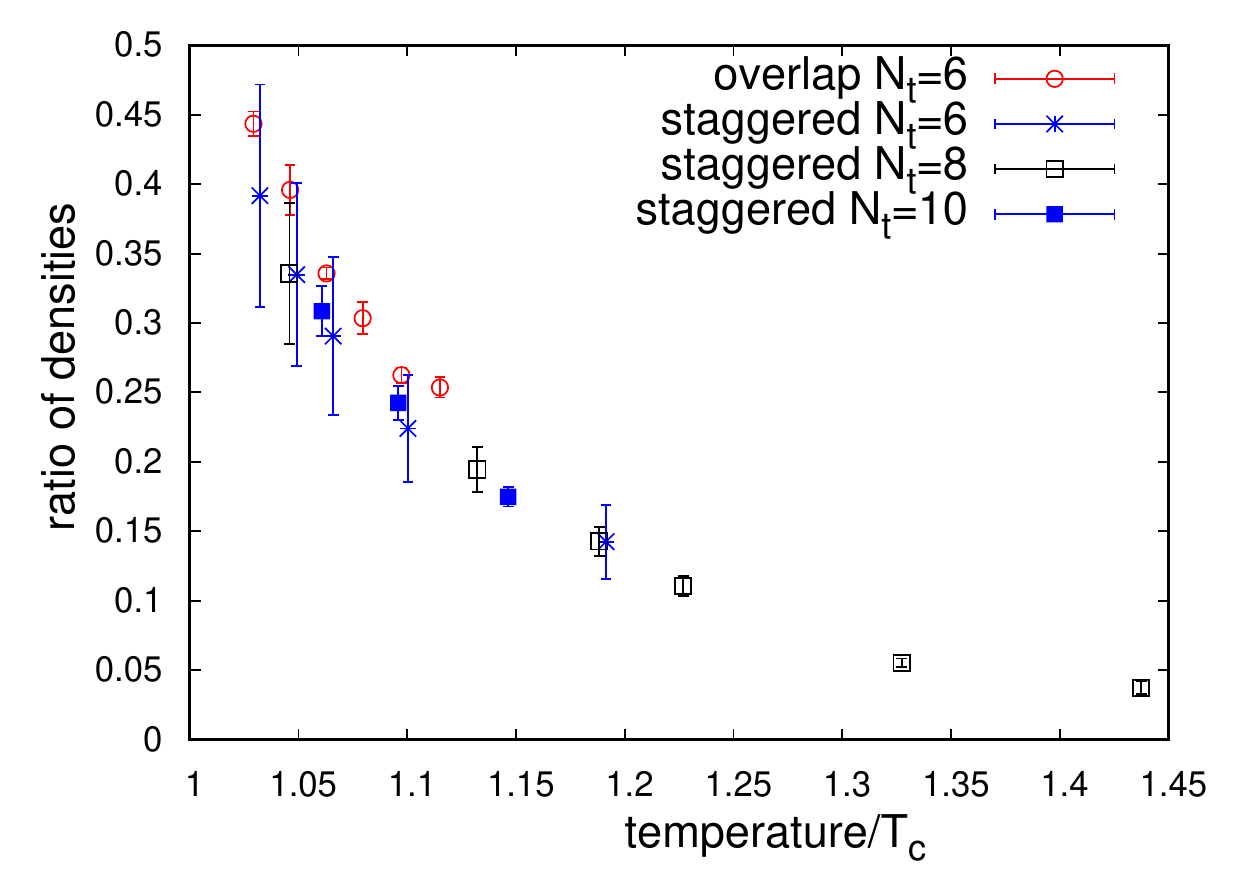}
    \caption{The fraction of localized modes in the quenched theory
      that are topology-related near-zero modes. The figure shows data
      calculated with the overlap Dirac operator on $N_t=6$ lattices,
      as well as with the staggered Dirac operator using three
      different lattice spacings corresponding to $N_t=6, 8$ and
      10. From Ref.~\cite{Kovacs:2019txb}.  \newline {\footnotesize
        Figure adapted from T.G.~Kov{\'a}cs and
        R.{\'A}.~Vig, % \textit{Localization and topology in high
        % temperature QCD},
        PoS LATTICE2018, 258 (2019), and used under a
        \href{https://creativecommons.org/licenses/by-nc-nd/4.0}{CC-BY-NC-ND
          4.0} license.}
 \label{fig:topvsloc}}
\end{figure}
\begin{paracol}{2}
\linenumbers
\switchcolumn

\subsection{The sea/islands picture}
\label{sec:mech_seaislands}

An alternative mechanism has been proposed in
Ref.~\cite{Bruckmann:2011cc}, and further elaborated in
Refs.~\cite{Giordano:2015vla,Giordano:2016cjs,Giordano:2016vhx}.  The
basic observation is that the eigenvalues of the untraced Polyakov
loop at a spatial site $\vec{x}$ effectively change the temporal
boundary condition felt locally by the quark eigenfunctions
$\psi(t,\vec{x})$.  Indeed, working for simplicity in the continuum,
if one fixes the gauge to the temporal gauge
$A_4(t,\vec{x})=\mathbf{0}$, the eigenvalue problem reduces to

\begin{equation}
  \label{eq:seaislandDirac}
  \slashed{D}\psi =  (\de_4 \gamma_4 + \slashed{D}_{(3)})\psi = i\lambda\psi\,, \qquad
  \slashed{D}_{(3)} = \sum_{j=1}^3\gamma_j (\de_j + i g A_j)\,, 
\end{equation}
while the effect of a nontrivial (untraced) Polyakov loop $P(\vec{x})$
is to change the temporal boundary condition from antiperiodic to

\begin{equation}
  \label{eq:seaislandbc}
 \psi(1/T,\vec{x})=-P(\vec{x})\psi(0,\vec{x})\,. 
\end{equation}
Eqs.~\eqref{eq:seaislandDirac} and \eqref{eq:seaislandbc} define the
eigenvalue problem in temporal gauge.  Clearly, the effective local
boundary condition Eq.~\eqref{eq:seaislandbc} affects the contribution
of site $\vec{x}$ to the Dirac eigenvalue,
$i\lambda = (\psi, \slashed{D}\psi)$.

To gain some insight on the effects of the Polyakov loop, it is useful
to study a family of configurations for which the eigenvalue problem
can be solved exactly, namely those with $A_\mu(t,\vec{x})=\mathbf{0}$
everywhere, and with a constant but nontrivial Polyakov loop.  This
can always be diagonalized by means of a global gauge transformation,
so without loss of generality we can take
$P(\vec{x})={\rm diag}(e^{i\phi_1}, e^{i\phi_2}, e^{i\phi_3})$, with
$e^{i(\phi_1+\phi_2+\phi_3)}=1$. On these configurations the
eigenfunctions of $-\slashed{D}^2$ are plane waves,

\begin{equation}
  \label{eq:seaislandbc2}
  \psi_c^{(a,k,\vec{p})}(t,\vec{x}) = \delta_{ca}e^{i(\omega_{ak} t + \vec{p}\cdot\vec{x})}\,,
\end{equation}
where $c$ is the color index,\footnote{Since $\slashed{D}^2$ is
  trivial in Dirac space in this case, the Dirac index is omitted.}
with temporal frequency ({\it effective
  Matsubara frequency}) given by

\begin{equation}
  \label{eq:seaislandbc3}
  \omega_{ak} = T[(2k+1)\pi + \phi_a]\,,\quad k\in\mathbb{Z}\,,
\end{equation}
and corresponding eigenvalues

\begin{equation}
  \label{eq:seaislandbc4}
  \lambda_{ak}(\vec{p})^2
  = \omega_{ak}^2 + \vec{p}^{\,2}\,.
\end{equation}
Restricting without loss of generality to $\phi_{1,2}\in ( %)
%[
-\pi,\pi]$, $\phi_1+\phi_2+\phi_3=0$, the lowest positive Dirac eigenvalue is
seen to be

\begin{equation}
  \label{eq:seaislandbc5}
  \lambda_{\rm min} = T(\pi - \max_a|\phi_a|)\,,
\end{equation}
i.e., it decreseas monotonically and symmetrically as one moves away
from $\phi_a=0~\forall a$, and vanishes when at least one of the
Polyakov loop eigenvalues equals $-1$.

While the configuration discussed above is obviously unrealistic, the
result Eq.~\eqref{eq:seaislandbc5} allows understanding qualitatively
which sites will be favored by a low Dirac eigenmode when the Polyakov
loop configuration is mostly ordered near
$P(\vec{x})\approx\mathbf{1}$, with ``islands'' of fluctuations in the
``sea'' of ordered Polyakov loops, as it happens at high
temperature. One can in fact interpret Eq.~\eqref{eq:seaislandbc5},
now with $\vec{x}$-dependent phases $\phi_a=\phi_a(\vec{x})$, as a
sort of three-dimensional local potential for the quarks, to which one
should add the appropriate ``hopping terms'' originating from the
spatial dependence of the Polyakov loops, as well as from the spatial
components of the gauge potential. From this point of view,
fluctuations of the Polyakov loop away from order provide regions of
lower potential that can ``trap'' the quarks.

More precisely, at high temperatures $\phi_a(\vec{x})\approx 0$ in an
extended region, and neglecting in a first approximation the effect of
the islands and of hopping, one finds fully delocalized modes.  In the
same approximation one finds a spectral gap, with the corresponding
(positive) eigenvalues starting at $T\pi$, i.e., the usual lowest
(fermionic) Matsubara frequency.  The presence of islands and the
effect of the interactions are expected to reduce this gap, but the
lowest eigenvalue that can be reached by a delocalized mode is
expected to remain separated from the origin.  On the other hand,
localizing on an island of fluctuations can be ``energetically'' more
favorable, and bring the corresponding eigenvalues inside the gap, as
long as the gain in potential energy achieved by avoiding the sea of
ordered Polyakov loops is sufficiently larger than the price paid for
localization in terms of spatial momenta.  This leads to expect the
following scenario, at least when the islands are sufficiently distant
from each other: a region of low spectral density, or {\it pseudogap},
opens between the origin and some point $\lambda_c$ in the spectrum,
or {\it mobility edge}, above which modes are extended throughout the
whole space; modes in the pseudogap can exist only if they are
localized on energetically convenient islands of fluctuations.

The scenario described above, which has been dubbed the sea/islands
picture of localization, shows a clear similarity with the
Anderson-type models of condensed matter physics, and the terminology
has been chosen precisely to reflect this similarity. From the point
of view of random Hamiltonians, the local fluctuations of the Polyakov
loop provide a three-dimensional source of on-site disorder. This
would naturally explain the fact that the critical behavior found at
the mobility edge in QCD is the same as that of the three-dimensional
unitary Anderson model.\footnote{Notice that the mobility edge is
  generally far from the near-zero zone where localized modes are of
  topological origin~\cite{Kovacs:2019txb}.}  There are, however,
important differences with the simple unitary Anderson model of
Eq.~\eqref{eq:uAM}. In that case, localization starts from the band
edges and moves towards the band center as the amount of disorder, as
measured by the width of its probability distribution, is increased.
In QCD, while localization may as well be present at the band edges
(cf.\ the results of Ref.~\cite{GarciaGarcia:2005vj} in the ILM model
and Ref.~\cite{Baranka:2021san} in $\mathbb{Z}_2$ gauge theory), it is
its appearance directly at the band center that characterizes the
deconfined phase. Moreover, the actual source of disorder are the
eigenvalues of the Polyakov loops, which are complex numbers lying on
the unit circle, and so the magnitude of the disorder is actually
bounded.

Another important difference, which is relevant also to the problem of
spontaneous chiral symmetry breaking, is the different structure of
the ``free'' Hamiltonian associated with the two cases in the absence of
fluctuations. For the Anderson model this is simply
$H^{\rm (AM)}_{\rm free} = \sum_{j=1}^3 T_j + T_j^\dag$ with $T_j$ the
translation operator in direction $j$, while for the Dirac operator
one has (for a naive lattice discretization)
$\slashed{D}_{\rm free} = \sum_{\mu=1}^4 \gamma_\mu (T_\mu -
T_\mu^\dag)$. Here periodic boundary conditions are understood in the
spatial directions; the effective boundary conditions
Eq.~\eqref{eq:seaislandbc} are assumed for the temporal direction.
While the spectrum of $H^{\rm (AM)}_{\rm free}$,
$E(\vec{p}) = \sum_j \cos p_j$, with $p_j= \f{2\pi k_j}{L}$, is dense
near the origin, the presence of the gamma matrices in
$\slashed{D}_{\rm free}$ leads to
$\lambda_{ak}(\vec{p}) = \sqrt{\omega_{ak}^2 + \vec{p}^{\,2}}$.  Even
in the case $\omega_{ak}=0$, in which there is no sharp spectral gap,
the spectral density near the origin is low and vanishes at
$\lambda=0$.

An important aspect of this scenario is that deconfinement is
naturally associated with the two effects that lead to localization of
the low modes in high-temperature QCD.  The first such effect is of
course the formation of a sea of ordered Polyakov loops close to the
identity, which can cause the opening of a spectral pseudogap and so
make modes that localize on the islands of fluctuations stable against
delocalization, as explained above.  However, the appearance of the
pseudogap requires also a second effect due to the ordering of
Polyakov loops at deconfinement, namely the increased correlation
between gauge fields on different time slices.  The discussion of this
effect requires a more detailed description of the Dirac operator in
the language of Anderson models, in what can be called the
``Dirac-Anderson approach''. Here we sketch the discussion in the
continuum in the temporal gauge; a more detailed and mathematically
more precise analysis is presented in Ref.~\cite{Giordano:2016cjs} for
staggered fermions on the lattice. 

Due to its compactness, the temporal direction can be treated as an
internal degree of freedom, in particular by expanding the quark
eigenfunctions on a complete basis of plane waves $e^{i\omega_{ak}t}$
obeying the appropriate effective boundary conditions,
Eq.~\eqref{eq:seaislandbc}.  Here $\omega_{ak}$ are the effective
Matsubara frequencies of Eq.~\eqref{eq:seaislandbc3}, now
$\vec{x}$-dependent, which provide a random on-site potential of the
form $\omega_{ak}(\vec{x})\gamma_4$.  For every color $a$ with
associated Polyakov-loop phase $\phi_a(\vec{x})$, in correspondence to
each wave number $k$ there is a different branch of the on-site
potential, and so a different associated three-dimensional
Anderson-type model, built by adding the on-site disorder to the
spatial part of the Dirac operator (projected on the $a,k$
subspace). We will refer to each of these models as a Dirac-Anderson
model. The full Dirac operator is obtained by putting the various
Dirac-Anderson models together, and by including their coupling
induced by the hopping terms (i.e., the spatial part of the operator).
The strength of the coupling among the Dirac-Anderson models turns out
to be inversely related to the correlation of the gauge fields on
different time slices.  At low temperatures this correlation is small,
the Dirac-Anderson models are strongly coupled, and the internal
degree of freedom is effectively one more direction in which the modes
can extend, thus facilitating their delocalization. This is in
agreement with the effectively four-dimensional nature of QCD in the
low temperature phase. As a matter of fact, the pseudogap does not
open at low temperatures, where the spectral density is finite near
zero, and this can only happen if the various Dirac-Anderson models do
mix with each other (see the discussion about the free Dirac
operator). In the absence of a pseudogap, localization of a mode is
generally unstable against mixing with modes of similar energies.  At
high temperature, instead, the Polyakov loops become ordered inducing
stronger correlations among different time slices, and the
Dirac-Anderson models decouple making the problem effectively
three-dimensional.  In particular, the pseudogap is now expected to
appear: it would be present for exactly decoupled Dirac-Anderson
models (see again the discussion about the free Dirac operator), and
their limited mixing is not sufficient to close it. Localized modes
near the band center can then be supported by Polyakov loop
fluctuations, as discussed above.  As shown in
Ref.~\cite{Giordano:2016cjs} in a toy model where ordering of the
Polyakov loop and correlation of the time slices can be varied
independently, both effects are required for localization to appear at
the band center.

An important aspect of the sea/islands picture is that it is not
incompatible with the growing density of localized modes observed in
QCD. Differently from the case of topological charge, Polyakov-loop
fluctuations are not quantized. As $T$ grows in the deconfined phase,
the volume $V_{\rm fluct}$ occupied by Polyakov-loop fluctuations is
expected to decrease, $V_{\rm fluct} \sim T^{-c_1}$, as the Polyakov
loop becomes more and more ordered. On the other hand, the typical size
$V_0$ of the islands of fluctuations is also expected to decrease,
$V_0 \sim T^{-c_2}$.  The number of localized modes is expected to be
directly related to the number of islands,
$V_{\rm fluct}/V_0 \sim T^{c_2-c_1} $, and whether this number
increases or decreases with temperature depends on the details of the
dynamics. For example, while increasing in QCD~\cite{Kovacs:2012zq} up
to $T\sim 5T_c$, it is seen to decrease in 2+1-dimensional SU(3) gauge
theory above $T\sim 1.1\div 1.2 T_c$~\cite{Giordano:2019pvc}.

Perhaps the most appealing feature of the sea/islands picture is its
simplicity: all that it needs to work is the ordering of the Polyakov
loop. This leads immediately to expect that localized modes will
appear at the low end of the spectrum whenever an ordering transition
takes place, independently of details such as the gauge group and its
representation, fermionic content, nontrivial topological features,
dimensionality,\footnote{Dimensionality should not matter as long as
  both deconfinement and localization are allowed. For example, no
  Anderson transition should be found in 1+1-dimensional gauge
  theories at finite temperature: no deconfinement transition is
  present there, and all modes are expected to be localized in one
  spatial dimension.} and so on. This is discussed in the next
Section.

\section{Localization in other gauge theories}
\label{sec:othergt}

% \end{paracol}
% \begin{figure}[t]
%   \widefigure
%   \centering
%   \includegraphics[width=0.45\textwidth]{figures/kv_lc_vs_beta4.pdf}\hfil  
%   \includegraphics[width=0.45\textwidth]{figures/vk_lc_plot.pdf}
%   \caption{Mobility edge in pure gauge SU(3) theory in the staggered
%     (\textbf{left}) and overlap (\textbf{right}) Dirac spectrum. From
%     Refs.~\cite{Kovacs:2017uiz,Vig:2020pgq}. \newline {\footnotesize
%       Figures adapted from T.G.~Kov{\'a}cs and
%       R.{\'A}.~Vig, % \textit{Localization transition in SU(3) gauge
%                     % theory},
%       Phys.\ Rev.\ D 97, 014502 (2018) (left), and from R.{\'A}.~Vig and
%       T.G.~Kov{\'a}cs, % \textit{Localization with overlap fermions},
%       Phys.\ Rev.\ D 101, 094511 (2020) (right), and used under a
%       \href{https://creativecommons.org/licenses/by/4.0}{CC-BY 4.0}
%       license.}
%     \label{fig:lc_SU3}}
% \end{figure}
% \begin{paracol}{2}
% \linenumbers
% \switchcolumn

In this Section, we discuss localization of the low Dirac modes in
gauge theories other than QCD. Some of the references have been
already discussed in Section \ref{sec:QCD} in connection with QCD, where
they where treated as approximations. Here they are briefly discussed
again, focussing more on the differences than on the similarities with QCD.

The main motivation in studying more general gauge theories is to
investigate further the extent of the connection between localization
on one side, and deconfinement and chiral restoration on the other.
In particular, studying localization in models with genuine
deconfining and/or chirally restoring phase transitions allows one to
investigate this connection in a more clear-cut setting than in QCD,
where it is somewhat blurred by the crossover nature of the
transition. Studying more general gauge theories also allows one to test
the sea/islands picture discussed in the previous Section, and its
generic prediction of localization of low modes in the
high-temperature, ``ordered'' phase.

Genuine deconfining phase transitions are found in pure gauge
SU($N_c$) theory. In 3+1 dimensions the transition is second order for
$N_c=2$ and first order for $N_c\ge 3$, while in 2+1 dimensions it is
second order for $N_c=2,3$ and first order for $N_c\ge 4$ (see, e.g.,
Refs.~\cite{Francis:2015lha,Lucini:2003zr,Lucini:2005vg,Liddle:2008kk}).
As already mentioned in Section \ref{sec:QCD}, localized low Dirac
modes have been found in pure gauge SU(2) theory in 3+1 dimensions,
both with the overlap~\cite{Kovacs:2009zj,Bruckmann:2011cc} and with
the staggered~\cite{Kovacs:2010wx,Bruckmann:2011cc} Dirac operator
(see Fig.~\ref{fig:ggl_kp}, right), above the deconfinement
temperature $T_c$. (Further details can be found in Section
\ref{sec:QCD} and will not be repeated here.)  No sign of localized
modes was found instead below $T_c$.  From the random-matrix point of
view, the SU(2) case differs from SU($N_c\ge 3$) as the symmetry class
is the symplectic instead of the unitary one. This is reflected in the
different behavior of the unfolded spectrum in the bulk, which agrees
with the symplectic Wigner surmise, Eqs.~\eqref{eq:WS} and
\eqref{eq:WS2}.  A detailed study of the Anderson transition was not
pursued.

\end{paracol}
\begin{figure}[t]
  \widefigure
  \centering
  \includegraphics[width=0.45\textwidth]{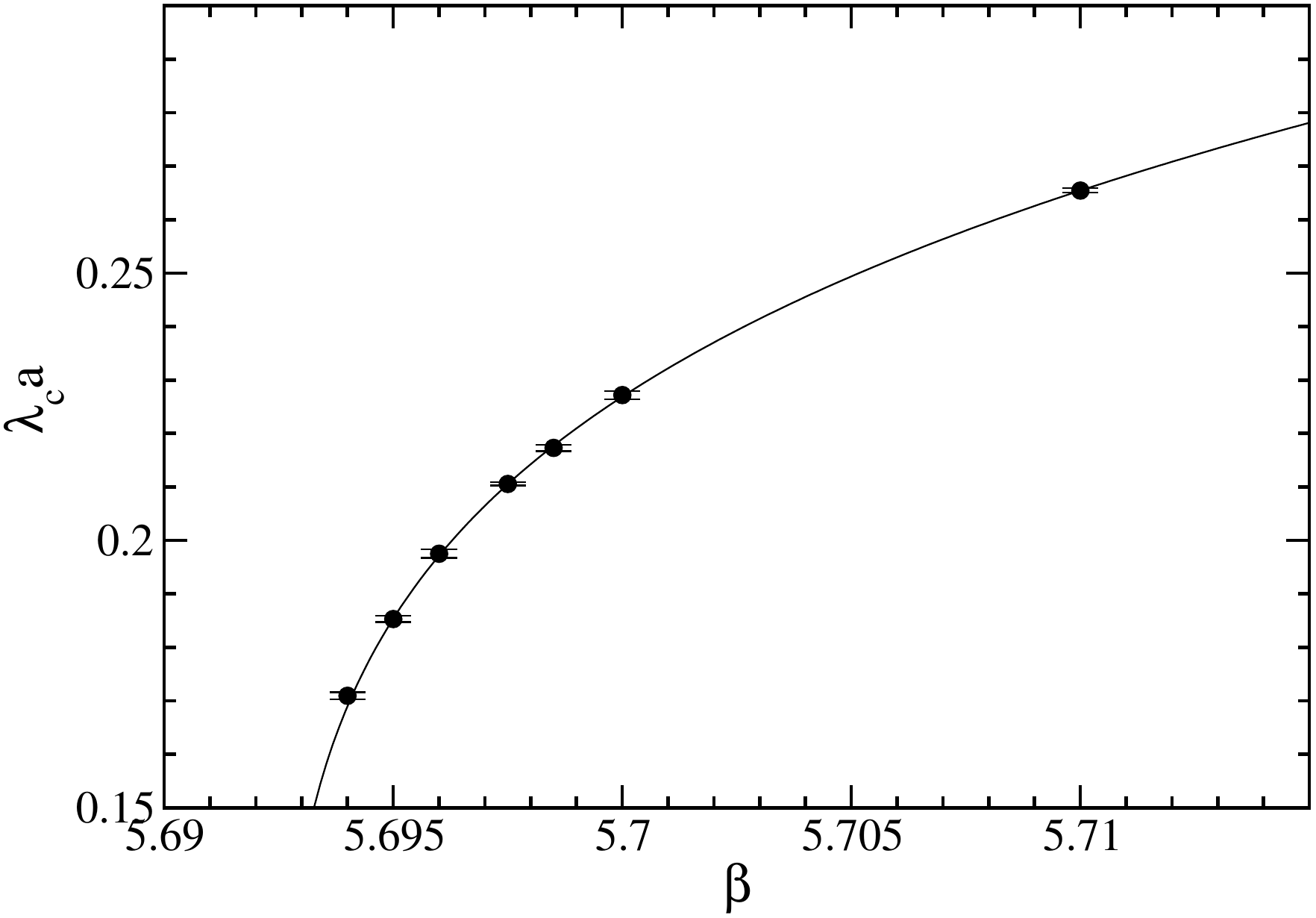}\hfil  
  \includegraphics[width=0.45\textwidth]{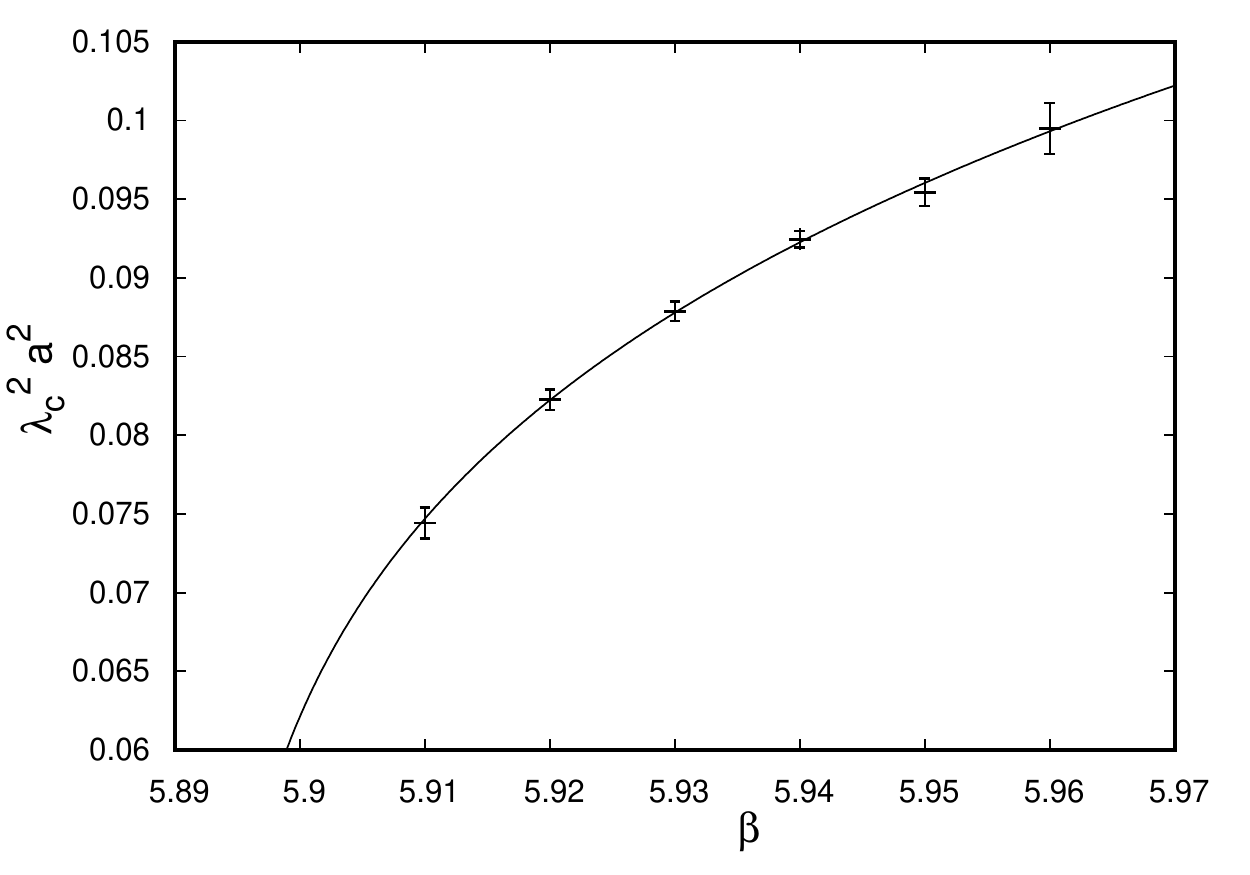}
  \caption{Mobility edge in pure gauge SU(3) theory in the staggered
    (\textbf{left}) and overlap (\textbf{right}) Dirac spectrum. From
    Refs.~\cite{Kovacs:2017uiz,Vig:2020pgq}. \newline {\footnotesize
      Figures adapted from T.G.~Kov{\'a}cs and
      R.{\'A}.~Vig, % \textit{Localization transition in SU(3) gauge
                    % theory},
      Phys.\ Rev.\ D 97, 014502 (2018) (left), and from R.{\'A}.~Vig and
      T.G.~Kov{\'a}cs, % \textit{Localization with overlap fermions},
      Phys.\ Rev.\ D 101, 094511 (2020) (right), and used under a
      \href{https://creativecommons.org/licenses/by/4.0}{CC-BY 4.0}
      license.}
    \label{fig:lc_SU3}}
\end{figure}
\begin{paracol}{2}
\linenumbers
\switchcolumn

Results for pure gauge SU(3) have been presented in
Refs.~\cite{Kovacs:2017uiz,Vig:2020pgq,Kovacs:2019txb} for the
3+1-dimensional case, and in Ref.~\cite{Giordano:2019pvc} for the
2+1-dimensional case. Localized low Dirac modes are found in both
cases in the deconfined phase. In 3+1 dimensions the temperature
dependence of the mobility edge $\lambda_c$ was studied using the
Wilson gauge action both with staggered~\cite{Kovacs:2017uiz} and
overlap~\cite{Vig:2020pgq} fermions (using in this case the magnitude
of the eigenvalues), smearing the gauge fields with two steps of stout
smearing~\cite{Morningstar:2003gk} in the staggered case, and two
steps of hex smearing~\cite{Capitani:2006ni} in the overlap case. The
integrated ULSD computed locally in the spectrum, $I_{s_0}(\lambda)$,
was used to determine $\lambda_c$ as the point where $I_{s_0}$ takes
its critical value $I_{s_0}^{(c)}$~\cite{Giordano:2013taa}, i.e.,
$I_{s_0}(\lambda_c)=I_{s_0}^{(c)}$. Here use was made of the
universality of the critical properties of the Anderson transition,
which should be shared by QCD and pure gauge SU(3) theory, as they are
both in the 3d unitary class. This was confirmed by the
volume-independence of the resulting $\lambda_c$. For both
discretizations, $\lambda_c$ is seen to extrapolate to zero at a
temperature which agrees with the deconfinement temperature (see
Refs.~\cite{Francis:2015lha,Lucini:2003zr} and references therein)
within numerical errors (see Fig.~\ref{fig:lc_SU3}).

The 2+1-dimensional case was studied in Ref.~\cite{Giordano:2019pvc}
using the Wilson gauge action and the staggered discretization
(without smearing). Universality arguments lead to expect that the
Anderson transition is of BKT type with exponentially divergent
correlation length, as found in Ref.~\cite{xie1998kosterlitz} for the
2d unitary Anderson model. The results of Ref.~\cite{Giordano:2019pvc}
support this scenario. In particular, spectral statistics are
critical, i.e., volume independent for all $\lambda$ above
$\lambda_c$, as expected for a BKT-type Anderson
transition~\cite{barber1983finite}, see Fig.~\ref{fig:lc_SU3_2+1d},
left.  The mobility edge was determined by means of a finite size
scaling study, and found to extrapolate to zero at a temperature
compatible with the deconfinement temperature~\cite{Liddle:2008kk}
(although with much larger numerical uncertainty), see
Fig.~\ref{fig:lc_SU3_2+1d}, right.  In the confined phase no
localization was found, but low modes were seen to display a
nontrivial fractal dimension $D_2< 2$ (see
Eq.~\eqref{eq:mfexp}). 

\end{paracol}
\begin{figure}[t]
  \widefigure
  \centering
  \includegraphics[width=0.46\textwidth]{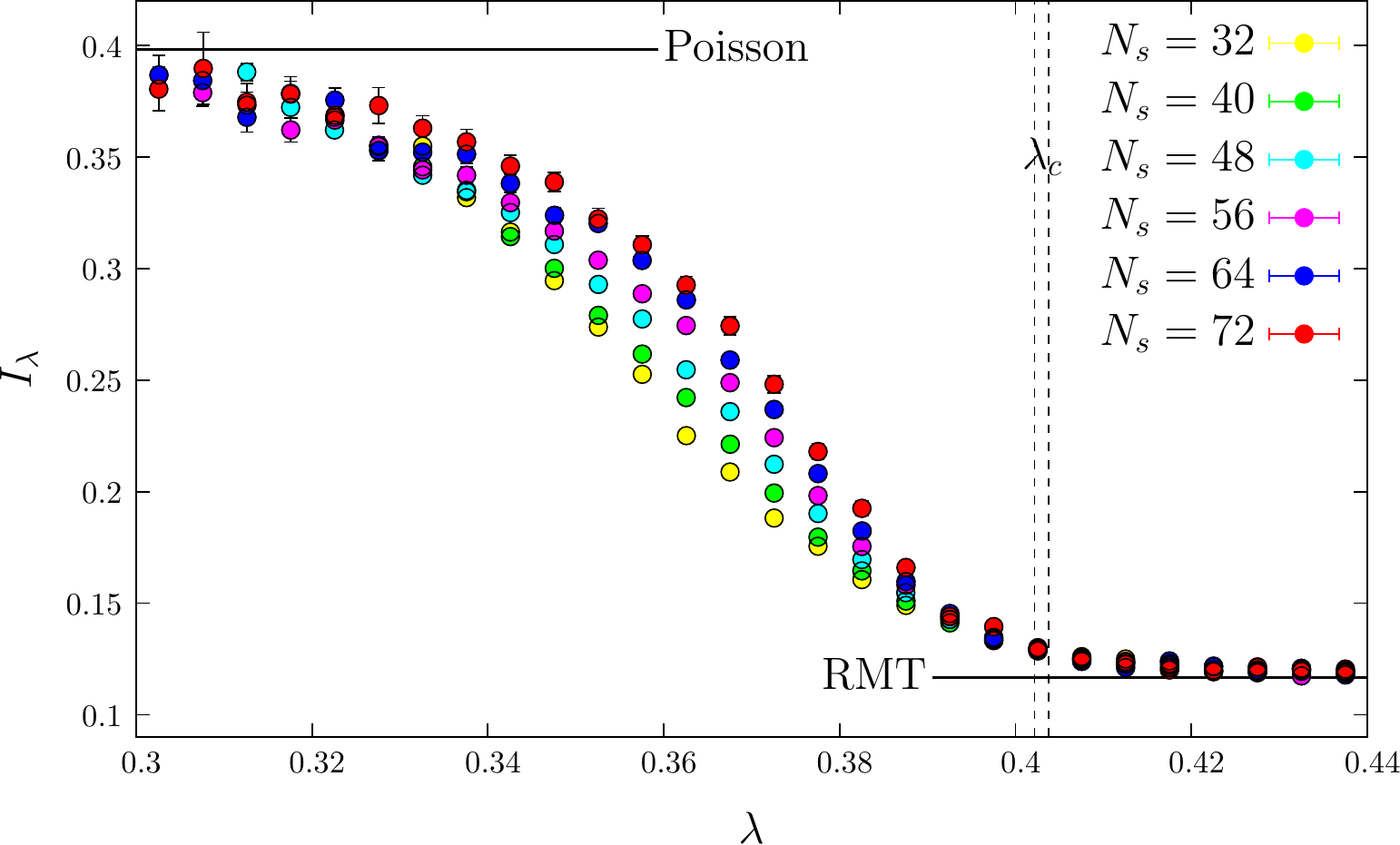}\hfil  
  \includegraphics[width=0.45\textwidth]{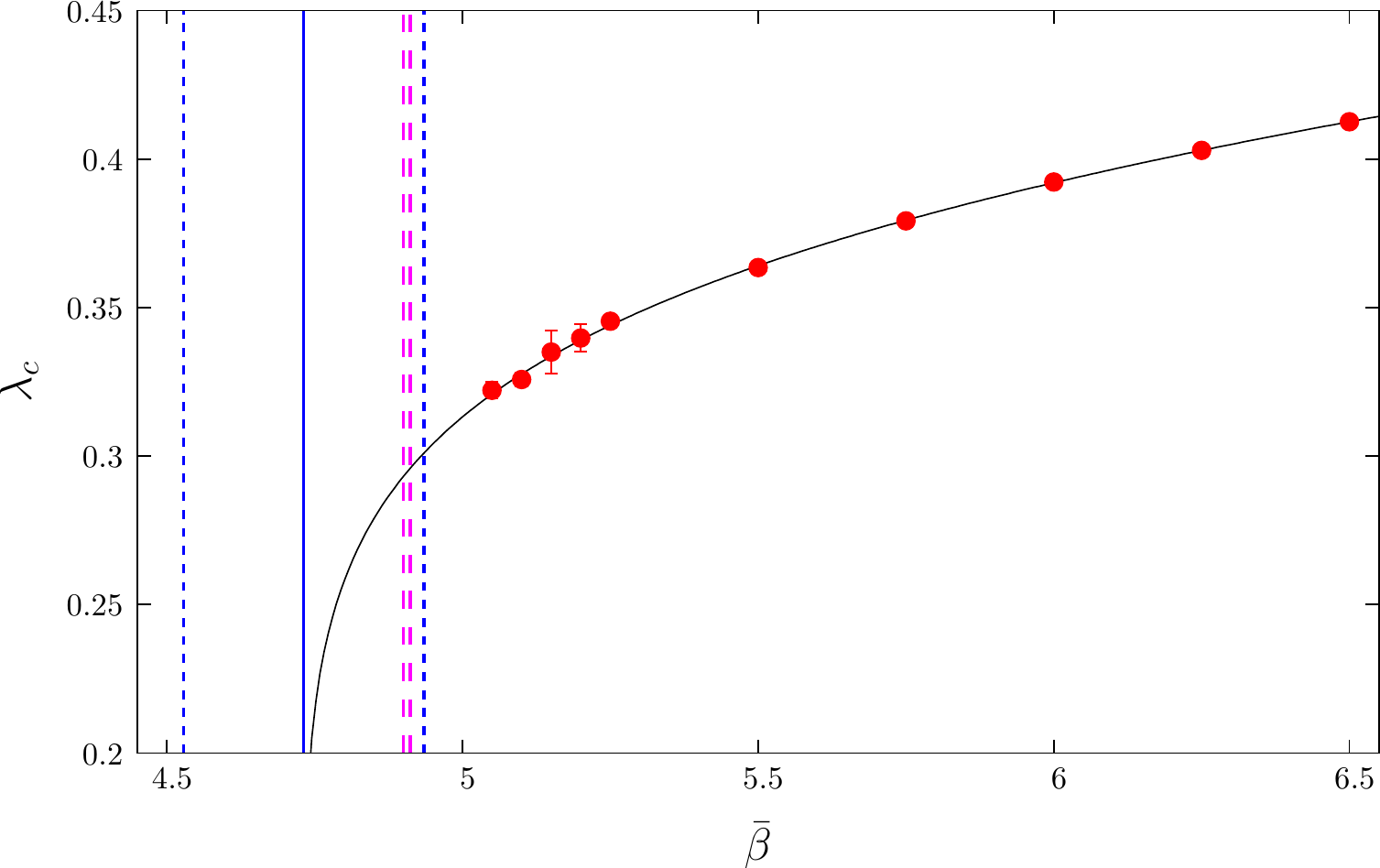}
  \caption{Spectral statistics $I_{s_0}$ along the spectrum for
    various volumes at coupling $\bar{\beta}=6.25$ (\textbf{left}), and
    mobility edge as a function of $\bar{\beta}$ (\textbf{right}) for the
    staggered operator in 2+1-dimensional pure gauge SU(3)
    theory. (Here $\bar{\beta}=\beta/3$ with $\beta$ the Wilson action
    coupling.) In the left panel, Poisson and RMT predictions and the
    position of the mobility edge are also shown. In the right panel,
    a power-law fit (black solid line) to $\lambda_c$, the position
    (blue solid line) and error band (blue dashed lines) of
    $\bar{\beta}_{\rm loc}$ at which $\lambda_c$ extrapolates to zero,
    and the error band of the critical $\bar{\beta}_c$ (magenta dashed
    lines) are also shown. From Ref.~\cite{Giordano:2019pvc}. \newline
    {\footnotesize Figures adapted from
      M.~Giordano, % \textit{Localisation in 2+1
      % dimensional
      % SU(3) pure gauge theory at finite temperature},
      J.\ High Energy Phys.\ 05, 204 (2019), and used under a
      \href{https://creativecommons.org/licenses/by/4.0}{CC-BY 4.0}
      license.}
    \label{fig:lc_SU3_2+1d}}
\end{figure}
\begin{paracol}{2}
\linenumbers
\switchcolumn

Localization of Dirac modes was studied in $\mathbb{Z}_2$ pure gauge
theory in 2+1 dimensions in Ref.~\cite{Baranka:2021san}, probed with
unimproved staggered fermions. This model has the simplest gauge
group, and the lowest dimensionality in which a deconfining transition
is found. Studying the fractal dimension $D_2$, it was shown that low
modes are localized ($D_2=0$) in the high-temperature, deconfined
phase of the theory in the positive center sector (i.e., positive
spatially averaged Polyakov loop), while they are delocalized (with
$D_2<2$) in the low-temperature, confined phase, and in the high
temperature phase in the negative center sector (i.e., negative
spatially averaged Polyakov loop). Localized modes are also found at
the high end of the spectrum, independently of the phase and of the
center sector. Significant correlation between localized modes and
both Polyakov loops and clusters of negative plaquettes was observed.

While a genuine phase transition is expected for SU(3) gauge group in
the presence of $N_f=3$, light enough dynamical
fermions~\cite{Pisarski:1983ms}, so far a critical point has been
observed only on coarse lattices, and disappears in the continuum
limit~\cite{Karsch:2001nf,deForcrand:2003vyj,deForcrand:2008vr}.
Although only a toy model for QCD, the SU(3) theory with $N_f=3$
flavors of unimproved staggered fermions on $N_t=4$ lattices is
nonetheless a well-defined statistical model with a genuine first
order transition, affecting both its chiral and confining properties,
despite the absence of exact chiral and center symmetries.  More
precisely, as the coupling $\beta$ crosses the critical value
$\beta_c$, the chiral condensate jumps downwards to a much smaller but
still finite value; and the average Polyakov loop jumps upwards from
its small but nonzero value to a considerably larger value.  Evidence
of localization of the low staggered Dirac modes was reported in
Ref.~\cite{Giordano:2016nuu} for bare fermion mass $m=0.01$, below the
critical value $m_c=0.0259$~\cite{deForcrand:2008vr}, where genuine
first-order phase transitions are present. A mobility edge was shown
to be present for $\beta>\beta_c$: it increases with $\beta$, and
extrapolates to zero close to $\beta_c$. The lowest mode was also seen
to turn from delocalized to localized at a coupling $\beta_{\rm loc}$
compatible with $\beta_c$, i.e., in correspondence with the
finite-temperature transition (see Fig.~\ref{fig:nt4_pr_trdef}, left).

\end{paracol}
\begin{figure}[t]
  \widefigure
  \centering
  \includegraphics[width=0.45\textwidth]{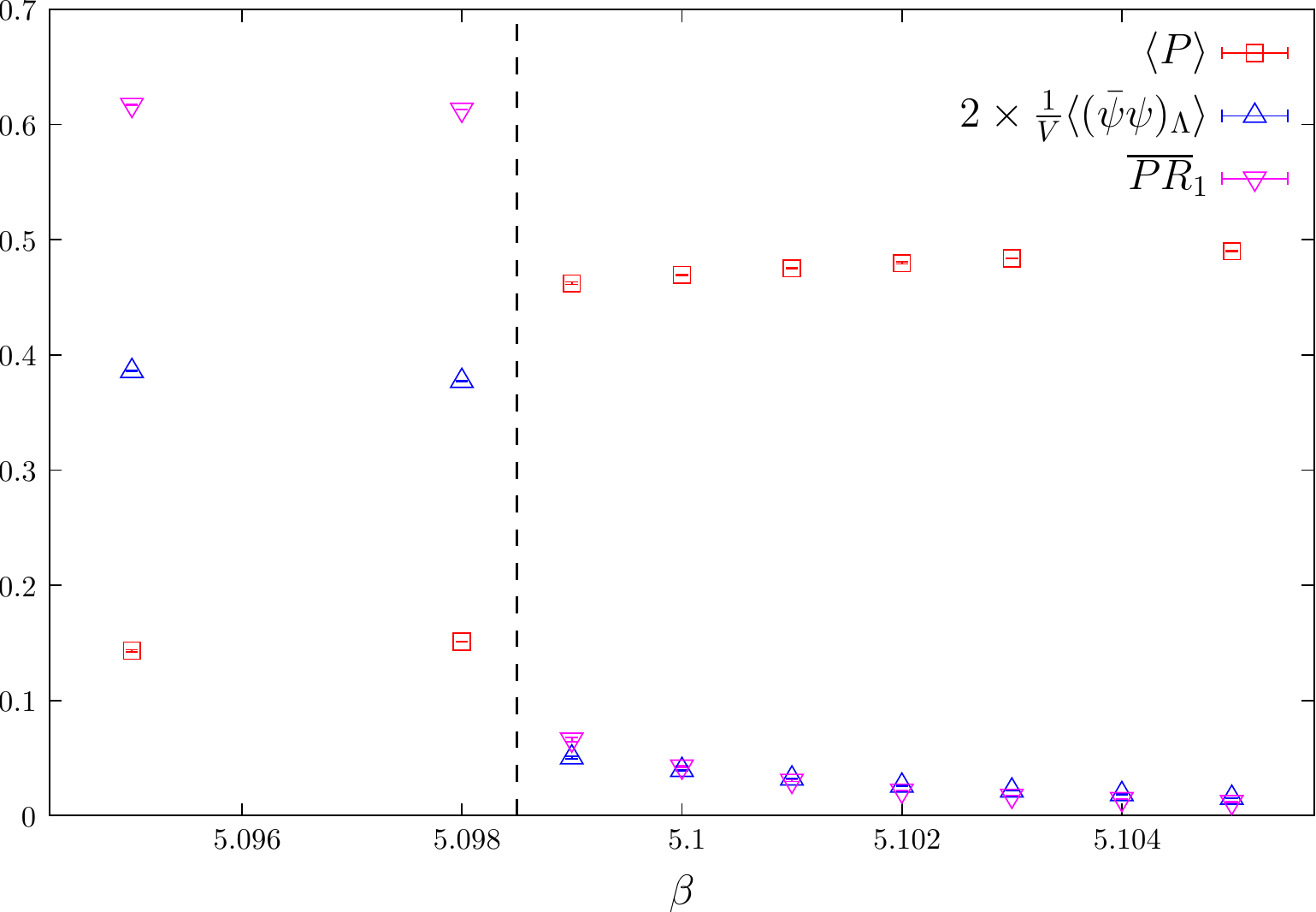}\hfil
  \includegraphics[width=0.45\textwidth]{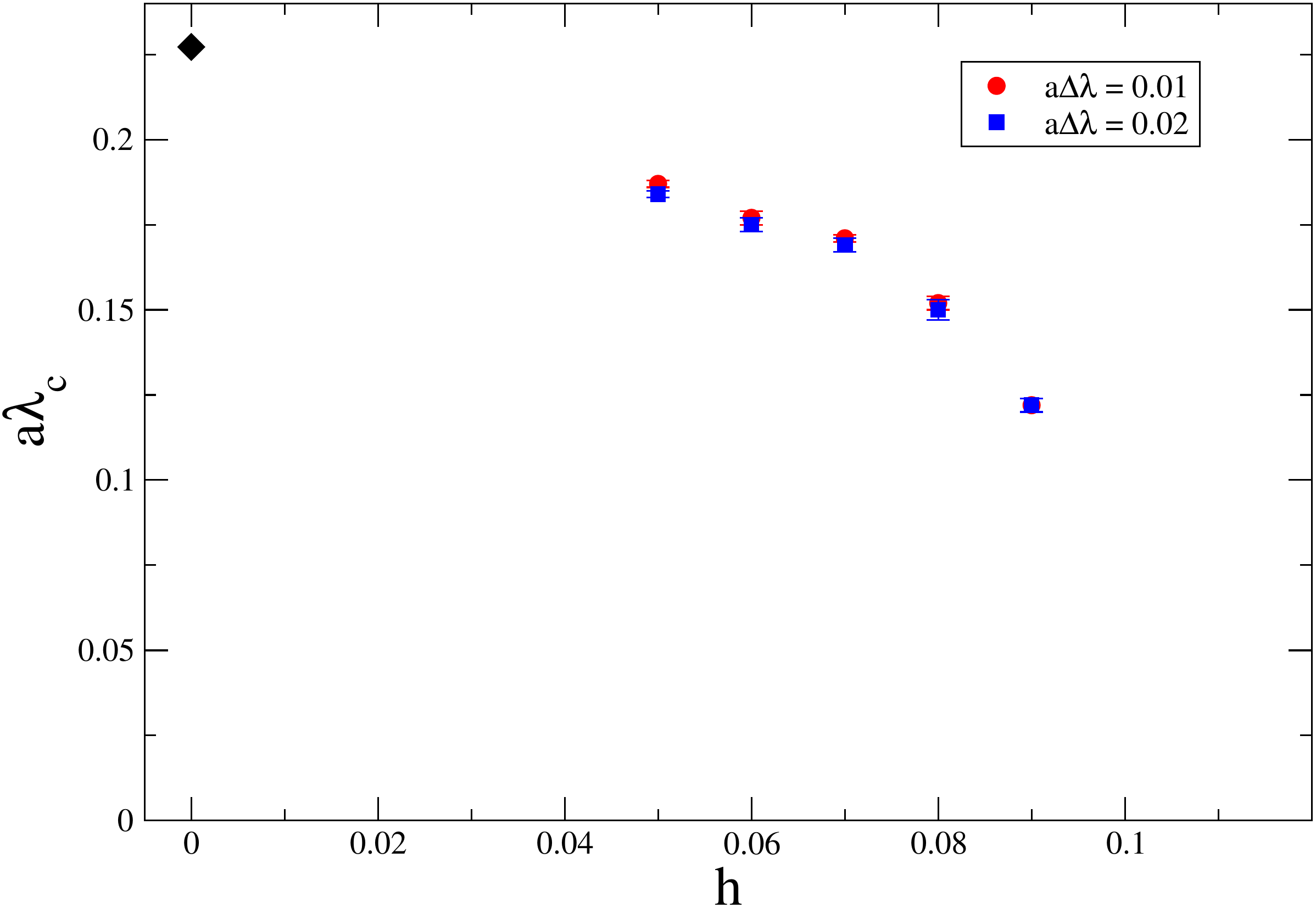}

  \caption{\textbf{Left}: average Polyakov loop (red squares), chiral
    condensate (upward blue triangles) and average PR of the lowest
    staggered mode (downward magenta triangle) for $N_f=3$ unimproved
    staggered fermions of bare mass $m=0.01$ on $N_t=4$ lattices. The
    critical coupling $\beta_c=5.0985$ is also shown. From
    Ref.~\cite{Giordano:2016nuu}.  \textbf{Right}: mobility edge as a
    function of the deformation parameter $h$ in trace-deformed SU(3)
    gauge theory in the high-temperature deconfined phase
    ($\beta=6.0$).  Here the critical deformation parameter for
    reconfinement is $h_c=0.1$.  The black diamond is the $h=0$ result
    of Ref.~\cite{Kovacs:2017uiz}. From Ref.~\cite{Bonati:2020lal}.
    \newline
    {\footnotesize Figure adapted from M.~Giordano,
    S.~D.~Katz, T.~G.~Kov{\'a}cs, and
    F.~Pittler, % \textit{Deconfinement, chiral transition and localisation in a
    % QCD-like model},
    J.\ High Energy Phys.\ 02, 055 (2017), and used under a
    \href{https://creativecommons.org/licenses/by/4.0}{CC-BY 4.0}
    license.  Figure adapted from C.~Bonati, M.~Cardinali, M.~D'Elia,
    M.~Giordano, and
    F.~Mazziotti, % \textit{Reconfinement, localization and thermal
      % monopoles in SU(3) trace-deformed Yang-Mills theory},
    Phys.\ Rev.\ D 103, 034506 (2021), and used under a
    \href{https://creativecommons.org/licenses/by/4.0}{CC-BY 4.0}
    license.}
    \label{fig:nt4_pr_trdef}}
\end{figure}
\begin{paracol}{2}
\linenumbers
\switchcolumn

The relation between localization and deconfinement was tested at a
different deconfinement phase transition in
trace-deformed~\cite{Myers:2007vc,Unsal:2008ch} pure gauge SU(3)
theory at finite temperature in Ref.~\cite{Bonati:2020lal}. In this
model a deformation term
$\Delta S = h\sum_{\vec{x}} |\tr P(\vec{x})|^2$ is added to the
action, which (for $h>0$) tends to locally suppress a nonzero trace
for the Polyakov loop $P(\vec{x})$. For temperatures above the
deconfinement temperature, $\Delta S$ pushes the theory towards a
``reconfined'' phase where $\tr P(\vec{x})\sim 0$. This happens when
the deformation parameter $h$ crosses a (temperature dependent)
critical value $h_c$. Ref.~\cite{Bonati:2020lal} studied the spectrum
of the two-stout smeared staggered spectrum at $\beta=6.0$ on $N_t=6$
lattices for various volumes and deformation parameters. Results
showed that localized modes are present for $h<h_c$, but disappear as
the system crosses over into the reconfined phase. The mobility edge
was determined by comparing the fractal dimension of the modes with
its value at criticality~\cite{ujfalusi2015finite}. While
monotonically decreasing with $h$, it is not clear whether it vanishes
continuously at $h_c$ or jumps to zero discontinuously (see
Fig.~\ref{fig:nt4_pr_trdef}, right).

Finally, the connection between localization and ordering of the
background configuration was studied in spin models in
Refs.~\cite{Giordano:2015vla,Giordano:2016cjs,Giordano:2016vhx,Bruckmann:2017ywh}.
Ref.~\cite{Giordano:2015vla} used a simple 3d Hamiltonian in the
orthogonal class with on-site disorder provided by the spins of a
continuous-spin Ising-type model. In the ordered phase of the spin
model, localization was observed for the low modes, with a mobility
edge separating them from higher modes, and critical behavior
compatible with that of the 3d orthogonal Anderson model.
Refs.~\cite{Giordano:2016cjs,Giordano:2016vhx} dealt with the
Dirac-Anderson form of the staggered operator (see Section
\ref{sec:mech_seaislands}), so in the 3d chiral unitary class, in the
case $N_t=2$ in the background of Polyakov loops constructed from a
spin model. Localized low modes are observed in the ordered phase of
the model~\cite{Giordano:2016cjs}, appearing at the critical
temperature~\cite{Giordano:2016vhx}.  Ref.~\cite{Bruckmann:2017ywh}
reports on the $CP^3$ model in 1+1 and 2+1 dimensions. While in 1+1
dimensions localized modes are found in both phases of the model, as
expected in one spatial dimension, in the 2+1 case localized modes are
found only in the ordered phase. This model belongs to the 2d chiral
unitary class.

\section{Conclusions and outlook}
\label{sec:concl}

The presence of localized modes in the spectrum of the Dirac operator
in the high-temperature phase of gauge theories is by now well
established. Numerical studies on the lattice have shown that above
the transition temperature the low-lying Dirac modes are spatially
localized on the scale of the inverse temperature, in QCD and QCD-like
theories, as well as in several pure gauge theories and related models
in 3+1 and 2+1
dimensions~\cite{Gockeler:2001hr,Gattringer:2001ia,GarciaGarcia:2005vj,
  GarciaGarcia:2006gr,Gavai:2008xe,Kovacs:2009zj,Bruckmann:2011cc,
  Kovacs:2010wx,Kovacs:2012zq,Giordano:2013taa,Nishigaki:2013uya,
  Giordano:2014qna,Ujfalusi:2015nha,Dick:2015twa,Cossu:2016scb,
  Holicki:2018sms,Kovacs:2017uiz,Vig:2020pgq,Kovacs:2019txb,
  Giordano:2019pvc,Baranka:2021san,Giordano:2016nuu,Bonati:2020lal,
  Giordano:2015vla,Giordano:2016cjs,Giordano:2016vhx,Bruckmann:2017ywh}.

The physical significance of localization has so far remained quite
elusive. First of all we should emphasize an important difference
between localization in electron systems and localization in QCD. In
the former case the mobility edge in the spectrum can be ``accessed''
by tuning a suitable control parameter, such as an electric field or
the density of electrons. As the Fermi energy crosses the mobility
edge, the system undergoes a genuine phase transition, with the
zero-temperature conductivity changing non-analytically. In contrast,
the mobility edge in the QCD Dirac spectrum cannot be directly
connected to a thermodynamic transition. This is because in that case,
in general, there is no control parameter that can be adjusted to make
the system sensitive to just the eigenmodes at the mobility edge in
the spectrum. The only exception is when the mobility edge is at zero,
which happens only at the critical temperature of localization. If at
the same time the quark masses are set to zero, the system becomes
most sensitive to the lowest Dirac eigenmodes, the ones closest to
zero. Thus, only in this double limit when the temperature tends to the
critical temperature of localization and the quark mass to zero can
one possibly directly connect the localization transition to a genuine
thermodynamic phase transition. Unfortunately, this limit is out of
the reach of present day lattice simulations and we have no numerical
evidence of what happens there.

On the other hand, some progress has been made to understand the
physical significance of localization in QCD. A clear connection with
deconfinement has emerged: in all the models investigated so far,
localization of the low modes shows up when the system transitions
from the confined, low-temperature phase to the deconfined,
high-temperature phase~\cite{GarciaGarcia:2006gr,Kovacs:2012zq,
  Kovacs:2017uiz,Vig:2020pgq,Giordano:2019pvc,Baranka:2021san,
  Giordano:2016nuu,Bonati:2020lal,Giordano:2016vhx}. Convincing
evidence has been presented for the crucial role played by the
ordering of the Polyakov loop and by its fluctuations in the formation
of a mobility edge in the Dirac spectrum, separating low-lying,
localized modes from the delocalized bulk
modes~\cite{Bruckmann:2011cc,Giordano:2016cjs,Cossu:2016scb,
  Holicki:2018sms,Baranka:2021san}. As the Polyakov loop is the
(approximate, in the case of QCD) order parameter for confinement, the
observed connection between localization and deconfinement has a
dynamical explanation, further backed by a viable mechanism (the
sea/islands picture~\cite{Bruckmann:2011cc,Giordano:2015vla,
  Giordano:2016cjs,Giordano:2016vhx}, see Section
\ref{sec:mech_seaislands}) relating the two phenomena. This raises the
hope that further studies can lead to a better understanding of
confinement, and possibly to the uncovering of the mechanism behind
this remarkable property of gauge theories. In this context, it would
be interesting to further elucidate the relation between localization
and center symmetry, since so far only models with nontrivial gauge
group center have been investigated.

Localization could also help in explaining the close relation observed
between deconfinement and restoration of chiral symmetry. These two
phenomena in fact take place at the same temperature, or in a
relatively narrow interval of temperatures, where also localized low
Dirac modes appear. Localization could then provide the key to
understanding this relation between in principle unrelated
phenomena. Unfortunately, while the connection between deconfinement
and localization can be easily studied in a clear-cut situation by
investigating pure gauge theories with a genuine deconfining phase
transition, studying the connection between chiral symmetry
restoration and localization by means of numerical lattice simulations
faces the considerable difficulties involved in taking the chiral
limit. Studies of this type would be of great interest, especially in
the light of the possible role played by localized modes in
suppressing the finite-temperature Goldstone excitations, suggested in
Ref.~\cite{giordano_GT_lett}. A particularly interesting case would be
that of adjoint massless fermions, for which both chiral and center
symmetries are exact, and an intermediate, deconfined but
chirally broken phase was observed on the lattice for two flavors
in Ref.~\cite{Karsch:1998qj}. This suggests that a nonzero density of
near-zero, localized modes is present in this phase, and that no
Goldstone excitation is present.

Nonetheless, even in theories such as QCD where chiral symmetry is only
approximate, the study of the relation between localized modes and
topological fluctuations of the gauge fields sheds indirectly some
light on the interplay of chiral symmetry and localization.  Indeed, a
peak of
near-zero~\cite{Alexandru:2015fxa,Ding:2020xlj,Kaczmarek:2021ser}
localized~\cite{Dick:2015twa} modes of topological origin appears
around the QCD pseudocritical temperature. These modes originate most
likely from the mixing of the localized zero modes associated with
isolated instantons and anti-instantons, which in the high-temperature
phase form a dilute gas of topological excitations (the disordered
medium scenario~\cite{Diakonov:1984vw,
  Diakonov:1985eg,Diakonov:1995ea,Smilga:1992yp,Janik:1998ki,
  Osborn:1998nm,Osborn:1998nf,GarciaGarcia:2003mn}, see Section
\ref{sec:mech_dms}).  In contrast, in the low temperature phase these
excitations form a dense medium, and the mixing of the associated zero
modes leads to a band of near-zero delocalized modes giving rise to a
nonzero spectral density near the origin, and so a large increase of
the chiral condensate.

An interesting observation is that in the quenched limit of QCD this
peak of near-zero modes can be accurately described in terms of a
non-interacting gas of topological
objects~\cite{Kovacs:2019txb,Vig:2021oyt}. The absence (or near
absence) of interactions could be related to why the associated zero
modes do not mix efficiently, thus remaining localized and failing to
spread in a near-zero band of eigenvalues. On the other hand, as this
occurs only in the high-temperature phase, it is natural to expect
that deconfinement is responsible for the radical change in the
behavior of topological excitations. This aspect surely deserves more
attention.

In this review, we summarized what is known (to us, at least) about
localization of Dirac modes in the deconfined phase of gauge theories,
and highlighted the connections between localized modes, ordering of
the Polyakov loop, density of low modes, and topological objects. We
hope that this will motivate further investigations of the interplay
of confinement, chiral symmetry, topology, and localization in
finite-temperature gauge theories.

%\acknowledgments{MG was partially supported by the NKFIH grant
%  KKP-126769.}

\authorcontributions{Writing---original draft preparation, M.G. and
  T.G.K; writing---review and editing, M.G. and T.G.K. All authors
  have read and agreed to the published version of the manuscript.}

\funding{M.G. was partially supported by the NKFIH grant KKP-126769.}

% \institutionalreview{Not applicable.}

% \informedconsent{Not applicable.}

\acknowledgments{We thank R.~Shindou and M.R.~Zirnbauer for useful
correspondence.}

\conflictsofinterest{The authors declare no conflict of interest.}

\end{paracol}
\reftitle{References}

\externalbibliography{yes}
\bibliography{references_review}

\end{document}